\documentclass[a4paper,fleqn,usenatbib]{mnras}

\usepackage[T1]{fontenc}
\usepackage{ae,aecompl}

\usepackage{amssymb}	% Extra maths symbols
\usepackage[font=small,labelfont=bf]{caption}
\usepackage{amssymb,amsmath}
\usepackage{afterpage} 
\usepackage{amsmath} 
\usepackage{booktabs} 
\usepackage{enumerate} 
\usepackage{rotating} 
\usepackage{graphicx} 
\usepackage{subcaption}
\usepackage{hyperref} 
\usepackage{url}
\usepackage{longtable}
\usepackage{makeidx}
\usepackage{natbib}
\usepackage{multirow}
\usepackage{epstopdf}
\usepackage{caption}
\usepackage[section]{placeins}
\usepackage{setspace}
\usepackage{xspace}
\DeclareCaptionFormat{cont}{#1 (cont.)#2#3\par}
\graphicspath{ {figures/} }

\newcommand\msun {M$_{\odot}$\xspace}

\def\lum{erg s$^{-1}$\xspace}
\def\vel{km s$^{-1}$\xspace}
\def\xmm{XMM-{\it Newton}\xspace}
\def\chan{{\it Chandra}\xspace}
\def\rosat{{\it ROSAT}\xspace}
\def\hst{{\it HST}\xspace}

\title[Final Search for NIR counterparts to ULXs]{NIR Counterparts to ULXs (III): Completing the photometric survey and selected spectroscopic results\thanks{Based on observations made with ESO Telescopes at the La Silla Paranal Observatory under programme 0102.D-0529(A).} }
\author[K. M. L\'{o}pez et al.]{
K. M. L\'{o}pez$^{1,2}$\thanks{Contact e-mail: \href{mailto:K.M.Lopez@sron.nl}{K.M.Lopez@sron.nl}},
M. Heida$^{3,4}$,
P. G. Jonker$^{1,2}$,
M. A. P. Torres$^{5,6,1}$
\newauthor
T. P. Roberts$^{7}$,
D. J. Walton$^{8}$,
D.-S. Moon$^{9}$,
F. A. Harrison$^{4}$
\\
\\
$^{1}$SRON Netherlands Institute for Space Research, 3584 CA Utrecht, The Netherlands
\\
$^{2}$Department of Astrophysics/IMAPP, Radboud University, P.O. Box 9010, 6500 GL Nijmegen, The Netherlands
\\
$^{3}$European Southern Observatory, Karl-Schwarzschild-Stra{\ss}e 2, 85748 Garching bei M\"unchen, Germany
\\
$^{4}$Space Radiation Laboratory, California Institute of Technology, Pasadena, CA 91125, USA
\\
$^{5}$Instituto de Astrof\'{i}sica de Canarias, E-38200 La Laguna, Tenerife, Spain
\\
$^{6}$Departamento de Astrof\'{i}sica, Universidad de La Laguna, Astrof\'{i}sico Francisco S\'{a}nchez s/n, E-38206 La Laguna, Tenerife, Spain
\\
$^{7}$Centre for Extragalactic Astronomy, Department of Physics, University of Durham, South Road, Durham DH1 3LE, United Kingdom
\\
$^{8}$Institute of Astronomy, Cambridge University, Madingley Road, Cambridge CB3 0HA, United Kingdom
\\
$^{9}$Department of Astronomy and Astrophysics, University of Toronto, Toronto, ON M5S 3H4, Canada
}

\date{Accepted 2020 May 31. Received 2020 May 27; in original form 2019 October 29}

\pubyear{2020}

\begin{document}
\label{firstpage}
\pagerange{\pageref{firstpage}--\pageref{lastpage}}
\maketitle

\begin{abstract}
We present results from the remaining sources in our search for near-infrared (NIR)
candidate counterparts to ultraluminous X-ray sources (ULXs) within $\simeq$ 10 Mpc.
We observed 23 ULXs in 15 galaxies and detected NIR candidate counterparts to
six of them. Two of these have an absolute magnitude consistent with a single red supergiant (RSG).
Three counterparts are too bright for a RSG and spatially extended, and thus we classify them as stellar clusters. The other
candidate is too faint for a RSG. Additionally, we present 
the results of our NIR spectroscopic follow-up of five sources: four originally classified as RSG and one as a stellar cluster
on the basis of previous photometry. The stellar cluster candidate is actually a nebula. Of the four RSGs candidates, 
one source has a broad H$\alpha$ emission line redshifted by $\sim z = 1$, making it a background AGN. Two other sources show stellar spectra
consistent with them being RSGs. The final RSG candidate is too faint to classify, but does not show strong (nebular) emission lines in its spectrum.
After our search for NIR counterparts to 113 ULXs, where we detected a candidate counterpart for 38 ULXs, we have spectroscopically 
confirmed the nature of 12: five sources are nebulae, one source is not classified, one source is an AGN and five are RSGs. These possible five ULX-RSG 
binary systems would constitute $\simeq$ $(4 \pm 2)\%$ of the observed ULXs, a fraction almost four times larger than what was predicted by binary 
evolution simulations.
\end{abstract}


%%%%%%%%%%%%%%%%%%%%%%%%%%%%%%%%%%%%%%%%%%%%%%%%%%

%%%%%%%%%%%%%%%%% BODY OF PAPER %%%%%%%%%%%%%%%%%%

\section{Introduction}
\label{intro}

\begin{table*}
\vspace{5mm}
\begin{center}
\caption{Journal of the imaging observations presented in this manuscript.}
\label{tab:galaxies}
\resizebox{\textwidth}{!}{\begin{tabular}{|lccccccccc|}
\hline\hline
Galaxy & Date & Effective & Total & Instrument/ & Band & WCS 1-$\sigma$& Average & Distance & Distance\\
 & observed & exposure & exposure & Telescope & & uncertainty$^c$ & seeing & & reference\\
 & & time$^a$ (sec) & time$^b$ (sec)  & & & ( \arcsec\ ) & ( \arcsec\ ) & (Mpc) & \\
\hline\hline
NGC 55 & 2018 October 22 & 5600 & 5600 & SOFI/NTT & $H$ & 0.30 & 0.8 & 1.96 $\pm$ 0.16 & A\\
NGC 253* & 2017 January 15 & 2920 & 3000 & LIRIS/WHT & $H$ & 0.50 & 2.0 & 3.56 $\pm$ 0.28 & A\\
NGC 253 & 2018 October 20 & 5400 & 5500 & SOFI/NTT & $H$ & 0.50 & 1.1 & 3.56 $\pm$ 0.28 & A\\
NGC 253 & 2018 October 21 & 5400 & 5500 & SOFI/NTT & $H$ & 0.51 & 0.7 & 3.56 $\pm$ 0.28 & A\\
NGC 598 & 2018 October 21 & 5500 & 5500 & SOFI/NTT & $H$ & 0.18  & 0.9 & 0.91 $\pm$ 0.05 & A\\
NGC 628* & 2019 January 24 & 2550 & 2550 & WIRC/Palomar & $K_s$ & 0.26 & 1.2 & 9.77 $\pm$ 0.32 & B\\
NGC 1291 & 2018 October 20 & 750 & 750 & SOFI/NTT & $H$ & 0.19 & 0.6 & 9.08 $\pm$ 0.29 & B\\
NGC 1313 & 2018 October 22 & 5500 & 5500 & SOFI/NTT & $H$ & 0.39 & 0.7 & 4.25 $\pm$ 0.34 & A\\
NGC 3623* & 2018 January 25 & 3380 & 4000 & LIRIS/WHT & $H$ & 0.28 & 1.2 & 12.19 $\pm$ 2.44 & A\\
NGC 3627* & 2019 January 27 & 1910 & 2510 & LIRIS/WHT & $H$ & 0.29 & 1.1 & 10.33 $\pm$ 2.07 & A\\
NGC 3628* & 2018 January 27 & 3690 & 3500 & LIRIS/WHT & $H$ & 0.20 & 1.3 & 10.30 $\pm$ 2.06 & A\\
NGC 4258* & 2018 January 25 & 1860 & 3500 & LIRIS/WHT & $H$ & 0.47 & 1.3 & 7.31 $\pm$ 0.37 & A\\
NGC 4258* & 2019 January 24 & 3870 & 7350 & WIRC/Palomar & $K_s$ & 0.46 & 1.6 & 7.31 $\pm$ 0.37 & A\\
NGC 4631* & 2019 January 25 & 6630 & 7200 & WIRC/Palomar & $K_s$ & 0.38 & 1.2 & 7.35 $\pm$ 0.74 & A\\
NGC 4594* & 2019 January 24 & 1700 & 2000 & LIRIS/WHT & $H$ & 0.20 & 1.3 & 11.27 $\pm$ 1.35 & A\\
NGC 5055* & 2018 January 26 & 2000 & 2000 & LIRIS/WHT & $H$ & 0.35 & 1.3 & 8.87 $\pm$ 0.29 & B\\
NGC 5194* & 2017 January 14 & 2380 & 2500 & LIRIS/WHT & $H$ & 0.77 & 1.2 & 9.05 $\pm$ 0.24 & C\\
NGC 5457 & 2019 January 24 & 3500 & 3500 & LIRIS/WHT & $H$ & 0.26 & 0.9 & 6.95 $\pm$ 0.42 & A\\
\hline\hline
\multicolumn{10}{l}{{\bf Notes:} $^a$:Effective exposure time, not necessarily equal to the total exposure time$^b$ (i.e., images taken during twilight, cloudy intervals were}\\
\multicolumn{10}{l}{rejected and are not counted in the effective exposure time, whereas they are included in the total exposure time). $^c$:Statistical uncertainty}\\
\multicolumn{10}{l}{with respect to the reference catalog, given by the {\scshape starlink gaia} tool. Objects marked with * indicate the images for which sky offsets with total}\\
\multicolumn{10}{l}{time equal to that of the on-source time were taken. References: A: \citet{2013AJ....146...86T}, B: \citet{2017AJ....154...51M} and C: \citet{2015AstL...41..239T}.}\\
\end{tabular}}
\end{center}
\end{table*}

Ever since their discovery in 1981 \citep{1981ApJ...246L..61L}, the nature of ultraluminous X-ray sources 
(ULXs) is puzzling, due to their high X-ray luminosities. ULXs are defined as point-like, off-nuclear 
sources with an X-ray luminosity $L_X > 10^{39}$ \lum, i.e. larger than the Eddington luminosity
for a 10 \msun black hole \citep{2017ARA&A..55..303K}. Several scenarios have been proposed to
explain the high luminosities of these sources. A first scenario considers the possibility of a ULX
being powered by a stellar-mass compact object either emitting anisotropically \citep{2001ApJ...552L.109K},
or accreting at super-Eddington rates \citep{2002ApJ...568L..97B,2003ApJ...586.1280M,2009MNRAS.397.1836G}. 
Examples of the latter include the several neutron star (NS) ULXs discovered in the past 6 years (e.g. 
\citealt{2014Natur.514..202B,2016ApJ...831L..14F,2017Sci...355..817I,2017MNRAS.466L..48I,2018MNRAS.476L..45C,2019MNRAS.488L..35S,2019arXiv190604791R}). 
These systems accrete from different types of stars, i.e. NGC 7793 P13, has a B9 supergiant donor \citep{2011AN....332..367M}; four NS ULXs most likeley accrete
from a young, massive main sequence star \citep{2014Natur.514..202B,2016ApJ...831L..14F,2017Sci...355..817I,2017MNRAS.466L..48I,2019MNRAS.488L..35S,2019arXiv190604791R};
and the remaining NS ULX was recently disovered to accrete from a RSG with an orbital period of $0.8-2.1$ years \citep{2019ApJ...883L..34H}.

The second scenario is that of a ULX powered by a BH more massive than 10 \msun (e.g. \citealt{2009MNRAS.400..677Z}), 
which owing to its higher Eddington luminosity, accretes at or below the Eddington limit. 
The existence of these more massive BHs was proven by the detection of gravitational waves
(e.g. \citealt{2016PhRvX...6d1015A,2016PhRvL.116f1102A,2018arXiv181112907T}), with BH masses between
10 \msun and 80 \msun. In a third scenario, the ULX harbors a BH more massive than the systems discussed
before, but less massive than the super-massive BHs in the centre of massive galaxies \citep{2013ARA&A..51..511K}. 
These intermediate systems, with 
masses between 10$^2$ and 10$^5$ \msun, are called intermediate mass black holes (IMBHs) and they would
accrete at sub-Eddington rates (e.g. \citealt{2010MNRAS.407..645J,2011AN....332..392F,2013MNRAS.436.3128M,2015MNRAS.454L..26H,2016AN....337..448E}).

The only ULXs for which we have reliable accretor mass constraints are the pulsars, which must contain 
neutron stars and therefore accretors with M $< 3$ \msun \citep{1996ApJ...470L..61K}.
Apart from the detection of pulsations, a way to identify the nature of the accretor powering the ULXs is dynamical mass measurements.
Detecting the donor star is the first step
(e.g. \citealt{2008MNRAS.386..543P,2016MNRAS.459..771H,2018ApJ...854..176V,2019ApJ...877...57Q}), as 
subsequent spectroscopic observations can be used to constrain the radial velocity amplitude of the donor star, and hence, 
the compact object mass function (e.g. \citealt{2014Natur.514..198M}). See \citealt{2014SSRv..183..223C}
for a detailed description of the application of this dynamical method to X-ray binaries.

\begin{table*}
\vspace{5mm}
\begin{center}
\caption{Complete sample of the 23 ULXs for which we took images in either $H$- or $K_s$-band.}
\label{tab:ulxs}
\resizebox{\textwidth}{!}{\begin{tabular}{|llccccc|}
\hline\hline
Galaxy & ULX name & \multicolumn{2}{c}{X-ray positional} & X-ray position & X-ray position & Positional \\
& (SIMBAD$^{\ddagger}$) & R.A. & Dec & uncertainty$^a$ & reference & uncertainty$^b$\\
& & (hh:mm:ss) & (dd:mm:ss) & (\arcsec) & &(\arcsec)\\ 
\hline\hline
NGC 55 & [BWE2015] NGC 55 119 & 00:15:28.9 & $-$39:13:18.8 & 0.5 & \citet{2011ApJS..192...10L} & 1.7\\
NGC 253 & [WMR2006] NGC 253 XMM7 & 00:47:09.2 & $-$25:21:23.3 & 0.2 & \citet{2012ApJ...756...27L} & 1.6\\
NGC 253 & RX J004717.4-251811 & 00:47:17.60 & $-$25:18:11.00 & 0.33 & \citet{2004ApJ...602..231C} & 1.8\\
NGC 253 & [WMR2006] NGC 253 XMM4 & 00:47:23.5 & $-$25:19:04.2 & 0.2 & \citet{2012ApJ...756...27L} & 1.6\\
NGC 253 & 2XMM J004820.0-251010 & 00:48:20.1 & $-$25:10:10.3 & 0.4 & \citet{2012ApJ...756...27L} & 1.9\\
NGC 598 & ChASeM33 J013350.89+303936.6 & 01:33:50.9 & 30:39:36.8 & 0.5 & \citet{2011ApJS..192...10L} & 1.6\\
NGC 628* & [KKG2005] M74 X-1 & 01:36:51.1 & 15:45:46.8 & 0.5 & \citet{2011ApJS..192...10L} & 2.7\\
NGC 1291 & CXO J031718.9-410627 & 03:17:18.54 & $-$41:06:29.26 & 0.29 & This work & 1.0\\
NGC 1313 & NGC 1313 X-2 & 03:18:22.27 & $-$66:36:03.72 & 0.23 & This work & 1.4\\
NGC 3623* & CXO J111858.4+130530 & 11:18:58.5 & 13:05:30.9 & 0.5 & \citet{2011ApJS..192...10L} & 1.7\\
NGC 3627* & [SST2011] J112020.90+125846.6 & 11:20:20.9 & 12:58:46.0 & 0.5 & \citet{2011ApJS..192...10L} & 1.7\\
NGC 3628* & CXOU J112037.3+133429 & 11:20:37.37 & 3:34:29.02 & 0.23 & This work & 0.9\\
NGC 4258 & [CHP2004] J121854.5+471649 & 12:18:54.50 & 47:16:49.00 & 0.33 & \citet{2004ApJ...602..231C} & 1.7\\
NGC 4258 & [CHP2004] J121855.8+471759 & 12:18:55.73 & 47:17:59.17 & 0.33 & \citet{Wang_2016} & 1.7\\
NGC 4258 & [CHP2004] J121856.4+472126 & 12:18:56.42 & 47:21:26.70 & 0.47 & \citet{2004ApJ...602..231C} & 2.0\\
NGC 4258 & CXO J121903.84+471832.4 & 12:19:03.87 & 47:18:32.29 & 0.33 & \citet{Wang_2016} & 1.7\\
NGC 4594 & [LB2005] NGC 4594 ULX2 & 12:39:48.62 & $-$11:37:13.10 & 0.47 & \citet{2010ApJ...721.1368L} & 1.5\\
NGC 4631** & [SST2011] J124155.56+323216.9 & 12:41:55.57 & 32:32:16.87 & 0.47 & \citet{2012MNRAS.419.2095M} & 1.8\\
NGC 4631** & CXO J124157.4+323202 & 12:41:57.42 & 32:32:02.98 & 0.47 & \citet{2012MNRAS.419.2095M} & 1.8\\
NGC 4631** & [WMR2006] NGC4631 XMM3 & 12:41:58.03 & 32:28:51.56 & 0.33 & \citet{Wang_2016} & 1.5\\
NGC 5055 & [SST2011] J131519.54+420302.3 & 13:15:19.55 & 42:03:02.30 & 0.47 & \citet{2004ApJS..154..519S} & 1.8\\
NGC 5194 & [CHP2004] J132940.0+471237 & 13:29:40.0 & 47:12:36.9 & 0.5 & \citet{2011ApJS..192...10L} & 2.8\\
NGC 5457 & 2XMM J140228.3+541625 & 14:02:28.30 & 54:16:26.69 & 0.47 & \citet{2012MNRAS.419.2095M} & 1.6\\
\hline\hline
\multicolumn{7}{l}{{\bf Notes:} $^a$1-$\sigma$ uncertainty. $^b$99.7$\%$ (3-$\sigma$) confidence radius of the position of the ULX within which we search for counterparts; this value}\\
\multicolumn{7}{l}{is calculated adding the uncertainties of the astrometric correction on the NIR images (see Table~\ref{tab:galaxies}) and the uncertainties of the X-ray}\\
\multicolumn{7}{l}{position. We added systematic uncertainties linearly and statistical uncertainties quadratically. $^{\ddagger}$Set of Identifications, Measurements and}\\
\multicolumn{7}{l}{Bibliography for Astronomical Data (SIMBAD; \citealt{2000A&AS..143....9W}). Sources analyzed by *\citet{2014MNRAS.442.1054H} and by **\citet{2017MNRAS.469..671L},}\\
\multicolumn{7}{l}{that we re-observe to provide deeper images with better seeing.}\\
\end{tabular}}
\end{center}
\end{table*}

\begin{table*}
\vspace{5mm}
\begin{center}
\caption{Log of the Keck/MOSFIRE spectroscopic observations. Apparent and absolute $H$-band magnitudes from \citet{2014MNRAS.442.1054H} and \citet{2017MNRAS.469..671L} are provided.}
\label{tab:ulxspec}
\resizebox{\textwidth}{!}{\begin{tabular}{|lccccccc|}
\hline\hline
Galaxy & ULX & H & M$_H$ & Observation date & Filter & Time on source & P.A. \\
 & & (mag) & (mag) & & & (h) & (deg) \\
\hline\hline
NGC 1058 & XMMU J024323.5+372038 & $19.7 \pm 0.4$ & $-10.1 \pm 0.4$ & 2019 January 17 & $J$ & 2.4 & 0\\
NGC 2403 & RX J073655.7+653542 & $17.46 \pm 0.02$ & $-10.05 \pm 0.26$ & 2019 January 17 & $J$ & 0.4 & 37\\
NGC 5457 & 2E 1402.4+5440 & $19.3 \pm 0.2$ & $-9.9 \pm 0.2$ & 2019 May 23 & $H$ & 2.4 & 35\\
NGC 5457 & [LB2005] NGC 5457 X26* & $17.78 \pm 0.01$ & $-11.44 \pm 0.22$ & 2019 May 23 & $H$ & 2.4 & 35\\
NGC 5457 & CXOU J140314.3+541807 & $18.35 \pm 0.10$ & $-10.69 \pm 0.10$ & 2019 July 23 & $H$ & 1.0 & 50\\
\hline\hline
\multicolumn{8}{l}{*Source originally classified as a stellar cluster. In our multi-object spectroscopic observations we took a spectrum of this source}\\
\multicolumn{8}{l}{at the same time as the spectrum of 2E 1402.4+5440 was obtained as it was in the field of view.}\\
\end{tabular}}
\end{center}
\end{table*}

Multi-wavelength searches for the donor stars have been pursued for more than a decade. Some have observed
ULXs in the optical range (e.g. \citealt{2006ApJS..166..154P,2006IAUS..230..310G,2011AN....332..398R,2013ApJS..206...14G,2015NatPh..11..551F}).
 As several of the ULXs are located in or near young star clusters (e.g. \citealt{2001ApJ...554.1035F,2002MNRAS.337..677R,2013MNRAS.432..506P}), 
some donor stars of ULXs might as well be red supergiants (RSGs, \citealt{2005MNRAS.362...79C,2007MNRAS.376.1407C,2008MNRAS.386..543P}), 
which are bright in the near-infrared (NIR) band.
In light of this, we started an observing campaign of ULXs in the NIR \citep{2014MNRAS.442.1054H,2017MNRAS.469..671L}, 
tailored to detect RSG donor stars ($-8 < H, K_s < -11$, \citealt{1985ApJS...57...91E,2000asqu.book..381D}). We targeted sources 
at distances up to 10 Mpc, within uncertainties. Of the 97 ULXs observed by \citet{2014MNRAS.442.1054H} and 
\citet{2017MNRAS.469..671L}, we detected a NIR candidate counterpart for 33 ULXs. Of these 33 NIR sources, 
19 had the absolute magnitude consistent with a RSG. So far, we performed spectroscopic follow-up observations for seven out of 
these 19 candidate RSGs and confirmed the RSG nature for three counterparts: ULX RX J004722.4-252051 
(in NGC 253, \citealt{2015MNRAS.453.3510H}), ULX J022721+333500 (in NGC 925) and ULX J120922+295559 
(in NGC 4136, \citealt{2016MNRAS.459..771H}). Additionally, we discovered that four counterparts, initially 
classified as RSG based on photometry, are actually nebulae partially powered by the X-ray emission of the ULX: ULX J022727+333443 
(in NGC 925), ULX J120922+295551 (in NGC 4136), ULX Ho II X-1 (in Holmberg II, \citealt{2016MNRAS.459..771H}) 
and [SST2011] J110545.62+000016.2 (in NGC 3521, \citealt{2019MNRAS.489.1249L}). This means that, at the time of
writing this manuscript, the photometric and spectroscopic observations provided three confirmed RSGs, 12 candidate RSGs, 
four confirmed nebulae and 14 candidate stellar clusters/AGNs. Recently, \citet{2019ApJ...878...71L} observed 96 ULX within 10 Mpc 
in the mid-IR with the {\it Spitzer Space Telescope}. They detected identified 12 ULXs with candidate counterparts whose absolute mid-IR 
magnitude is consistent with being RSGs. 

In this manuscript we present the final results of our imaging campaign as well as results of our spectroscopic follow-up with Keck/MOSFIRE 
of four candidate RSGs identified earlier in our campaigns by \citet{2014MNRAS.442.1054H} and \citet{2017MNRAS.469..671L}. 
We describe the observed sample in Section~\ref{Sample}, 
the NIR (imaging and spectral) observations in Section~\ref{nirobs} and data reduction/photometry in Section~\ref{datared}. 
Our results are presented and discussed in Section~\ref{results}. We summarize the results of the
8-year-long systematic photometric search in Section~\ref{recap} and we end with the conclusions of our work in Section~\ref{conclusions}.

\section{Sample}
\label{Sample}

For our imaging observations, our sample consists of 23 ULXs located in 15 different galaxies 
within 10 Mpc from our Galaxy (see Table~\ref{tab:galaxies}).
This sample completes the NIR imaging campaign started by \citet{2014MNRAS.442.1054H}
and continued by \citet{2017MNRAS.469..671L}. Four ULXs were observed before
by \citet{2014MNRAS.442.1054H} and three ULXs were observed before by \citet{2017MNRAS.469..671L},
but we observe them again under better sky conditions to make the survey as homogeneous in depth as possible.
Several of the sources were originally discovered with the {\it R\"ontgensatellit}
(\rosat), three sources ([WMR2006] NGC253 XMM7, [WMR2006] NGC253 XMM4 and 2XMM J004820.0-251010)
have been re-observed by the X-ray Multi-Mirror Mission (\xmm) and the rest by the \chan X-ray Observatory,
offering an improved X-ray position (see Table~\ref{tab:ulxs}). Note that we report the 3-$\sigma$ confidence radius
of the position of the ULX which takes into account the X-ray positional uncertainty of the ULX and the WCS uncertainty
of the NIR image (see Section~\ref{imaging}).

Additionally, we obtain NIR spectra of five ULX candidate counterparts (see Table~\ref{tab:ulxspec}). Three sources 
(XMMU J024323.5+372038 in NGC 1058, and 2E 1402.4+5440 and CXOU J140314.3+541807 in NGC 5457) were 
photometrically identified and classified by \citet{2014MNRAS.442.1054H} as potential RSGs. The remaining two sources were 
photometrically identified and classified by \citet{2017MNRAS.469..671L}: one as a potential RSG (RX J073655.7+653542 in NGC 2403) 
and one as a stellar cluster ([LB2005] NGC 5457 X26). The latter is located in the field of view from the ULX 2E 1402.4+5440.

\section{Observations}
\label{nirobs}

\subsection{NIR Imaging}
\label{nirobsim}

$H$-band images of regions of galaxies containing ULXs were obtained with the 
Long-slit Intermediate Resolution Infrared Spectrograph (LIRIS, \citealt{1998SPIE.3354..448M}) 
mounted on the 4.2-m William Herschel Telescope (WHT) and the Son OF Isaac (SOFI, 
\citealt{1998Msngr..91....9M}) mounted on the 3.6-m New Technology Telescope (NTT); whereas 
$K_s$-band imaging was obtained with the Wide Field Infrared Camera (WIRC) mounted 
on the Palomar Hale 5-m Telescope. LIRIS has a field of view of 
4.27\arcmin\ $\times$ 4.27\arcmin\ and a pixel scale of 0.25\arcsec\ /pixel.
SOFI has a field of view of 4.92\arcmin\ $\times$ 4.92\arcmin\ and a pixel 
scale of 0.288\arcsec\ /pixel and WIRC has a field of view of 8.7\arcmin\ $\times$ 8.7\arcmin\
and a pixel scale of 0.2487\arcsec\ /pixel.

The LIRIS observations were performed using 7 or 8 repetitions of a 5-point dither 
pattern where 5 images (10 or 20 seconds exposure per image) were taken at each point.
For the observations done with SOFI, we used the template SOFI\_img\_obs\_AutoJitter,
in which an observing cycle of 10 images (5 or 10 seconds exposure per image) is
repeated 55 (or 110) times. The WIRC observations consisted of a 5-point dither pattern
where 6 images of 5 seconds exposure each were taken at each point. This pattern was repeated
17 times for NGC 628, 48 times for NGC 4631 and 49 times for NGC 4258.
Sky offsets were taken if the field of view was crowded (see Table~\ref{tab:galaxies}, where we also
indicate the filter and instrument/telescope used, the average seeing during the observation
and the distance to each galaxy).

\begin{table}
\vspace{5mm}
\begin{center}
\caption{The ULX NIR candidate counterparts for which we analysed archival \hst observations.}
\label{tab:hstobsids}
%\resizebox{\textwidth}{!}{
\begin{tabular}{|lcl|}
\hline\hline
 ULX name & \hst  & Filter\\
 & obs. ID &  \\
\hline\hline
 2XMM J004820.0-251010 & 10915 & ACS/WFC/F606W\\
 & & ACS/WFC/F814W\\
 NGC 1313 X-2 & 14057 & WFC3/UVIS/F555W \\
 & & WFC3/UVIS/F814W \\ 
 & & WFC3/IR/F125W \\
 RX J073655.7+653542 & 11719 & WFC3/IR/F110W\\
 & & WFC3/IR/F160W\\
 CXOU J140314.3+541807 & 9490 & ACS/WFC/F555W\\
 & & ACS/WFC/F814W\\
\hline\hline
\end{tabular}%}
\end{center}
\end{table}

\subsection{NIR Spectra}
\label{nirobsspec}

We obtained $J$- and/or $H$-band spectroscopy of five ULX counterparts with the Multi-Object 
Spectrometer for Infra-Red Exploration (MOSFIRE, \citealt{2010SPIE.7735E..1EM,2012SPIE.8446E..0JM}), 
mounted on the 10-m Keck I telescope on Mauna Kea. MOSFIRE has a field of view of 6.1\arcmin\ $\times$ 6.1\arcmin, a pixel
scale of 0.18\arcsec\ /pixel and a robotic slit mask system with 46 reconfigurable slits. 
XMMU J024323.5+372038 and RX J073655.7+653542 were observed on 2019 January 17 with an 
0.7\arcsec\ slit in the $J$-band, with seeing varying between $0.4-0.7$\arcsec. The ULXs in NGC 5457 were observed with an 
0.7\arcsec\ slit in the $H$-band; 2E 1402.4+5440 and [LB2005] NGC 5457 X26 on 2019 May 23 
with a seeing of $\sim 0.5$\arcsec, and CXOU J140314.3+541807 on 2019 July 13 with a seeing of $\sim 0.8$\arcsec .
For all sources we used an ABBA nodding pattern with a nod amplitude along the slit chosen to avoid nearby sources. 
The integration time per exposure was 119.3 s for all observations. The spectral resolution R and spectral coverage $\Delta \lambda$ for 
the $J$-band observation are R $= 3100$ and $\Delta \lambda = 11530-13520$ \AA; for the $H$-band observations, 
R $= 3660$ and $\Delta \lambda = 14500-18220$ \AA . The total exposure times are listed in Table~\ref{tab:ulxspec}. We observed 
A0V-type telluric standard stars at similar airmass as the science targets before and after every series of exposures. These were also 
used for flux calibration. For every slit mask configuration, we obtained flat-field and comparison lamp spectra in the afternoon before 
the observing night.

\section{Data reduction}
\label{datared}

\subsection{NIR Imaging}
\label{imaging}

The data reduction was performed using the {\scshape theli} pipeline \citep{2013ApJS..209...21S}.
To flat-field correct the NIR data, {\scshape theli} produces a master flat by median combining
sky images taken during twilight. Then, median combining the data frames without correcting for 
the offsets introduced by the dithering, it generates a sky background model which is subtracted
from each individual image. After this, {\scshape theli} detects sources in each image using
{\scshape SExtractor} \citep{1996A&AS..117..393B} and matches them to sources from the 2 Micron All Sky Survey (2MASS; 
\citealt{2006AJ....131.1163S}) with {\scshape scamp} \citep{2006ASPC..351..112B}. This creates an astrometric solution
which then is used for the coaddition of all the data frames using {\scshape swarp} 
\citep{2002ASPC..281..228B}.

\begin{table*}
\vspace{5mm}
\begin{center}
\caption{NIR candidate counterparts to the ULXs listed in Table~\ref{tab:ulxs} and our preliminary classification as candidate RSG, stellar cluster (SC) or (N/A) without classification. The preliminary classification is based on their absolute magnitude, WISE colours, visual inspection of the NIR image and/or spatial extent of the candidate counterpart.}
\label{tab:candidates}
\resizebox{\textwidth}{!}{\begin{tabular}{|llccccccc|}
\hline\hline
Galaxy & ULX name & R.A & Dec. & Positional$^a$ & Filter & Apparent & Absolute$^b$ & Preliminary\\
& & & & uncertainty & &  magnitude & magnitude & classification\\
& & (hh:mm:ss) & (dd:mm:ss) & ( \arcsec\ ) & & (mag) & (mag) &\\ 
\hline\hline
NGC 55 & $[$BWE2015$]$ NGC 55 119 & $-$ & $-$ & $-$ & $H$ & $> 20.5$ & $> -5.9$ & $-$\\
NGC 253 & $[$WMR2006$]$ NGC 253 XMM7 & $-$ & $-$ & $-$ & $H$ & $> 18.7$ & $> -9.4$ & $-$\\
NGC 253 & RX J$004717.4-251811$ & $-$ & $-$ & $-$ & $H$ & $> 17.3$ & $> -10.8$ & $-$\\
NGC 253 & $[$WMR2006$]$ NGC 253 XMM4& $-$ & $-$ & $-$ & $H$ & $> 16.1$ & $> -12.0$ & $-$\\
NGC 253 & 2XMM J$004820.0-251010$ & 00:48:19.9 & $-$25:10:10.3 & 1.6 & $H$ & 20.4 $\pm$ 0.1 & $-$7.4 $\pm$ 0.2 & N/A\\
NGC 598 & ChASeM33 J$013350.89+303936.6$ & 01:33:50.9 & +30:39:36.6 & 0.6 & $H$ & 11.8 $\pm$ 0.1 & $-$13.0 $\pm$ 0.1 & SC\\
NGC 628 & $[$KKG2005$]$ M74 X-1& $-$ & $-$ & $-$ & $K_s$ & $> 19.5$ & $> -10.4$ & $-$\\
NGC 1291 & CXO J$031718.9-410627$ & 03:17:18.6 & $-$41:06:28.9 & 0.6 & $H$ & 11.3 $\pm$ 0.1 & $-$18.4 $\pm$ 0.1 & SC\\
NGC 1313 & NGC 1313 X-2 & 03:18:22.4 & $-$66:36:04.0 & 1.2 & $H$ & 21.1 $\pm$ 0.2 & $-$7.1 $\pm$ 0.2 & RSG\\
NGC 3623 & CXO J$111858.4+130530$ & $-$ & $-$ & $-$ & $H$ & $> 18.9$ & $> -11.5$ & $-$\\
NGC 3627 & $[$SST2011$]$ J$112020.90+125846.6$ & $-$ & $-$ & $-$ & $H$ & $> 19.0$ & $> -11.1$ & $-$\\
NGC 3628 & CXOU J$112037.3+133429$ & 11:20:37.3 & +13:34:28.9 & 0.7 & $H$ & 20.4 $\pm$ 0.2 & $-$9.7 $\pm$ 0.5 & RSG\\
NGC 4258 & $[$CHP2004$]$ J$121854.5+471649$ & $-$ & $-$ & $-$ & $H$ & $> 19.6$ & $> -9.7$ & $-$\\
NGC 4258 & $[$CHP2004$]$ J$121855.8+471759$ & $-$ & $-$ & $-$ & $H$ & $> 18.2$ & $> -11.1$ & $-$\\
NGC 4258 & $[$CHP2004$]$ J$121856.4+472126$ & $-$ & $-$ & $-$ & $K_s$ & $> 19.5$ & $> -9.8$ & $-$\\
NGC 4258 & CXO J$121903.84+471832.4$ & $-$ & $-$ & $-$ & $H$ & $> 19.0$ & $> -10.3$ & $-$\\
NGC 4594 & $[$LB2005$]$ NGC 4594 ULX2 & $-$ & $-$ & $-$ & $H$ & $> 19.3$ & $> -10.9$ & $-$\\
NGC 4631 & $[$SST2011$]$ J$124155.56+323216.9$ & $-$ & $-$ & $-$ & $K_s$ & $> 18.8$ & $> -10.5$ & $-$\\
NGC 4631 & CXO J$124157.4+323202$ & $-$ & $-$ & $-$ & $K_s$ & $> 17.9$ & $> -11.4$ & $-$\\
NGC 4631 & $[$WMR2006$]$ NGC4631 XMM3 & $-$ & $-$ & $-$ & $K_s$ & $> 20.1$ & $> -9.2$ & $-$\\
NGC 5055 & $[$SST2011$]$ J$131519.54+420302.3$ & $-$ & $-$ & $-$ & $H$ & $> 20.1$ & $> -9.6$ & $-$\\
NGC 5194 & $[$CHP2004$]$ J$132940.0+471237$ & $-$ & $-$ & $-$ & $H$ & $> 19.8$ & $> -10.0$ & $-$\\
NGC 5457 & 2XMM J$140228.3+541625$ & 14:02:28.2 & +54:16:26.3 & 0.8 & $H$ & 16.0 $\pm$ 0.1 & $-$13.2 $\pm$ 0.1 & SC\\
\hline\hline
\multicolumn{9}{l}{{\bf Notes:} $^a$:99.7$\%$ (3-$\sigma$) confidence radius of the position of the NIR candidate counterpart; this value is calculated taking into account the}\\
\multicolumn{9}{l}{uncertainties of the astrometric correction on the NIR images and the uncertainties given by the source detection by {\scshape SExtractor}. We added}\\
\multicolumn{9}{l}{systematic uncertainties linearly and statistical uncertainties quadratically. $^b$:Values calculated using the distance from Table~\ref{tab:galaxies} and the}\\
\multicolumn{9}{l}{apparent magnitude, with a 1-$\sigma$ uncertainty.}\\
\end{tabular}}
\end{center}
\end{table*}

We improved the accuracy of the global astrometric solution of the coadded NIR images using the {\scshape starlink}
tool {\scshape gaia}, fitting several isolated point sources from Gaia Data Release 2 (DR2, \citealt{2018A&A...616A...1G}). We only used stars 
with a proper motion pm $< 5$ mas yr$^{-1}$ to build a local astrometric solution around the position of the ULX.
The rms errors of these fits are listed in Table~\ref{tab:galaxies}, indicated as WCS (World Coordinate System) uncertainties.

\subsection{NIR Spectra}
\label{spectra}

We reduced the MOSFIRE spectra with the MOSFIRE data reduction pipeline (version 2018)\footnote{\url{https://github.com/Keck-DataReductionPipelines/MosfireDRP}}.
The pipeline outputs co-added, rectified 2D-spectra that are wavelength calibrated with an RMS $< 0.1$ \AA. We used the {\sc Figaro} package \citep{1993ASPC...52..219S} to optimally extract the spectra and the {\sc Molecfit} software \citep{2015A&A...576A..78K,2015A&A...576A..77S} to correct for telluric absorption. With {\sc Molecfit}, we performed fits of regions with strong telluric features in the high signal-to-noise ratio (S/N) spectra of our A0V telluric standard stars to determine the atmospheric conditions and used those solutions to correct the science spectra, only fitting for the local continuum and correcting for changes in airmass. We used the same A0V telluric standards to correct for the instrument response and to provide an approximate flux calibration. We did not correct for slit losses, but expect these to be small as our slit width matched or exceeded the seeing during most of our observations. 

\subsection{Photometry}
\label{Photometry}

\subsubsection{NIR images}
\label{nirphotometry}

We performed aperture photometry to determine instrumental magnitudes. 
We used {\scshape SExtractor} for the source detection and photometry, 
making sure that each detection was more than 3-$\sigma$ above 
the local background. As the aperture radius, we use the average full-width at half maximum (FWHM) 
of the point-like objects in each image determined with the {\scshape starlink} tool {\scshape gaia}. 
Instrumental magnitudes for all detected sources were converted to apparent magnitudes using
photometric zero points that we measured using isolated 2MASS objects in the field of view.
Then, we determined the absolute magnitudes of candidate counterparts to ULXs using the distances given in Table~\ref{tab:galaxies}.
The (1-$\sigma$) uncertainties on these values are estimated taking into account the distance uncertainty, the uncertainty
in the determination of the zero point magnitude and the uncertainty in the instrumental magnitude given by {\scshape SExtractor}.
For the images in which we did not detect a counterpart, we estimated the limiting magnitude by
simulating 10,000 stars at the position of the ULX using the {\scshape Iraf}\footnote{IRAF is distributed by the National Optical Astronomy Observatories, which are operated by the Association of Universities for Research in Astronomy, Inc., under cooperative agreement with the National Science Foundation.} 
task {\scshape Mkobjects} and then detecting them with the task {\scshape Daofind}. Our limiting magnitude is the magnitude
at which 99.7$\%$ of the simulated stars are detected.

\subsubsection{\hst images}
\label{hstphotometry}

For the four sources for which we analyse archival \hst data (see Table~\ref{tab:hstobsids}), 
we also improved the accuracy of the astrometry using the {\scshape starlink} tool {\scshape gaia} and isolated Gaia DR2 sources.
Additionally, we performed aperture photometry on these \hst images with {\scshape SExtractor}. To derive 
the magnitude we transform the instrumental magnitudes into apparent magnitudes using the 
{\scshape vegamag} zero points of the corresponding \hst filter as advised by \citet{2005PASP..117.1049S} 
and \citet{2009wfc..rept...21K}. To correct for extinction on these data, we assumed $R_V = 3.1$ 
\citep{Fitzpatrick_1999}, $R_R = 2.31$, $R_I = 1.71$, $R_J = 0.72$, $R_H = 0.306$ \citep{10.1093/mnras/stt039} and 
N$_H = 1.87 \times 10^{21}$ atoms cm$^{-2}$ mag$^{-1} A_V$ \citep{1978ApJ...224..132B}.

\subsection{X-ray astrometry}
\label{xrayastro}

For all the ULXs in our sample accurate positions exist in the literature (see Table~\ref{tab:ulxs}). However,
we were able to improve the accuracy of the positions of 3 ULXs (CXO J031718.9-410627, NGC 1313 X-2 and CXOU J112037.3+133429)
using archival Chandra/ACIS observations. For CXO J031718.9-410627 we used observation ID 2059, for NGC 1313 X-2 we used
observation ID 4750 and for CXOU J112037.3+133429 we used observation ID 2919.

We used the task {\scshape acis-process-events} in CIAO \citep{ciao} to reprocess the event files with the
calibration files (CALDB version 4.9.0) taking into account whether the observations were made in the "Faint" or "Very Faint" mode.
We then produced images from data in the $0.3-7$ keV energy range, on which we run the {\scshape wavdetect} task \citep{ciao} to
establish accurate positions of all X-ray sources in the field of view of the Chandra ACIS CCDs. With this procedure we
improved the X-ray position of ULXs NGC 1313 X-2 and CXOU J112037.3+133429 (see Table~\ref{tab:ulxs}).

For the ULX in NGC 1291, CXO J031718.9-410627, we were able to further improve the knowledge of the location of
the source by applying a boresight correction (e.g. \citealt{2010MNRAS.407..645J,2017MNRAS.469..671L}). For this, we select only X-ray sources
detected with more than 20 X-ray counts and that lie within $3$\arcmin of the optical axis of the satellite. 
When found, we investigated whether counterparts in the Gaia DR2 catalogue exist. We take the Gaia DR2 source as a counterpart if it lies within 1\arcsec from
the position of the X-ray source. Subsequently, we determine the offsets between the RA and Dec. of the new and old source and we then apply shifts in RA and Dec. 
to the X-ray coordinates using the {\scshape wcs-update} tool. The updated coordinates are indicated in Table~\ref{tab:ulxs}.

\section{Results and discussion}
\label{results}

Of the 23 ULXs we observed, we detect a NIR candidate counterpart
for six of them (about 30$\%$). Their apparent and absolute magnitudes are detailed in 
Table~\ref{tab:candidates}. For the cases where no counterpart is detected, 
the apparent (absolute) limiting magnitude of the NIR image at the position of the ULX 
is indicated, and it ranges from $16.1$ to $20.5$ ($-12.0$ to $-5.9$) mag.

Out of the six NIR candidate counterparts, three have absolute magnitudes in the $H$-band from 
$-13.0 \pm 0.1$ to $-18.4 \pm 0.1$, i.e. too bright to be a single star. Hence, these 
sources are most likely unresolved groups of stars, perhaps a cluster of stars. Even though we do not know whether
these clusters are gravitationally bound, i.e. whether they are stellar clusters, we do refer to them this way. One source
is too faint to be a RSG so we do not give any classification to it.
The remaining two sources have absolute magnitudes in the $H, K_s$  consistent with those of a RSG, within uncertainties 
\citep{1985ApJS...57...91E,2000asqu.book..381D}. Below we discuss in detail each of the detected counterparts, providing 
some clues on their possible nature, based on their photometry and their spatial properties.

On the other hand, of our five sources spectroscopically observed, we confirm two RSG candidates as RSGs based on their
stellar spectra and we reveal one RSG candidate to be an AGN. The SC candidate turns out to be a nebula and the last RSG
candidate cannot be classified based on its spectrum, as no significant absorption or emission lines were detected.

\begin{figure*}
  \begin{minipage}{0.333\textwidth}
  \includegraphics[width=0.98\textwidth]{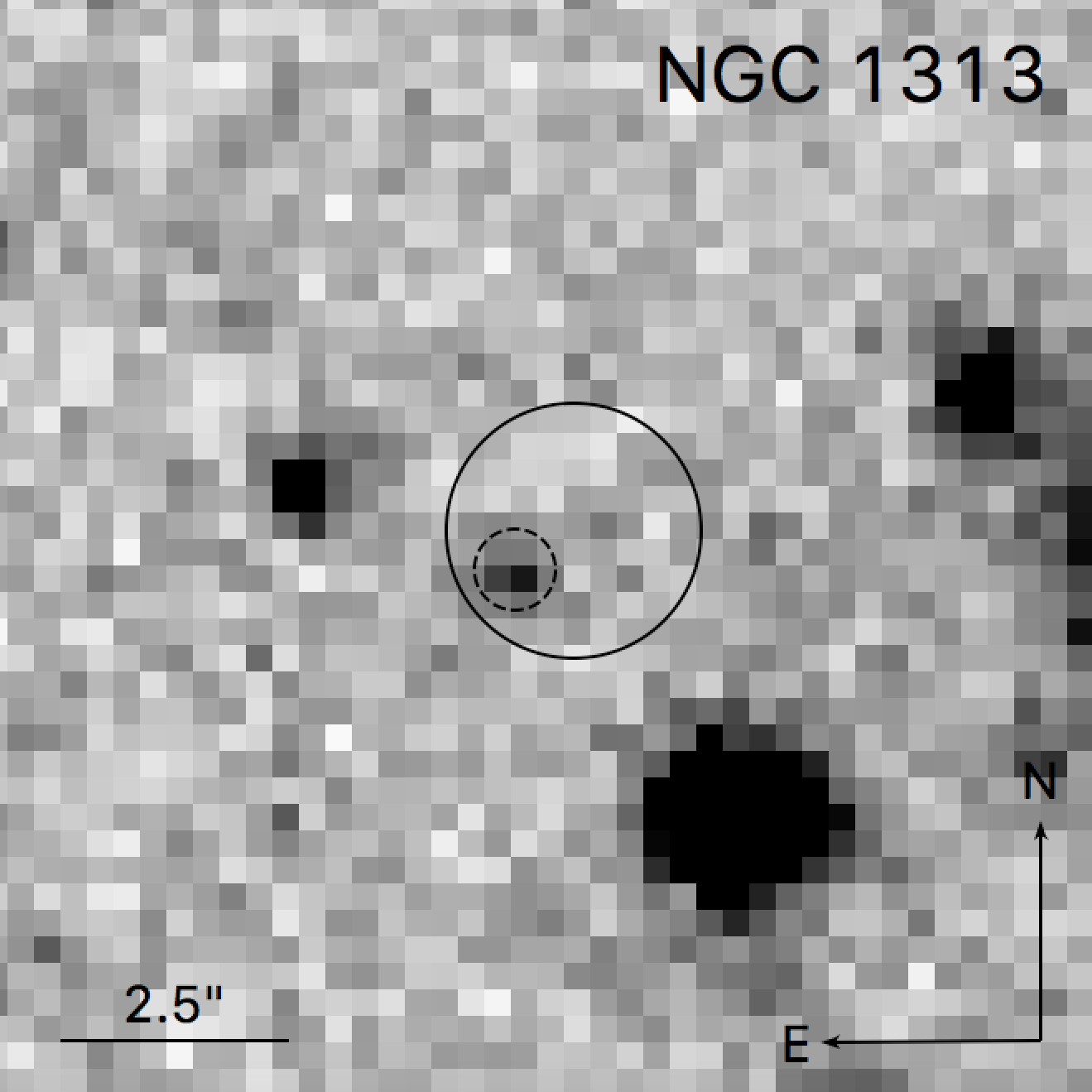}%
  \subcaption{NGC 1313 X-2} 
  \vspace{0.5cm}
  \label{fig:ngc1313-2}   
 \end{minipage}%
   \begin{minipage}{0.333\textwidth}
  \includegraphics[width=0.98\textwidth]{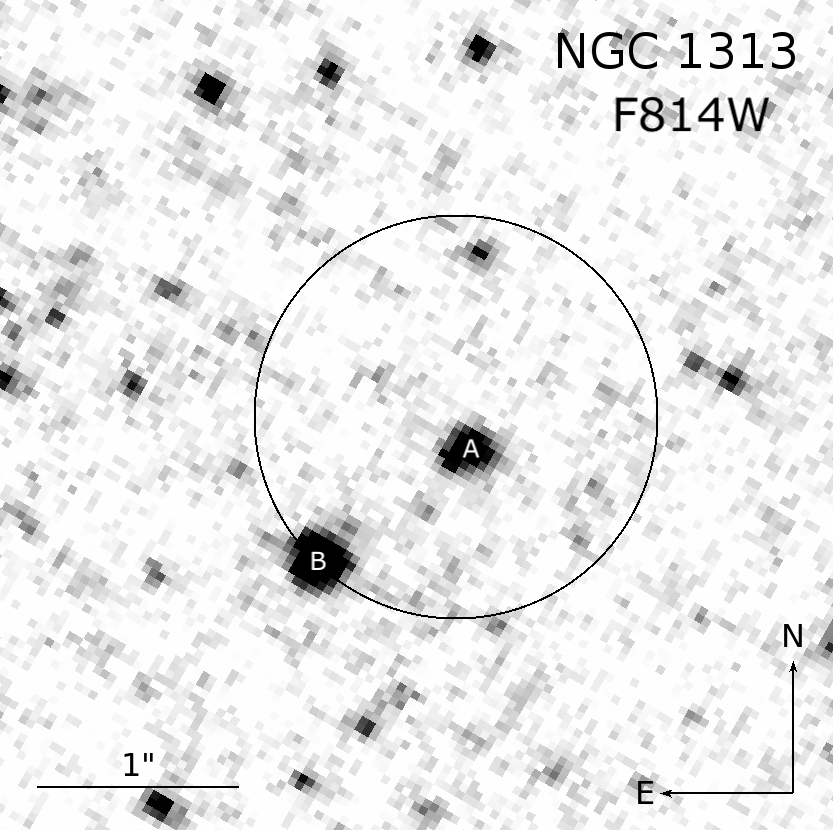}%
  \subcaption{NGC 1313 X-2 (\hst)} 
  \vspace{0.5cm}
  \label{fig:1313hst}   
 \end{minipage}%
 \begin{minipage}{0.333\textwidth}
  \includegraphics[width=0.98\textwidth]{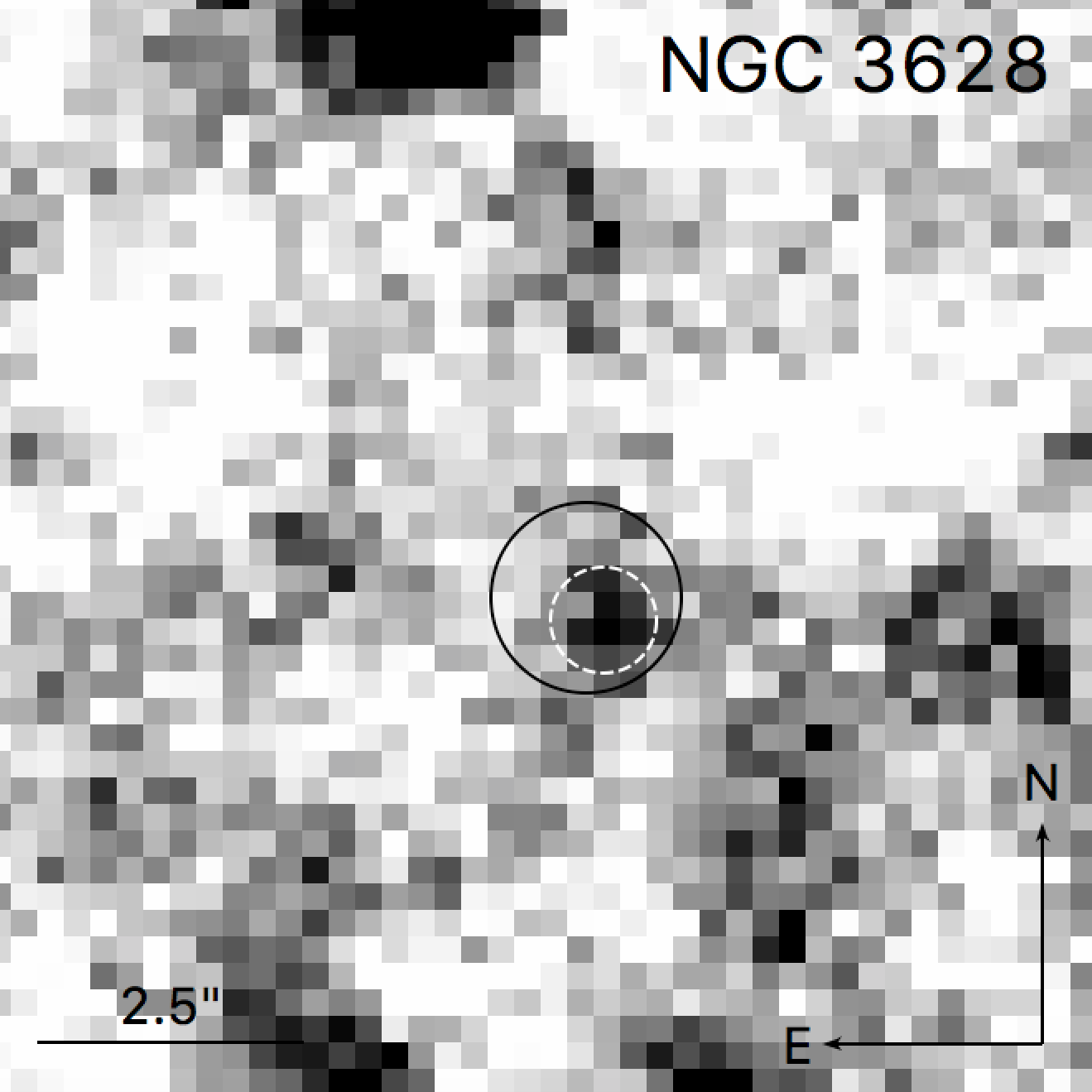}%
  \subcaption{CXOU J112037.3+133429}
  \label{fig:ngc3628}
  \vspace{0.5cm}   
 \end{minipage}
 \begin{minipage}{0.333\textwidth}
  \includegraphics[width=0.98\textwidth]{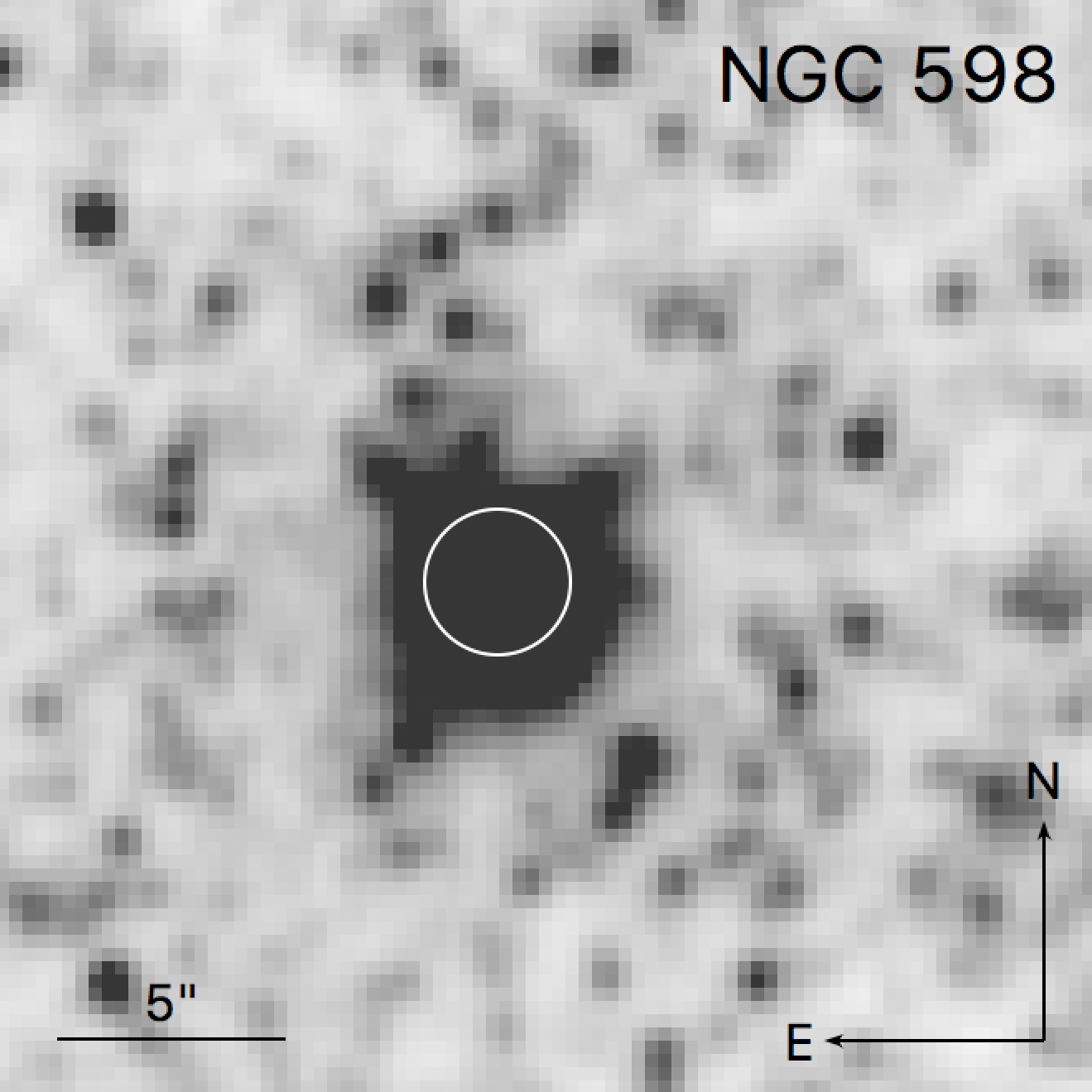}%
  \subcaption{ChASeM33 J013350.89+303936.6} 
  \label{fig:ngc598}
  \vspace{0.5cm}   
 \end{minipage}%
 \begin{minipage}{0.333\textwidth}
  \includegraphics[width=0.98\textwidth]{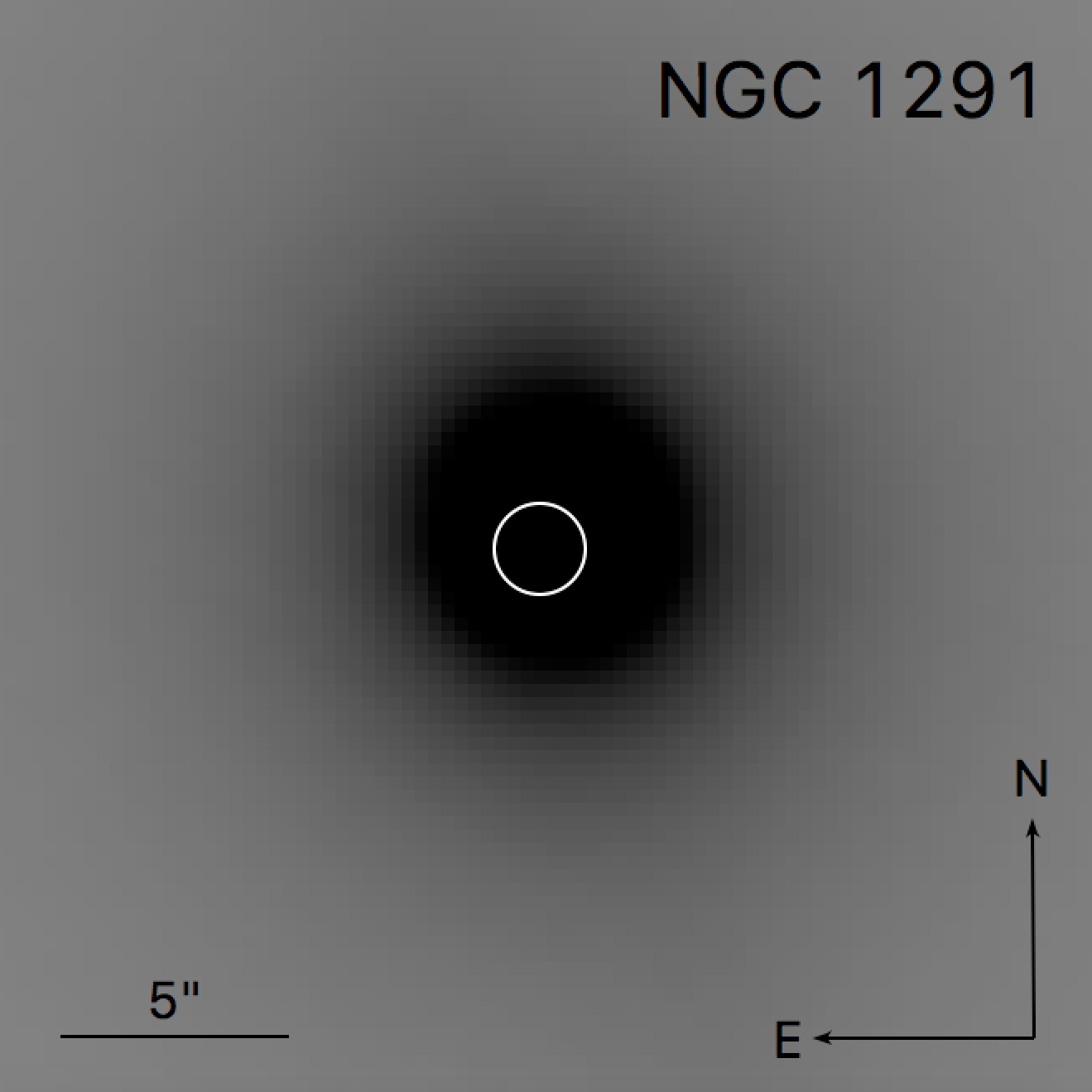}%
  \subcaption{CXO J031718.9-410627}
  \vspace{0.5cm} 
  \label{fig:ngc1291}   
 \end{minipage}%
  \begin{minipage}{0.333\textwidth}
  \includegraphics[width=0.98\textwidth]{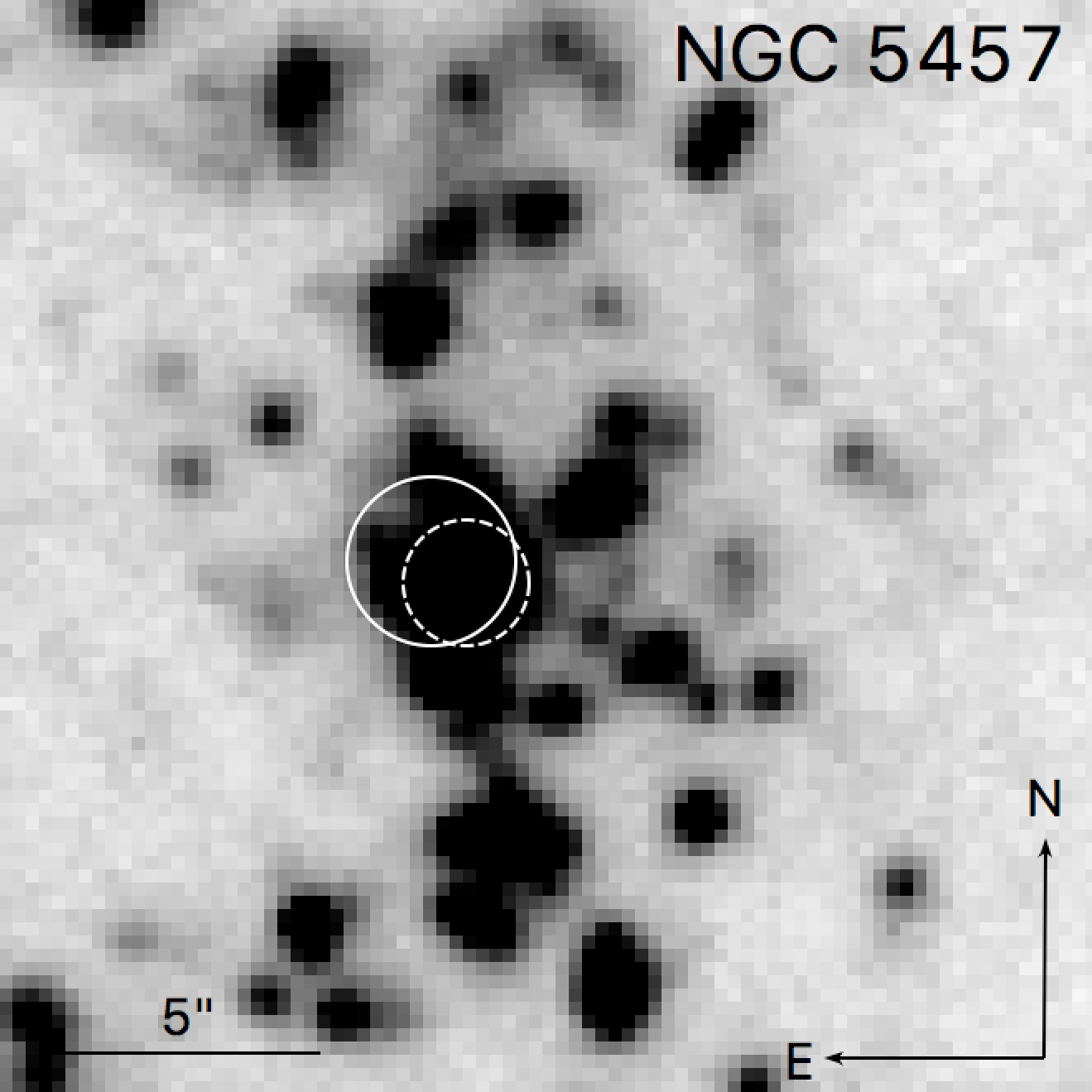}%
  \subcaption{2XMM J140228.3+541625}
  \label{fig:ngc5457} 
  \vspace{0.5cm}   
 \end{minipage}
  \begin{minipage}{0.333\textwidth}
  \includegraphics[width=0.98\textwidth]{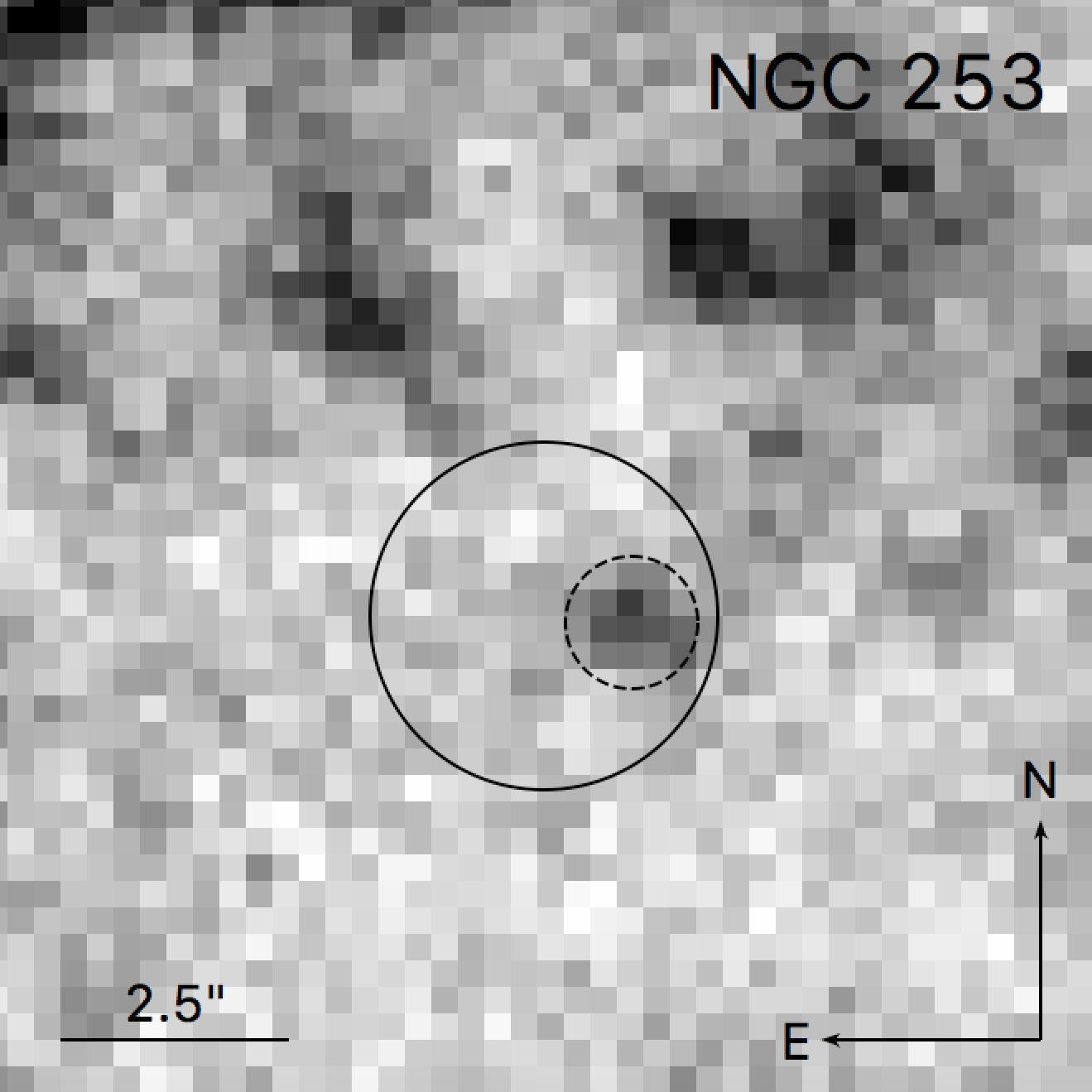}%
  \subcaption{2XMM J004820.0-251010} 
  \label{fig:ngc253}
  \vspace{0.5cm}  
 \end{minipage}%
   \begin{minipage}{0.333\textwidth}
  \includegraphics[width=0.98\textwidth]{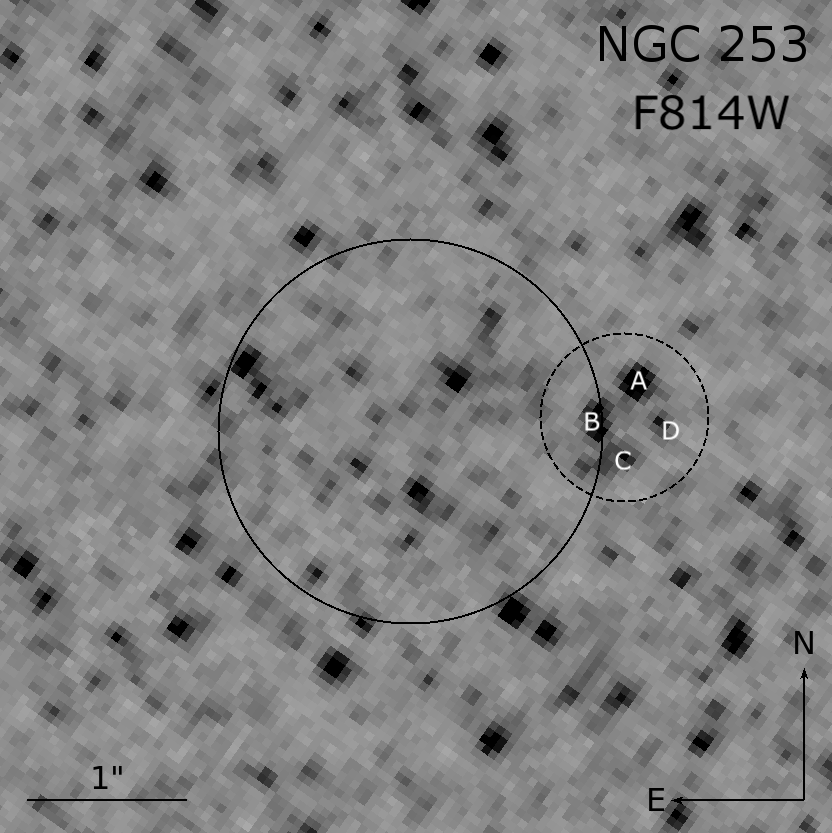}%
  \subcaption{2XMM J004820.0-251010 (\hst)} 
  \label{fig:253hst}
  \vspace{0.5cm}  
 \end{minipage}%
	\caption{Finder charts of the ULXs with a NIR candidate counterpart. The black/white solid circles correspond to the 99.7$\%$ confidence region of the position of the ULXs, whereas the black/white dashed circles mark the candidate counterpart as detected with {\scshape SExtractor}. Images are in the $H$-band, unless indicated otherwise.}
 \label{fig:images} 
\end{figure*}

\subsection{NIR Red supergiant candidates}
\label{RSG}

\subsubsection{NGC 1313 X-2}
\label{ulx10}

This ULX has been observed several times, both in X-ray and in the optical (e.g.
\citealt{1995AJ....109.1199S,2004ApJ...603..523Z,2005ApJ...633L.101M,2007ApJ...661..165L}).
It is now one of the six NS ULX  on the basis of the detection of pulsations \citep{2019MNRAS.488L..35S}.
\citet{2004ApJ...603..523Z} first identified an optical counterpart for this source and classified it as an evolved 
OB supergiant. On the other hand, \citet{2005ApJ...633L.101M} resolved this counterpart into two sources 
(0.75\arcsec apart) and classified them as a B0-O9 main sequence star and a G supergiant. Moreover, 
they concluded that NGC 1313 X-2 could be associated with either one of them. While \citet{2006ApJ...641..241R} 
identified the B type star as the counterpart to the ULX using \hst data, \citet{2007ApJ...661..165L} argued that the 
spectral type of the counterpart is actually O7V. Additionally, they found a third source within the $1-\sigma$ confidence radius that was then 
ruled out as a cosmic ray. Although this source has been spectroscopically followed-up, no radial
velocity measurement or mass function has been established 
(see \citealt{2006IAUS..230..293P,2009AIPC.1126..201G,2011AN....332..398R}.) In fact, it was suggested that the optical 
emission may not come from the counterpart, but from the accretion disk \citep{2011AN....332..398R}. We analyse the archival 
\hst images of this source in the F555W, F814W and F125W filters. At the position of the ULX, we detect the two sources reported 
by \citet{2005ApJ...633L.101M} (see Figure~\ref{fig:1313hst}). Their apparent, absolute Vega magnitudes 
and extinction are indicated in Table~\ref{tab:candidateshst} and from them, we can see that the source labeled B has absolute $V$-, $I$- and $J$-band magnitudes consistent with being 
a faint RSG, whereas the source we labeled as A is bluer and consistent with it being an O-B type star, as previously reported.
On our $H$-band image, we only detect one candidate counterpart within the 99.7$\%$ confidence radius (see Figure~\ref{fig:ngc1313-2}). 
It has an absolute magnitude of $H = -7.1 \pm 0.2$, i.e. this source is also consistent with being a faint RSG.
Alternatively, (part of) the NIR emission could be due to nebular emission lines. NGC 1313 X-2 is surrounded by a bright optical nebula (e.g. 
\citealt{2008AIPC.1010..303P}) and these ULX nebula often show strong emission lines in the NIR 
(e.g. \citealt{2016MNRAS.459..771H,2019MNRAS.489.1249L}, this work, see Section~\ref{speccp}). 
In conclusion, neither source A nor source B have definite characteristics that prove whether one of them is the donor star to the ULX.

\begin{table*}
\vspace{5mm}
\begin{center}
\caption{Counterparts detected in archival \hst images for four of our observed ULXs. We provide their apparent and absolute Vega magnitudes, and their respective intrinsic colours. The labels A, B, C and D are the same as used in Figures~\ref{fig:1313hst} and~\ref{fig:253hst} to mark the counterparts.}
\label{tab:candidateshst}
\resizebox{\textwidth}{!}{\begin{tabular}{|lcccccc|}
\hline\hline
ULX name & \multicolumn{2}{c}{Apparent magnitude} & \multicolumn{2}{c}{Absolute magnitude} & Extinction* & Colour\\
 & \multicolumn{2}{c}{(mag)} & \multicolumn{2}{c}{(mag)} & (mag) & (mag) \\
\hline\hline
NGC 1313 X-2 & $V_A = 23.58 \pm 0.33$ & $V_B = 24.66 \pm 0.55$ & $V_A = -4.88 \pm 0.37$ & $V_B = -3.80 \pm 0.58$ & $A_V = 0.32$ & $(V - I)_A= 0.13 \pm 0.69$ \\
 & $I_A = 23.56 \pm 0.56$ & $I_B = 22.84 \pm 0.40$ & $I_A = -4.76 \pm 0.59$ & $I_B = -5.48 \pm 0.44$ & $A_I = 0.18$ & $(V - I)_B = 1.68 \pm 0.72$ \\
 & $J_A = 24.01 \pm 0.41$ & $J_B = 22.38 \pm 0.19$ & $J_A = -4.28 \pm 0.48$ & $J_B = -5.91 \pm 0.31$ & $A_J = 0.08$ &  \\
 \hline
2XMM J004820.0 & $R_A = 25.59 \pm 0.43$ & $I_A = 23.74 \pm 0.34$ & $R_A = -2.27 \pm 0.46$ & $I_A = -4.09 \pm 0.38$ & $A_R = 0.10$ & $(R - I)_A = 1.83 \pm 0.60$ \\
-251010 & $R_B = 27.22 \pm 0.95$ & $I_B = 25.71 \pm 0.85$ & $R_B = -0.64\pm 0.97$ & $I_B = -2.12 \pm 0.87$ & $A_I = 0.08$ & $(R - I)_B = 1.49 \pm 1.30$ \\
& $R_C = 26.00 \pm 0.55$ & $I_C = 24.68 \pm 0.53$ & $R_C = -1.85 \pm 0.58$ & $I_C = -3.16 \pm 0.56$ & & $(R - I)_C = 1.30 \pm 0.80$ \\
& $R_D = 26.64 \pm 0.74$ & $I_D = 25.28 \pm 0.69$ & $R_D = -1.22 \pm 0.76$ & $I_D = -2.56 \pm 0.71$ & & $(R - I)_D = 1.34 \pm 1.04$ \\
 \hline
RX J073655.7& $J_A = 20.56 \pm 0.12$ & $H_A = 19.68 \pm 0.15$ & $J_A = -7.10 \pm 0.22$ & $H_A = -7.90 \pm 0.24$ & $A_H = 0.06$ & $(J - H)_A = 0.72 \pm 0.32$ \\
+653542 & $J_B = 20.49 \pm 0.12$ & $H_B = 19.50 \pm 0.12$ & $J_B = -7.17 \pm 0.22$ & $H_B = -8.07 \pm 0.22$ & $A_J = 0.15$ & $(J - H)_B = 0.96 \pm 0.31$ \\
 \hline
CXOU J140314.3 & $V_A = 24.76 \pm 0.53$ & $I_A = 22.39 \pm 0.18$ & $V_A = -4.90 \pm 0.56$ & $I_A = -7.07 \pm 0.26$ & $A_V = 0.25$ & $(V - I)_A = 2.17 \pm 0.62$ \\
+541807 & $V_B = 24.84 \pm 0.54$ & $I_B = 23.82 \pm 0.36$ & $V_B = -4.82 \pm 0.57$ & $I_B = -5.64 \pm 0.40$ & $A_I = 0.46$ & $(V - I)_B = 0.82 \pm 0.69$ \\
& $V_C = 23.13 \pm 0.23$ & $I_C = 23.06 \pm 0.25$ & $V_C = -6.54 \pm 0.28$ & $I_C = -6.40 \pm 0.31$ & & $(V - I)_C = -0.14 \pm 0.43$ \\
& $V_D = 23.11 \pm 0.23$ & $I_D = 22.86 \pm 0.23$ & $V_D = -6.55 \pm 0.30$ & $I_D = -6.60 \pm 0.29$ & & $(V - I)_D = 0.05 \pm 0.42$ \\
\hline\hline
\multicolumn{7}{l}{{\bf Notes:} *Calculated using N$_H = 6.04 \times 10^{20}$ atoms cm$^{-2}$ for NGC 1313 X-2, N$_H = 2.55 \times 10^{20}$ atoms cm$^{-2}$ for 2XMM J004820.0-251010,}\\
\multicolumn{7}{l}{N$_H = 1.19 \times 10^{21}$ atoms cm$^{-2}$ for RX J073655.7+653542 and N$_H = 8.52 \times 10^{20}$ atoms cm$^{-2}$ for CXOU J140314.3+541807 (values taken from}\\
\multicolumn{7}{l}{\citealt{2016AA...594A.116H}).}\\
\end{tabular}}
\end{center}
\end{table*}

\begin{figure*}
  \begin{minipage}{0.5\textwidth}
    \includegraphics[width=0.95\textwidth]{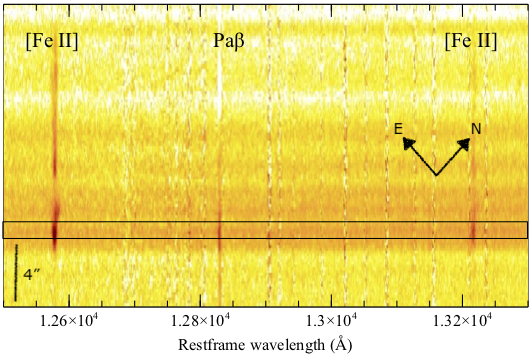}
    \subcaption{RX J073655.7+653542}
    \label{fig:ngc2403_2d}
  \vspace{0.5cm}   
 \end{minipage}%
  \begin{minipage}{0.5\textwidth}
    \includegraphics[width=0.95\textwidth]{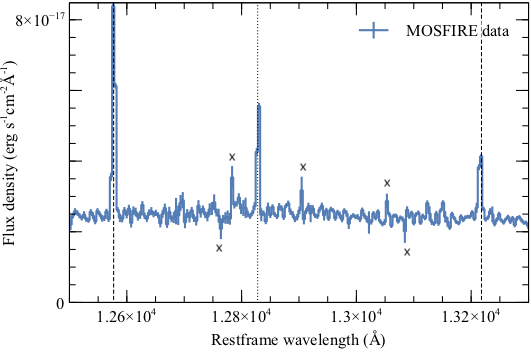}
    \subcaption{RX J073655.7+653542}
     \label{fig:ngc2403full}
  \vspace{0.5cm}   
 \end{minipage}% 
\caption{(a) Part of the 2-D $J$-band spectrum of RX J073655.7+653542, centered on the two $[$Fe {\scshape II}$]$ emission lines. The horizontal black lines indicate the location of the continuum source at the X-ray position of the ULX. The $[$Fe {\scshape II}$]$ emission extends for $\sim 3$\arcsec around the ULX, with a larger region extending out to $\sim 12$\arcsec towards the north-east. (b) MOSFIRE $J$-band spectrum of RX J073655.7+653542 (blue, with errorbars). The $[$Fe {\scshape II}$]$ emission lines are indicated by dashed lines, the dotted line shows the position of the strongest Paschen line in the band. All wavelengths are in vacuum.}
 \label{fig:2403} 
\end{figure*}

\subsubsection{CXOU J112037.3+133429}
\label{ulx18}

This ULX in NGC 3628 was observed by \citet{2014MNRAS.442.1054H} in the $K_s$-band, where they did not detect
a counterpart, to a limiting magnitude of $-11.2$. We re-observe it in the $H$-band and detect 
one candidate counterpart within its 99.7$\%$ confidence radius (see Figure~\ref{fig:ngc3628}). 
The detected source has an absolute magnitude of $H = -9.7 \pm 0.5$, consistent with the absolute
magnitude of a RSG. We can estimate a $H - K_s$ color from these two observations, although we must
note that they were performed 5.5 years apart. The counterpart would have $H - K_s < 1.5 \pm 0.5$ which 
is consistent within uncertainties with the values for a typical RSG \citep{1985ApJS...57...91E,2000asqu.book..381D}.

\subsection{Possible NIR star cluster candidates}
\label{star}

As we give this classification to any source whose absolute magnitude is too bright to be a RSG
and unlikely to be an AGN as it is spatially extended, it is a preliminary, rough categorization. Considering that the
only SC candidate that we have spectroscopically followed-up turned out to be a nebula (see Section~\ref{x26}),
we cannot ignore that there is a possibility that the sources below might as well also be nebulae.

\subsubsection{ChASeM33 J013350.89+303936.6}
\label{ulx6}

This source was detected in NGC 598 in early X-ray observations
(e.g. \citealt{1981ApJ...246L..61L,1983ApJ...275..571M}).
Its \chan X-ray position coincides with the nucleus of NGC 598 \citep{2002MNRAS.336..901D}.
No sign of nuclear activity at other wavelengths was detected \citep{2004A&A...425...95D}, thus
it was classified as a ULX. We detect one NIR candidate counterpart (see Figure~\ref{fig:ngc598}) 
within the 99.7$\%$ confidence radius, with absolute magnitude of $H = -13.0 \pm 0.1$ mag. 
However, this source location is consistent with that of the nucleus of NGC 598 and thus,
the NIR emission is dominated by the nucleus of the galaxy. We cannot resolve the counterpart,
and moreover, it is likely that we are detecting a nuclear star cluster.

\subsubsection{CXO J031718.9-410627}
\label{ulx8}

This potential ULX in NGC 1291 has one NIR candidate counterpart (see Figure~\ref{fig:ngc1291})
and, similarly to the case of the ULX in NGC 598, its position is consistent with that of the nucleus of the 
galaxy. Hence, we cannot distinguish if the emission comes from an individual source or if its contaminated/dominated
by the nucleus of NGC 1291. We deem it likely that we are detecting a nuclear star cluster.

\subsubsection{2XMM J140228.3+541625}
\label{ulx30}

This ULX has one NIR candidate counterpart (see Figure~\ref{fig:ngc5457}) with an absolute
magnitude of $H = -13.2 \pm 0.1$, too bright to be a single star. The FWHM of this extended source
is 0.9\arcsec\, which is set by the seeing. At the distance of NGC 5457 this is equivalent to 30 pc. 
The size is therefore smaller than 30 pc, but we cannot rule out that it is a stellar cluster (see \citealt{2009ApJ...699.1690M}).

\subsection{NIR candidate without classification}
\label{noclue}

\subsubsection{2XMM J004820.0-251010}
\label{ulx5}

For this ULX in NGC 253, we detect a counterpart inside the 99.7$\%$ confidence radius of its
X-ray position, (see Figure~\ref{fig:ngc253}) with an absolute magnitude of $H = -7.4 \pm 0.2$,
consistent with the faint end of a RSG. However, when we analyse archival \hst images 
of this source in the F606W and F814W filters (observation ID 10915), taken from the 
Barbara A. Mikulski Archive for Space Telescopes (MAST)\footnote{\url{https://mast.stsci.edu/portal/Mashup/Clients/Mast/Portal.html}},
we find something different. At the position of our NIR counterpart detection (the dashed circle in Figures~\ref{fig:ngc253} and~\ref{fig:253hst}), 
the \hst images allow us to resolve it into at least four point-like sources. {\scshape SExtractor} detected only four sources, for which
we calculate apparent and absolute $R$ and $I$ band Vega magnitudes (see Table~\ref{tab:candidateshst}). These sources are all too faint for a RSG,
even after we consider possible extinction in both bands (see Table~\ref{tab:candidateshst}).
There are also other point-like sources inside the 99.7$\%$ confidence radius of the X-ray position which we do not detect in our $H$-band image. 
Therefore, none of these sources has a color red enough for it to be a RSG, and we conclude that 
the source we detected in the NIR observation is in fact a blend of the resolved source detected in the HST image.

\subsection{Spectroscopy}
\label{speccp}

\subsubsection{RX J073655.7+653542: a RSG in NGC 2403}
\label{2403}

This X-ray source is located inside a large extended nebula. Our 2-D $J$-band spectrum reveals strong $[$Fe {\scshape II}$]$ $\lambda1257$ emission around the ULX with an extent of $\sim 3''$, as well as emission extending up to $\sim 12''$ along the slit offset from the position of the ULX (see Figure \ref{fig:ngc2403_2d}). At the distance to NGC 2403, these values correspond to physical sizes of $\sim 35$ and $\sim 240$ pc. This same structure is not visible in the Pa$\beta$ line, possibly due to strong emission in that line on the other side of the ULX that shows up in absorption at the location of the extended $[$Fe {\scshape II}$]$ emission due to our nodding pattern. The $[$Fe {\scshape II}$]$ $\lambda1257$ line is redshifted by $175 \pm 5$ \vel, consistent with the radial motion of NGC 2403. In addition, there is a continuum source at the location of the ULX that is very likely stellar in origin (see Figure \ref{fig:ngc2403full}). 

\begin{figure*}
  \begin{minipage}{0.5\textwidth}
    \includegraphics[width=0.95\textwidth]{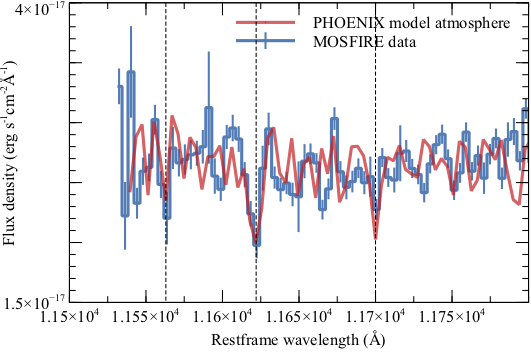}
    \subcaption{RX J073655.7+653542}
    \label{fig:ngc2403zoom}
  \vspace{0.5cm}   
 \end{minipage}%
  \begin{minipage}{0.5\textwidth}
    \includegraphics[width=0.95\textwidth]{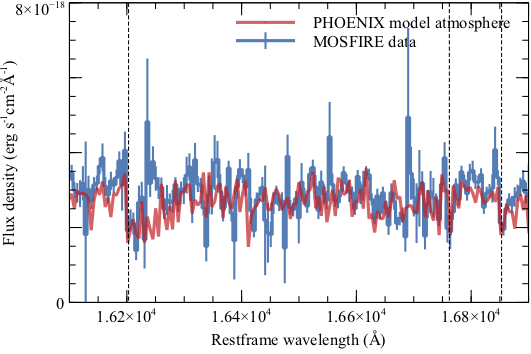}
    \subcaption{CXOU J140314.3+541807}
    \label{fig:xmm1spec}
  \vspace{0.5cm}   
 \end{minipage}% 
\caption{(a) Part of the MOSFIRE $J$-band spectrum of RX J073655.7+653542 (blue, with errorbars) is shown along with the 4000 K PHOENIX model atmosphere (red), redshifted by 160 \vel to match the radial velocity of the star. Dashed lines indicate the positions of strong Fe {\scshape I} absorption features. (b) Part of the MOSFIRE $H$-band spectrum of CXOU J140314.3+541807 (blue, with errorbars) is shown along with the 3000 K PHOENIX model atmosphere (red), redshifted by 190 \vel to match the radial velocity of the star. Dashed lines indicate the positions of some strong absorption features found in RSGs. All wavelengths are in vacuum.}
 \label{fig:stars} 
\end{figure*}

Three Fe {\scshape I} absorption lines are visible in the 1.15-1.17 $\mu$m region of the spectrum (see Figure \ref{fig:ngc2403zoom}). We cross-correlated our spectrum with {\sc PHOENIX} model atmospheres for RSGs with temperatures of 3000, 4000 and 5000 K and $\log(g) = -0.5$ \citep{2007A&A...468..205L} in the 1.15-1.17 $\mu$m region. From the three model atmospheres that we tested, the 4000 K model best matches our spectrum; the 3000 K model contains a strong CO bandhead at 1.24 $\mu$m that is not present in our data, while the slope of the 5000 K model is too blue. The 4000 K model moreover yields the strongest cross-correlation signal. The RVs found through cross-correlating the three different templates are all consistent. We therefore conclude that this star most likely has an effective temperature in the $3500-4500$ K range. For the RV, we adopt the value found through cross-correlating with the 4000 K model: $160 \pm 12$ \vel. This is consistent (within 2-$\sigma$) with the RV of the emission lines and with the radial velocity of NGC 2403 at the position of the ULX ($\sim 170$ \vel, \citealt{2002AJ....123.3124F}).

\begin{figure*}
  \begin{minipage}{0.5\textwidth}
  \begin{center}
    \includegraphics[width=0.6\textwidth]{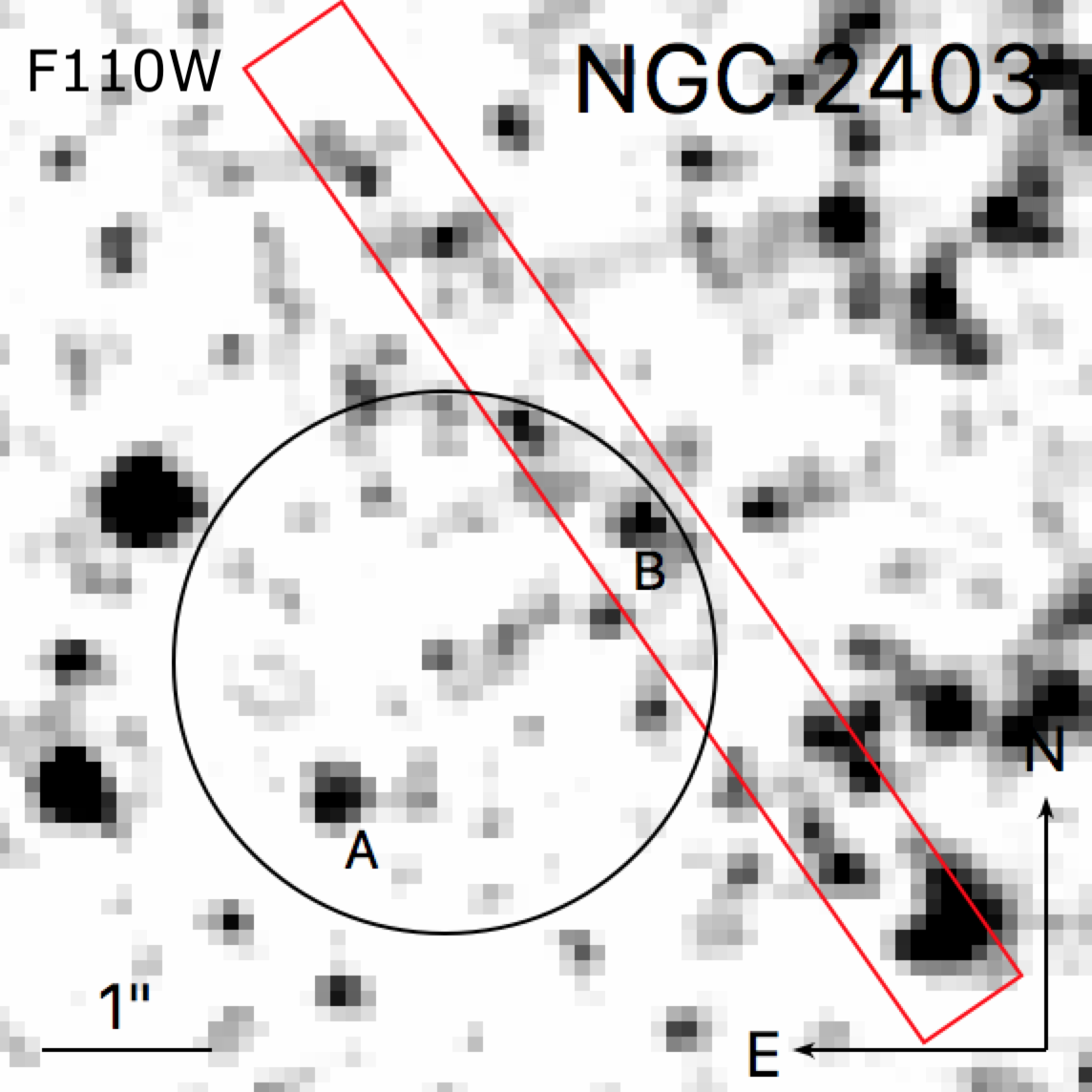}
    \subcaption{RX J073655.7+653542}
    \label{fig:ngc2403hst}
  \vspace{0.5cm}  
  \end{center} 
 \end{minipage}%
  \begin{minipage}{0.5\textwidth}
  \begin{center}
    \includegraphics[width=0.6\textwidth]{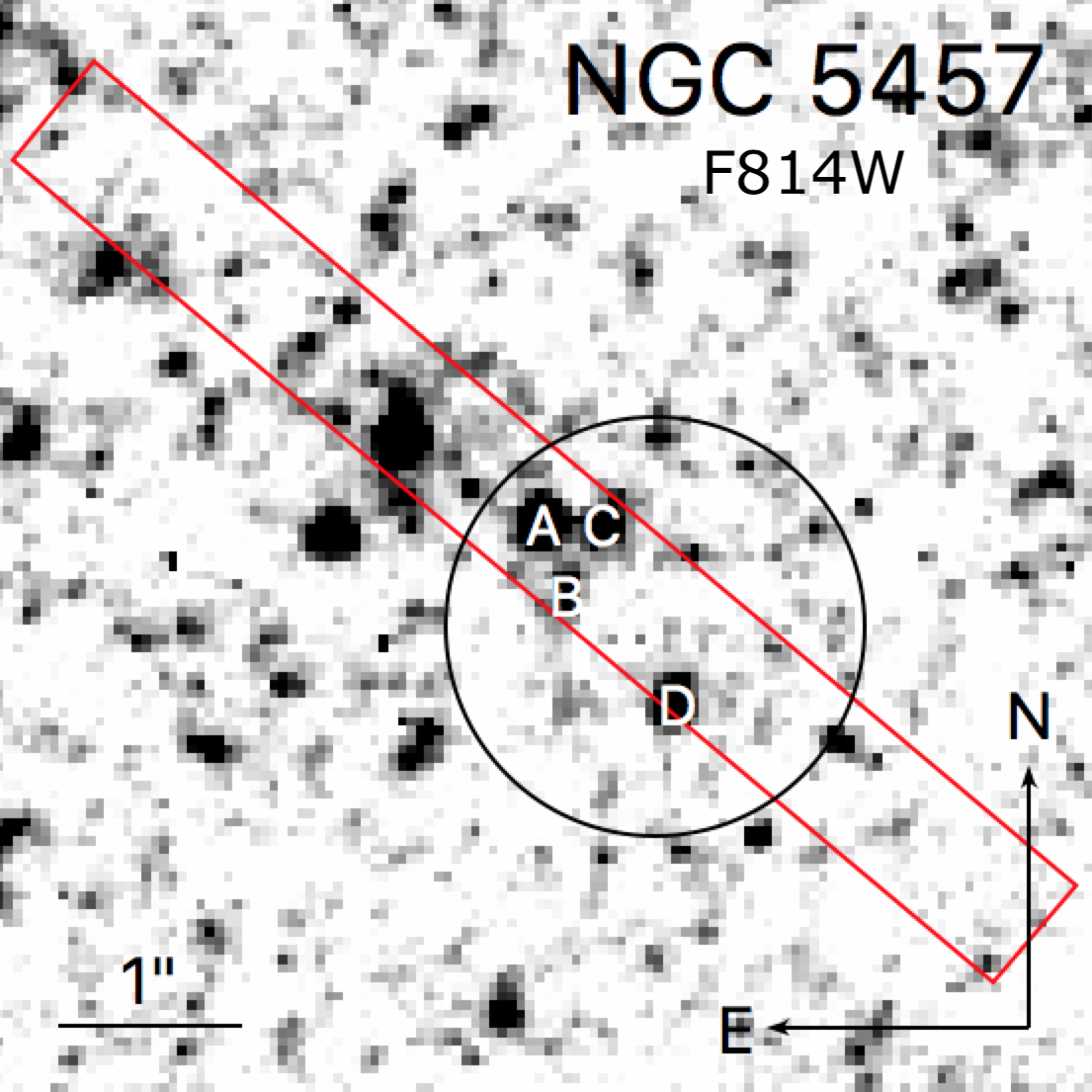}
    \subcaption{CXOU J140314.3+541807}
    \label{fig:xmm1hst}
  \vspace{0.5cm}
  \end{center}   
 \end{minipage}% 
\caption{\hst finder charts for the ULXs (a) RX J073655.7+653542 and (b) CXOU J140314.3+541807. The solid circles correspond to the 99.7$\%$ confidence region of the position of the ULXs, the letters mark the sources detected with {\scshape SExtractor}, and the 0.7\arcsec -MOSFIRE-slit-width is indicated in red. For presentation purposes, the slit length is set to 7\arcsec ; however, a 0.7\arcsec\ $\times$ 110.7\arcsec\ slit was used for (a) and a 0.7\arcsec\ $\times$ 118.7\arcsec\ slit was used for (b).}
 \label{fig:starshst} 
\end{figure*}

We further examine these results by analysing the archival \hst images of this source in the F110W and F160W filters. At the position of the ULX, we detect two 
bright sources, the one labeled as B is at the position reported by \citet{2017MNRAS.469..671L} (see Figure~\ref{fig:ngc2403hst}). Their apparent, absolute Vega 
magnitudes and extinction are indicated in Table~\ref{tab:candidateshst} and from them, we can see that the source we labeled B has absolute magnitudes 
consistent with being a RSG, if in the faint end. However, we can see in the image that there are several other unresolved sources, which opens the possibility of
both sources A and B being contaminated by the light of these groups of stars. This could result in fainter $J$- and $H$-band magnitudes, weakening the claim
that RX J073655.7+653542 is a RSG. An adaptive optics assisted spectrum, reducing as much as possible the potential contamination of the spectrum by light from 
currently unresolved nearby stars, will provide important new information on the nature of the candidate counterpart to ULX RX J073655.7+653542.

\subsubsection{CXOU J140314.3+541807: a RSG in NGC 5457}
\label{xmm1}

Our $H$-band spectrum of this source shows a continuum with absorption features, most notably several CO-bandheads. We cross-correlate this spectrum with PHOENIX model atmospheres with effective temperatures of 3000, 4000 and 5000 K, using the spectral range from 1.615--1.69 $\mu$m, where several strong absorption features are seen (see Figure~\ref{fig:xmm1spec}). The cross-correlation with the 3000 K model yields the strongest signal; the cross-correlation with the 4000 K model gives a consistent result but weaker signal, and the cross-correlation with the 5000 K model does not show significant peaks. We adopt the radial velocity obtained from the cross-correlation with the 3000 K model, of $193 \pm 17$ \vel. This is consistent with the radial velocity of NGC 5457 at the position of the ULX (i.e. $\sim 200$ \vel \citealt{2018ApJ...854...68H}). We conclude that this source is most likely a red supergiant with an effective temperature in the $3000-4000$ K range, matching the classification based on mid-IR photometry presented in \citet{2019ApJ...878...71L}. 

We also examine these results by analysing the archival \hst images of this source in the F555W and F814W filters. At the position of the ULX, we detect four 
bright sources, one of them (A) at the position reported by \citet{2014MNRAS.442.1054H} (see Figure~\ref{fig:xmm1hst}). Their apparent, absolute Vega 
magnitudes and extinction are indicated in Table~\ref{tab:candidateshst} and from them, we can see that the source labeled A has absolute magnitudes 
consistent with being a RSG. However, we can see in the image that inside the 3-$\sigma$ confidence position radius, there are three bright bluer sources. There
is no physical reason that rules out these sources as possible donor stars to the ULX. A second epoch spectrum from our NIR counterpart would confirm
its association with the ULX, if radial velocity variations are detected.

\subsubsection{XMMU J024323.5+372038: a likely AGN}
\label{ngc1058}

The spectrum of this counterpart in NGC 1058 is dominated by a broad emission line, reminiscent of the broad H$\alpha$
emission profile often seen in AGNs (see e.g. \citet{2016MNRAS.461..647C}). Hence, we deem it reasonable to fit 
Gaussian profiles to the structure, fixing the relative spacing between H$\alpha$ and $[$N{\scshape II}$]$
a line ratio of 1:3 for the $[$N{\scshape II}$]\lambda$6549:$[$N {\scshape II}$]\lambda$6583 lines \citep{1989agna.book.....O}, with 
redshift, amplitude and FWHM as free parameters. As a result, we see a broad component and the three narrow lines: H$\alpha$ and 
$[$N {\scshape II}$]\lambda \lambda$6548, 6583 (see Figure~\ref{fig:ngc1058}). This would yield a redshift of $z = 1.0301 \pm 0.0001$, 
making XMMU J$024323.5+372038$ a likely background AGN. 

\subsubsection{[LB2005] NGC 5457 X26: a nebula}
\label{x26}

This source was preliminarily classified as a stellar cluster by \citet{2017MNRAS.469..671L} based on its absolute magnitude
($-11.44 \pm 0.22$) and spatial extent. Its $H$-band spectrum is dominated by a strong $[$Fe {\scshape II}$]$ emission line.
Several other $[$Fe {\scshape II}$]$ weaker lines and two Bracket lines are detected as well (see Figure~\ref{fig:x26}).
After fitting Gaussian profiles to these lines, we calculated the radial velocity of the source to be $295 \pm 16$ \vel, consistent 
with the radial velocity of NGC 5457 at the position of the ULX \citep{2018ApJ...854...68H}.

\subsubsection{2E 1402.4+5440: an unknown source}
\label{xmm3}

\begin{figure*}
  \begin{minipage}{0.5\textwidth}
  \includegraphics[width=0.98\textwidth]{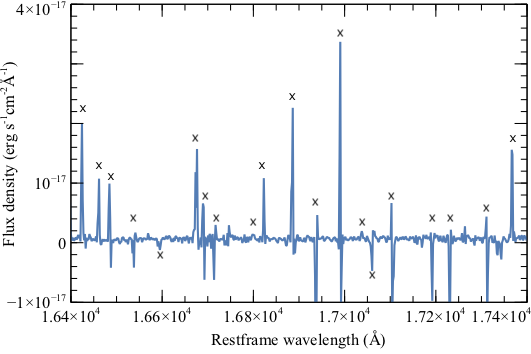}%
  \subcaption{2E 1402.4+5440}  
  \label{fig:xmm3} 
  \vspace{0.5cm}   
 \end{minipage}%
   \begin{minipage}{0.5\textwidth}
  \includegraphics[width=0.98\textwidth]{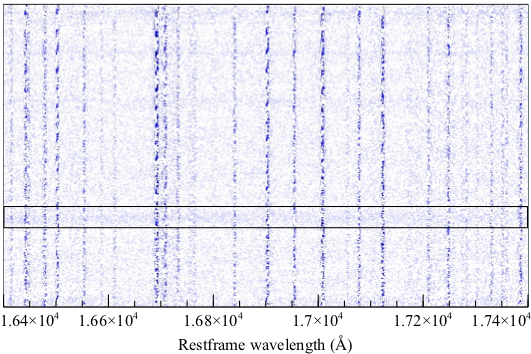}%
  \subcaption{2E 1402.4+5440}  
  \label{fig:xmm32d} 
  \vspace{0.5cm}   
 \end{minipage} 
\caption{ (a) Part of the NIR spectrum of 2E $1402.4+5440$. (b) Part of the 2-D spectrum of 2E $1402.4+5440$ showing the same wavelength range as in (a). The faint continuum is inside the solid rectangle. Residuals of sky lines subtraction or spurious lines are marked with an x. All wavelengths are in vacuum.}
 \label{fig:spectra2} 
\end{figure*}

This source in NGC 5457 has an apparent and absolute $H$-band magnitude of $19.3 \pm 0.2$ and $-9.9 \pm 0.2$,
respectively. Its spectrum shows a weak continuum (see Figure~\ref{fig:xmm3}) and some emission lines appear to be present. 
However, these apparent emission lines are residuals from the subtraction of the sky emission as they lie close to strong
strong emission lines (see Figure~\ref{fig:xmm32d}). No clear absorption or emission lines are detected. Therefore, we conclude 
that this source does not host a strong nebula, but the continuum emission could be coming from an accretion disk or a RSG 
(or a combination of the two).

\section{Recapitulation of our NIR imaging survey}
\label{recap}

\subsection{Host galaxies to observed ULXs}
\label{galaxies}

In the span of 8 years we have studied 113 ULXs in 52 different galaxies, up to a distance of $\simeq 10$ Mpc
(see Figure~\ref{fig:dist}), within uncertainties. These 113 ULXs were located at a distance range of 
$0.5$ pc to $29$ kpc from the centre of their respective host galaxies (see Figure~\ref{fig:center}). 

\begin{figure*}
 \begin{minipage}{0.5\textwidth}
  \includegraphics[width=0.98\textwidth]{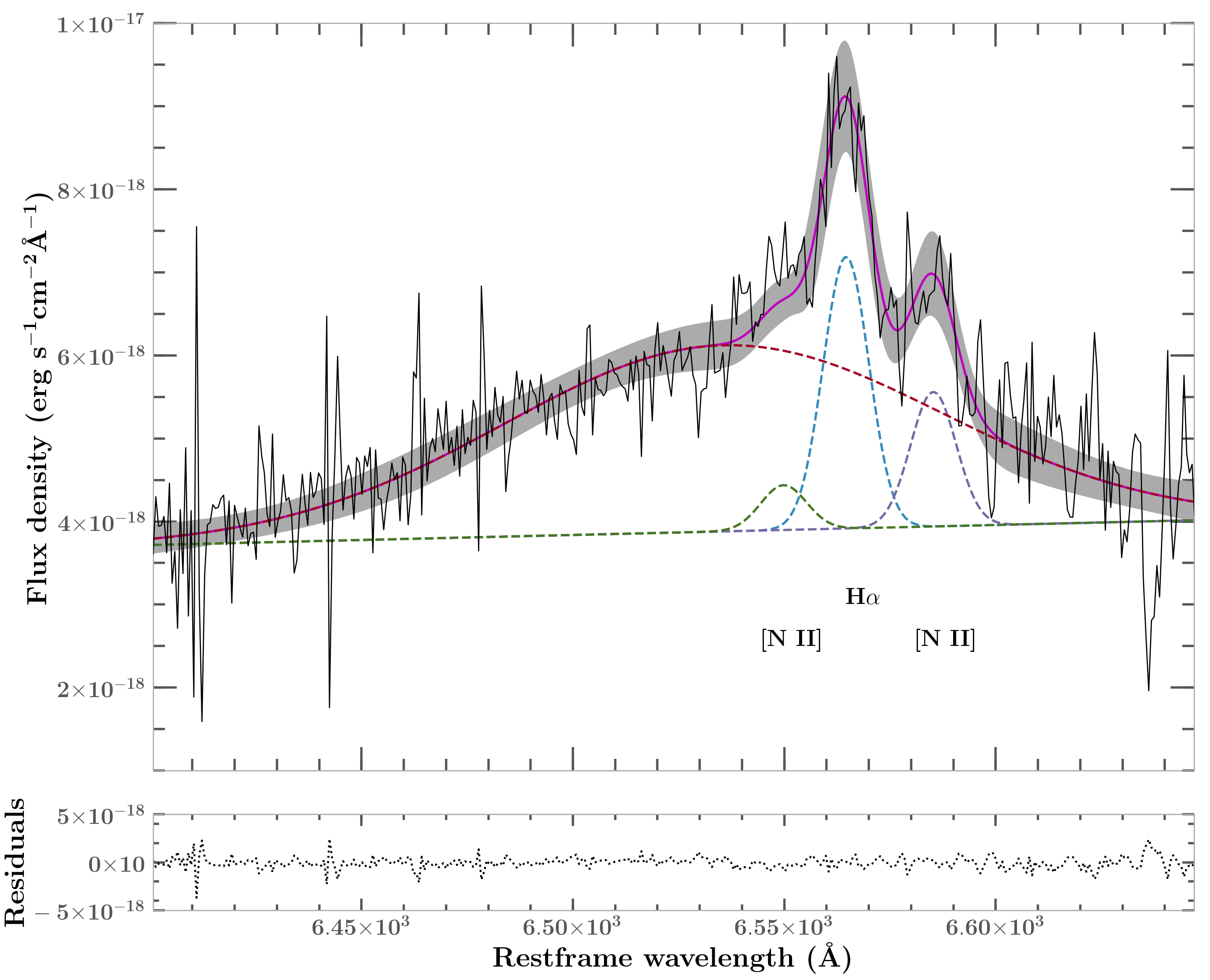}%
  \subcaption{XMMU J024323.5+372038}  
  \label{fig:ngc1058} 
  \vspace{0.5cm}   
 \end{minipage}%
   \begin{minipage}{0.5\textwidth}
  \includegraphics[width=0.98\textwidth]{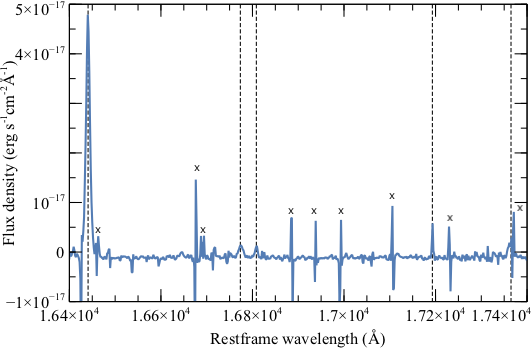}%
  \subcaption{[LB2005] NGC 5457 X26}  
  \label{fig:x26} 
  \vspace{0.5cm}   
 \end{minipage} 
\caption{(a) Part of the NIR spectrum of XMMU J$024323.5+372038$ showing the broad and narrow components of the emission line. The Gaussian profiles are shown indicating the position of the putative $[$N {\scshape II}$]\lambda \lambda 6548,6583$ and H$\alpha$ lines. The shaded area shows the 3-$\sigma$ confidence region around the fit. The residuals of the fit are shown in the bottom panel. The lines are redshifted by $z = 1.0301 \pm 0.0001$. (b) Part of the NIR spectrum of $[$LB2005$]$ NGC 5457 X26 where the dashed lines indicate the positions of the nebular $[$Fe {\scshape II}$]$ and Bracket emission lines, redshifted by 295 \vel. Residuals of sky lines subtraction or spurious lines are marked with an x. All wavelengths are in vacuum.}
 \label{fig:spectra} 
\end{figure*}

Of the 52 galaxies, we observed two or more ULXs in 21 galaxies and only one ULX in each of the remaining 31 
galaxies. We made use of 5 different telescopes and the $H, K_s$ photometric bands. We tailored the survey to detect RSGs by aiming to achieve
a depth of 20 mag in $H$ and $K_s$. Considering that the distances to the observed sources range from $0.91 \pm 0.5$
to $14.19 \pm 2.84$ Mpc, our expected depth was $-10.8 \pm 0.4 < H, K_s < -4.8 \pm 0.1$ mag.

We could not observe all ULXs located within $10$ Mpc, as the total number of ULX within $10$ Mpc is 170. 
This was due to several factors such as not being able to use the telescope due to adverse weather conditions.

\begin{table*}
\vspace{5mm}
\begin{center}
\caption{The 38 identified NIR candidate counterparts to ULXs. The preliminary classification of the NIR candidate counterparts is based on their absolute magnitudes, WISE colours, spatial extent and/or visual inspection of the NIR image. The spectroscopic classification of 12 sources is based on the spectra that we obtained for them, reported in \citet{2015MNRAS.453.3510H,2016MNRAS.459..771H} and \citet{2019MNRAS.489.1249L}.}
\label{tab:allcp}
\resizebox{\textwidth}{!}{\begin{tabular}{|llccccc|}
\hline\hline
Galaxy & ULX name & Filter & Apparent & Absolute & Preliminary & Spectroscopic \\
 & (in SIMBAD) & & magnitude & magnitude & classification & classification \\
 & & & (mag) & (mag) & & \\ 
\hline\hline
NGC 253 & RX J$004722.4-252051$ & $K_s$ &17.2 $\pm$ 0.5 & $-$10.5 $\pm$ 0.1 & RSG & RSG\\
NGC 253 & 2XMM J$004820.0-251010$ & $H$ & 20.4 $\pm$ 0.1 & $-$7.4 $\pm$ 0.2 & N/A & $-$\\
NGC 598 & ChASeM33 J$013350.89+303936.6$& $H$ & 11.8 $\pm$ 0.1 & $-$13.0 $\pm$ 0.1 & SC & $-$\\
NGC 925 & $[$SST2011$]$ J$022721.52+333500.7$ & $H$ & 18.7 $\pm$ 0.2 & $-$10.6 $\pm$ 0.4 & RSG &RSG\\
NGC 925 & $[$SST2011$]$ J$022727.53+333443.0$ & $H$ & 20.1 $\pm$ 0.2 & $-$9.2 $\pm$ 0.4 & RSG & Nebula\\
NGC 1058 & XMMU J$024323.5+372038$ & $H$ & 20.8 $\pm$ 0.3 & $-$9.0 $\pm$ 0.4 & RSG & AGN\\
NGC 1291 & CXO J$031718.9-410627$& $H$ & 11.3 $\pm$ 0.1 & $-$18.4 $\pm$ 0.1 & SC & $-$\\
NGC 1313 & NGC 1313 X-2 & $H$ & 21.1 $\pm$ 0.2 & $-$7.1 $\pm$ 0.2 & RSG & $-$\\
NGC 1637 & $[$IWL2003 68$]$ & $K_s$ &16.3 $\pm$  0.5 & $-$13.7 $\pm$ 0.5 & SC/AGN & $-$\\
NGC 2403 & RX J$073655.7+653542$ & $H$ &17.46 $\pm$ 0.02 & $-$10.1 $\pm$ 0.2 & RSG & RSG\\
NGC 2500 & CXO J$080157.8+504339$ & $K_s$ &15.7 $\pm$ 0.2 & $-$14.1 $\pm$ 0.4 & AGN & $-$\\
Holmberg I & $[$WMR2006$]$ Ho I XMM1 & $H$ &17.1 $\pm$ 0.1 & $-$10.14 $\pm$ 0.1 & AGN & $-$\\
Holmberg II & Holmberg II X-1 & $H$ &20.6 $\pm$ 0.1 & $-$7.1 $\pm$ 0.1 & RSG & Nebula\\
NGC 3521 & $[$SST2011$]$ J$110545.62+000016.2$ & $H$ &19.8 $\pm$ 0.1 & $-$10.9 $\pm$ 0.9 & RSG & Nebula\\
NGC 3627 & $[$SST2011$]$ J$112018.32+125900.8$ & $K_s$ & 20.6 $\pm$ 1.9 & $-$9.1 $\pm$ 1.9 & RSG & $-$\\
NGC 3628 & CXOU J$112037.3+133429$ & $H$ & 20.4 $\pm$ 0.2 & $-$9.7 $\pm$ 0.5 & RSG & $-$\\
NGC 4136 & CXOU J$120922.6+295551$ & $H$ &19.1 $\pm$ 0.1 & $-$10.78 $\pm$  0.4 & RSG & Nebula\\
NGC 4136 & $[$SST2011$]$ J$120922.18+295559.7$  & $H$ & 19.2 $\pm$ 0.1 & $-$10.75 $\pm$ 0.4 & RSG& RSG\\
NGC 4258 & RX J$121844.0+471730$ & $H$ & 17.8 $\pm$ 0.1 & $-$11.5 $\pm$ 0.1 & SC & $-$\\
NGC 4258 & 3XMM J$121847.6+472054$ & $H$ & 20.1 $\pm$ 0.1 & $-$9.3 $\pm$ 0.2 & RSG & $-$\\
NGC 4258 & $[$LB2005$]$ NGC 4258 X9 &  $H$ &15.86 $\pm$ 0.001 & $-$13.5 $\pm$ 0.2 & AGN & $-$\\
NGC 4485 & RX J$1230.5+4141$ & $H$ &19.0 $\pm$ 0.1 & $-$10.7 $\pm$ 0.4 & RSG & $-$\\
NGC 4490 & $[$SST2011$]$ J$123029.55+413927.6$  & $H$ & 16.05 $\pm$ 0.01 & $-$13.4 $\pm$ 0.3 & SC & $-$\\
NGC 4490 & CXO J$123038.4+413831$ & $H$ &16.69 $\pm$ 0.01 & $-$12.8 $\pm$ 0.3 & RSG & $-$\\
NGC 4490 & 2XMM J$123043.1+413819$ &  $H$ &18.41 $\pm$ 0.04 & $-$11.0 $\pm$ 0.3 & SC & $-$\\
NGC 4559 & $[$SST2011$]$ J$123557.79+275807.4$ & $H$ &17.33 $\pm$ 0.01 & $-$11.99 $\pm$ 0.90 & SC & $-$\\
NGC 4559 & RX J$123558+27577$ & $H$ &18.95 $\pm$ 0.05 & $-$10.4 $\pm$ 0.9 & RSG & $-$\\
NGC 4594 & $[$LB2005$]$ NGC 4594 X5 & $H$ &18.65 $\pm$ 0.02 & $-$11.6 $\pm$ 0.6 & AGN & $-$\\
NGC 4631 & $[$SST2011$]$ J$124211.13+323235.9$ & $H$ &14.48 $\pm$ 0.001 & $-$14.85 $\pm$ 0.49 & SC & $-$\\
NGC 5194 & XMMU J$132953.3+471040$  & $H$ & 15.7 $\pm$ 0.1 & $-$13.88 $\pm$ 0.2 & SC & $-$\\
NGC 5194 & RX J$132947+47096$ & $H$ &18.60 $\pm$ 0.04 & $-$11.18 $\pm$ 0.04 & RSG & $-$\\
NGC 5408 & NGC 5408 X-1 & $K_s$ & 20.3 $\pm$ 0.2 & $-$8.1 $\pm$ 0.8 & RSG & $-$\\
NGC 5457 & 2E $1402.4+5440$ & $H$ & 19.3 $\pm$ 0.2 & $-$9.7 $\pm$ 0.2 & RSG & N/A\\
NGC 5457 & 2XMM J$140248.0+541350$ & $H$ & 17.7 $\pm$ 0.1 & $-$11.3 $\pm$ 0.1 & RSG & RSG\\
NGC 5457 & CXOU J$140314.3+541807$ & $H$ & 18.4 $\pm$ 0.1 & $-$10.7 $\pm$ 0.1 & AGN & $-$\\
NGC 5457 & $[$LB2005$]$ NGC 5457 X32 & $H$ &18.67 $\pm$ 0.02 & $-$10.5 $\pm$ 0.1 & AGN & $-$\\
NGC 5457 & $[$LB2005$]$ NGC 5457 X26 & $H$ &17.78 $\pm$ 0.01 & $-$11.4 $\pm$ 0.2 & SC & Nebula\\
NGC 5457 & 2XMM J$140228.3+541625$ & $H$ & 16.0 $\pm$ 0.1 & $-$13.2 $\pm$ 0.1 & SC & $-$\\
\hline\hline
\end{tabular}}
\end{center}
\end{table*}

\subsection{Detected NIR counterparts}
\label{nircounterparts}

Out of the 113 observed ULXs, we detected a NIR candidate counterpart for 38 ULXs, 
i.e. about a third of the total sample. Based only on photometry, of the 38 detected NIR counterparts, 20 have NIR colours and/or magnitudes consistent 
with a RSG, 11 are preliminarily classified as stellar clusters, six are consistent with being an AGN and one is too faint to be either a RSG or a SC/AGN (see Table~\ref{tab:allcp}). 
After spectroscopic follow-up of 12 out of these 38 counterparts, we confirmed five RSGs (\citealt{2015MNRAS.453.3510H,2016MNRAS.459..771H} and this work), while
four sources photometrically classified as RSG turned out to be nebulae (\citealt{2016MNRAS.459..771H,2019MNRAS.489.1249L} and this work).
For the remaining 3 followed-up sources, one RSG candidate is a background AGN (this work), one stellar cluster candidate turned out to be a nebula (this work, see Table~\ref{tab:allcp})
and the last RSG candidate is too faint to be classified. This means that currently we have 5 confirmed RSGs and ten RSG candidates (photometrically classified).

The distribution of the detected counterparts in terms of their distance to our Galaxy (see Figure~\ref{fig:dist}) most likely follows the distribution of 
observed ULXs, based on the Kolmogorov--Smirnov (K--S) test between them (p-value = $0.88$). For the distribution of detected
counterparts in terms of their distance to the centre of their host galaxies (see Figure~\ref{fig:center}), we find that it also most likely follows the
distribution of observed ULXs, with a p-value = $0.67$. In addition, we detect a counterpart to all observed ULXs located 
at $> 15$ kpc from their host galaxy centre, which could be due to the ULXs having no contamination from their host galaxy, making the detection 
of a counterpart easier.

The detected counterparts in the $H$-band have an apparent magnitude ranging from 12 to 22 and an absolute magnitude from $-18$ to $-7$, those 
detected in the $K_s$-band have an apparent magnitude ranging from 15 to 22 mag and an absolute magnitude from $-15$ to $-8$ (see Figure~\ref{fig:histomags}).
We compare the distribution of our 15 detected RSGs (five confirmed and 10 photometrically classified) with the $K_s$ absolute magnitudes distribution of 85 RSGs in 
the Milky Way (MW) and the Large Magellanic Cloud (LMC) \citep{2005ApJ...628..973L}, and with the $K_s$ absolute magnitudes distribution of 245 RSGs in the 
Andromeda galaxy (M31) \citep{2016ApJ...826..224M}. A K--S test between these two distributions show that they are not similar (p-value = $6.6 \times 10^{-9}$).
We also compare our RSGs distribution with all 330 RSGs in the MW, LMC and M31 (see Figure~\ref{fig:both}). We assume a colour of $H - K_s \simeq 0$, typical 
for RSGs \citep{1985ApJS...57...91E,2000asqu.book..381D}. A K--S test between ours and the distribution of RSGs in the MW and the LMC shows that our detected 
RSGs are most likely drawn from the distribution of RSGs in the MW/LMC (p-value = $0.54$). We reach the same conclusion when comparing our RSG distribution with 
that of M31 (p-value = 0.08) and that of MW/LMC/M31 (p-value = $0.38$).

\subsection{Probability of chance superposition}
\label{superposition}

More than 50$\%$ and 80$\%$ of the observed ULXs are located at less than 5 and 10 kpc, respectively, from the nucleus of their host galaxies (see 
Figure~\ref{fig:dist}). This could result in ULXs being in crowded environments (i.e. close to the nucleus, in spiral arms of the galaxy, etc), so the probability of a 
chance superposition of an unrelated object may not be negligible. Using {\scshape SExtractor}, we search for point sources with an absolute magnitude in the RSG 
range ($-11<H, K_s<-8$) in all the images of our 113 ULXs. We then measure the total area of each image in arcsec$^2$ and the area of the error circle around the 
X-ray position of each ULX (also in arcsec$^2$). 

\begin{figure}
  \includegraphics[width=0.5\textwidth]{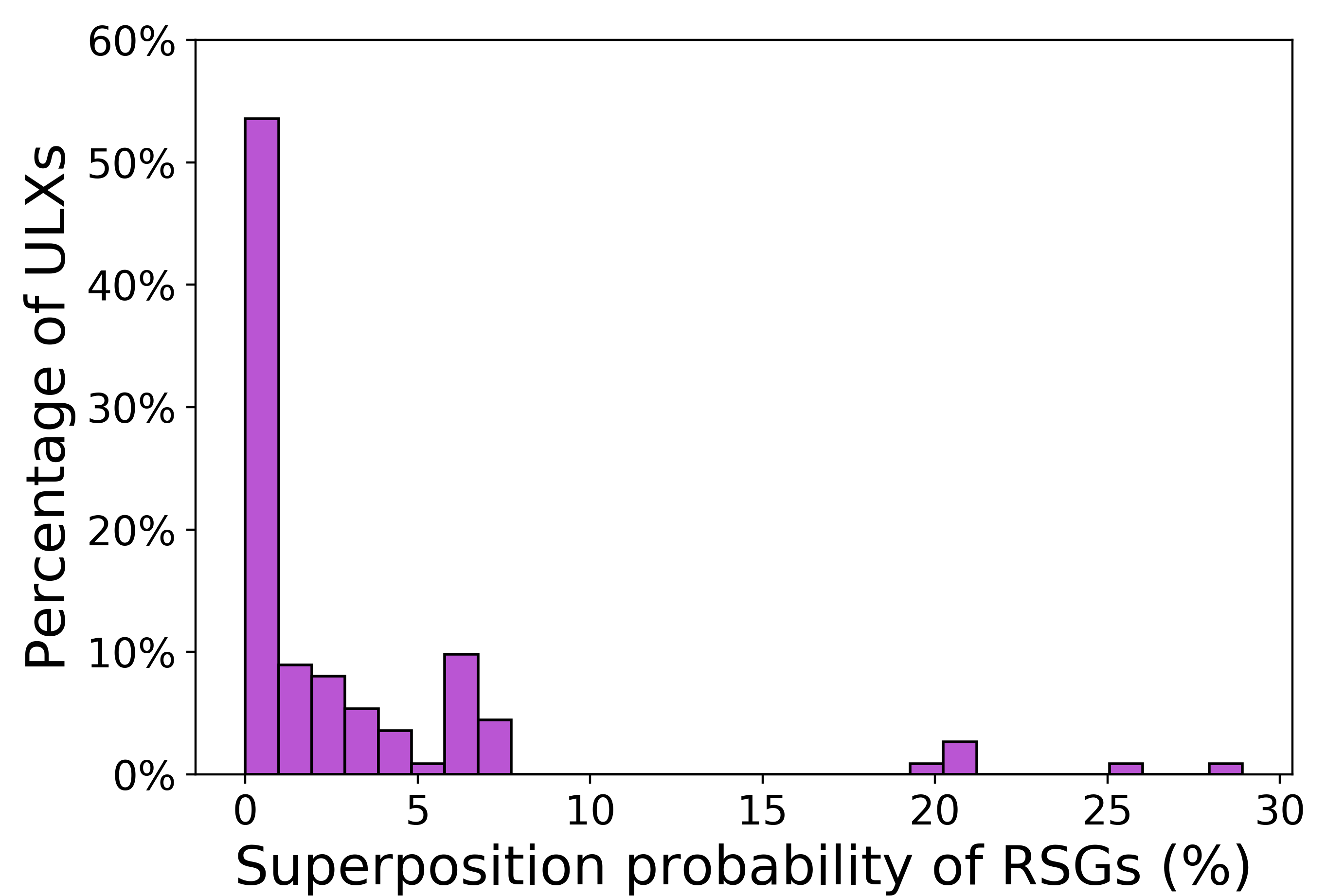}%
\caption{Histogram of the distribution of probabilities of a chance superposition of a source with an absolute magnitude in the RSG range for all 113 observed ULXs. The bins have a 1$\%$ probability width, for visualization purposes.}
\label{fig:histo_probs_rsg}
\end{figure}

These values, along with the detected point sources, allow us to estimate the probability of finding a source with an 
absolute magnitude in the RSG range for each of the 113 ULXs. See the probability distribution in Figure~\ref{fig:histo_probs_rsg}. More than 90 out of the 113
observed ULXs have $\leq 5\%$ chance superposition probability, whereas six ULXs have a probability between 20--30$\%$. For only one of these six 
ULXs a candidate RSG counterpart was detected (NGC 5408 X-1, \citealt{2014MNRAS.442.1054H}).

We calculate the total probability of a chance superposition for the sample of 113 ULXs as 3.3$\%$, which means that we could have 3.7 positives in our detections. 
We note that the distribution of stars in a galaxy is inhomogenenous, and the value of a chance superposition depends on the 
part of the image selected for the observation, for which we cannot calculate an accurate value for this probability, only provide a rough estimate. 
Based on this, we could expect that up to four of our 38 counterparts may be chance superpositions.

\subsection{Non-detections}
\label{nondet}

For the 75 ULXs for which we did not detect a candidate counterpart, we calculated the limiting magnitude at the position
of the ULX. In the $H$-band, the limiting apparent (absolute) magnitude ranges from 16
to 22 ($-13$ to $-6$), whereas in the $K_s$ band, the limiting apparent (absolute) magnitude ranges from 18 to 22 ($-12$ to $-7$). 
The distribution of these values can be seen in Figure~\ref{fig:limhistomags}. 

Examining Figure~\ref{fig:limhistomags} we can see that at the position of 67 out of these 75 ULXs, we reached the depth necessary to be
able to detect the bright end of the distribution of the absolute magnitude of RSGs, i.e. $H, K_s > -11$ mag. Of these 67 sources, for seven ULXs the limiting 
absolute magnitude at their position is fainter than the faintest absolute magnitude for a RSG ($H, K_s > -8$). We can draw one conclusion 
for these seven ULXs: the donor star is certainly not a RSG (see e.g. \citealt{2005ApJ...628..973L,2006MNRAS.372.1721K,2016ApJ...826..224M}). 
This means that for 60 ULXs, the limiting absolute magnitude at their position is within the range of a typical RSG, i.e. $-11<H, K_s<-8$. Of these 60 ULX, 
for 18 ULXs we probe the brightest end for a RSG ($-11<H, K_s<-10$, M3 spectral type or later). For 28 ULXs we probe 
the intermediate RSG range ($-10<H, K_s<-9$, K5 spectral type or later); and for 14 ULXs , we probe the faint end ($-9<H, K_s<-8$, 
K0 spectral type or later). 

We compare this distribution to the $K_s$ absolute magnitudes distribution of the 330 RSGs in the MW, the LMC, and M31. As can be seen from Figure~\ref{fig:both}, 
about 12$\%$ of these have absolute magnitudes brighter than $-11$ mag; almost one-third of the RSGs are in the brightest end for a RSG, i.e. $-11<H, K_s<-10$; 
$\simeq 40\%$ of the RSGs are between $-10$ and $-9$ mag and about 15$\%$ are in the faint end ($-9<H, K_s<-8$). In light of this, we can only rule out the presence of 
a bright RSG ($-11<H, K_s<-10$) for the 42 ULXs for which the limiting magnitude is $-10<H, K_s<-8$. For the remaining eight ULXs out of the 75 with non-detections, 
our limiting absolute magnitude at the position of the ULX is $H, K_s < -11$ mag, due to crowding or high background from the host galaxies. This prevents us from 
actually concluding if there is a RSG counterpart (or not) to these eight ULXs, as only 12$\%$ of RSGs are that bright.

\subsection{Spectroscopically observed NIR counterparts}
\label{followup}

At the time of writing this manuscript, only seven counterparts had been
spectroscopically followed-up by \citet{2015MNRAS.453.3510H}, \citet{2016MNRAS.459..771H} and 
\citet{2019MNRAS.489.1249L}. All seven sources have an absolute magnitude consistent with being RSGs.
\citet{2015MNRAS.453.3510H} confirmed the nature of a RSG in the galaxy NGC 253. \citet{2016MNRAS.459..771H} 
observed 5 sources and confirmed the nature of two more RSGs, in the galaxies NGC 925 and NGC 4136. 
Three sources turned out to be nebulae (in NGC 925, in Holmberg II and NGC 4136). 
\citet{2019MNRAS.489.1249L} followed-up a candidate RSG in NGC 3521 identified by \citet{2017MNRAS.469..671L}.
Its spectral signatures single this source out as a nebula. In this manuscript, we analyse the nature of five more sources
out of those 38 counterparts: four candidate RSGs and one candidate stellar cluster (see Table~\ref{tab:allcp}). Of the four candidate RSGs, 
one cannot be classified, the other one is most likely a background AGN and the other two sources show absorption lines consistent with them 
being RSGs. Additionally, we show that a source originally classified as a stellar cluster by \citet{2017MNRAS.469..671L} is actually a nebula (see Section~\ref{fig:x26}).

In summary, we have confirmed the RSG nature for five sources. We need to additionally corroborate if these confirmed RSGs are the actual donor 
stars of the ULXs by detecting radial velocity variations. If we corroborate this for our five confirmed RSGs, this would imply that  $\simeq (4 \pm 2)\%$ of the observed ULXs have a RSG 
donor star, about four times more than predicted by binary evolution models (see e.g. \citealt{2017ApJ...846...17W}).

Of our 11 followed-up candidate RSGs, $(36 \pm 18)\%$ are actually nebulae and $(45 \pm 20)\%$ are in fact RSGs, 
which potentially implies that of our remaining ten RSG candidate counterparts, half could be confirmed as RSG and the other half
as nebulae (i.e., we cannot distinguish a RSG from a nebulae based only on our broadband photometry). If these numbers hold, it means that
we will have ten confirmed RSGs for a sample of 113 observed ULXs. If we further corroborate their donor star nature, this would mean that
$\simeq (10 \pm 3)\%$ of the ULXs would have a RSG donor star, a number 10 times larger than predicted by \citet{2017ApJ...846...17W}.

\section{Conclusions}
\label{conclusions}

This is the third and last paper on the systematic photometric search for NIR counterparts to ULXs. We observed
23 ULXs in 15 galaxies and detected a candidate counterpart for six of them. For the ULXs with non-detections, 
we report limiting apparent and absolute magnitudes. Two ULXs have candidate counterparts with
absolute magnitudes consistent with a RSG. Three candidates are too bright to be single RSGs, and after visual 
inspection of their NIR images and/or determination of their size, we conclude that they are more likely to be star clusters. 
The remaining candidate counterpart is too faint to either be a RSG or a star cluster.
We also spectroscopically followed-up four RSG candidates identified by \citet{2014MNRAS.442.1054H} and a stellar cluster candidate identified by \citet{2017MNRAS.469..671L}. 
The stellar cluster candidate turned out to be a nebula, showing strong $[$Fe {\scshape II}$]$ emission. As for the RSG candidates, 
one source cannot be classified as significant absorption or emission lines were not detected, 
and another one has a spectrum dominated by a broad putative H$\alpha$ emission line, making it likely a background AGN at $z = 1.0301 \pm 0.0001$. 
However, the other two sources show stellar spectra consistent with them being RSGs.

Evaluating the complete search presented in \citet{2014MNRAS.442.1054H} and \citet{2017MNRAS.469..671L} and this work, 
we found that we have observed 113 ULXs in 52 different galaxies using 5 different telescopes and the $H, K_s$-bands. 
The ULXs are located at a distance range $0.5$ to $29$ kpc from the centre of their respective host galaxies. These galaxies 
are up to a distance of $\simeq 10$ Mpc.
We detected a NIR candidate counterpart for 38 ULXs out of the 113 ULXs, i.e. roughly one-third of the total sample.
In addition to these detections, we reached our expected depth, i.e. $H, K_s < -11$ at the position of 67 ULXs,
although we did not detect a counterpart. Of the 38 detected counterparts, we have 27 sources preliminarily classified based only on
broadband photometry. Ten of these candidate counterparts are candidate RSGs, 10 are candidate stellar clusters, six are candidate 
AGN and one is too faint to be either a RSG or a SC/AGN. We estimate the probability of a chance superposition for the whole sample of
113 ULXs as 3.3$\%$, which means that up to four of our detections could be false positives.

We have confirmed the nature of 11 sources by spectroscopic observations. For these, we have five RSGs, one AGN and five nebulae. As four of the 
five nebulae were originally classified as RSGs, it shows that RSGs and nebulae are difficult to distinguish in terms of broadband photometry. Moreover, 
if we can determine that our five spectroscopically confirmed RSGs are in fact the donor stars of the ULXs, via detection of radial variations, these possible ULX-RSG binary systems are in contrast with the predictions from the binary evolution simulations from \citet{2017ApJ...846...17W}.
Our results could be at least four times (and maybe even ten times) larger than predicted.

\section*{Acknowledgements}
We are infinitely grateful to the anonymous referee for her/his thorough comments in greatly improving this manuscript.
This research is based on observations made with: (a) the William Herschel Telescope operated on the island of La Palma by the Isaac Newton Group in the Spanish Observatorio del Roque de los Muchachos of the Instituto de Astrof\'{i}sica de Canarias; (b) ESO Telescopes at the La Silla Paranal Observatory; (c) the 5 m Hale Telescope at the Palomar Observatory; and (d) the Keck I Telescope at the W. M. Keck Observatory. We wish to recognize and acknowledge the very significant cultural role and reverence that the summit of Mauna Kea has always had within the indigenous Hawaiian community. We are most fortunate to have the opportunity to conduct observations from this mountain.
KML would like to thank Mischa Schirmer for his invaluable help with the data reduction pipeline {\scshape theli} and the setting/adding of new instruments to it.
We have made use of the SIMBAD database, operated at CDS, Strasbourg, France; of the NASA/IPAC Extragalactic Database (NED) which is operated by the Jet Propulsion Laboratory, California Institute of Technology, under contract with the National Aeronautics and Space Administration; and of data obtained from the Chandra Data Archive and the Chandra Source Catalog. PGJ and KML acknowledge funding from the European Research Council under ERC Consolidator Grant agreement no 647208. MH acknowledges the ESO fellowship program. MAPT acknowledges support via a Ram\'on y Cajal Fellowship (RYC-2015-17854) and support by the Spanish Ministry of Economy, Industry and Competitiveness under grant AYA2017-83216-P. TPR acknowledges funding from Science and Technology Facilities Council (STFC) as part of the consolidated grant ST/L00075X/1. DJW acknowledges support from an STFC Ernest Rutherford Fellowship.

%%%%%%%%%%%%%%%%%%%%%%%%%%%%%%%%%%%%%%%%%%%%%%%%%%



\bibliographystyle{mnras}
\bibliography{library}

\begin{thebibliography}{}
\makeatletter
\relax
\def\mn@urlcharsother{\let\do\@makeother \do\$\do\&\do\#\do\^\do\_\do\%\do\~}
\def\mn@doi{\begingroup\mn@urlcharsother \@ifnextchar [ {\mn@doi@}
  {\mn@doi@[]}}
\def\mn@doi@[#1]#2{\def\@tempa{#1}\ifx\@tempa\@empty \href
  {http://dx.doi.org/#2} {doi:#2}\else \href {http://dx.doi.org/#2} {#1}\fi
  \endgroup}
\def\mn@eprint#1#2{\mn@eprint@#1:#2::\@nil}
\def\mn@eprint@arXiv#1{\href {http://arxiv.org/abs/#1} {{\tt arXiv:#1}}}
\def\mn@eprint@dblp#1{\href {http://dblp.uni-trier.de/rec/bibtex/#1.xml}
  {dblp:#1}}
\def\mn@eprint@#1:#2:#3:#4\@nil{\def\@tempa {#1}\def\@tempb {#2}\def\@tempc
  {#3}\ifx \@tempc \@empty \let \@tempc \@tempb \let \@tempb \@tempa \fi \ifx
  \@tempb \@empty \def\@tempb {arXiv}\fi \@ifundefined
  {mn@eprint@\@tempb}{\@tempb:\@tempc}{\expandafter \expandafter \csname
  mn@eprint@\@tempb\endcsname \expandafter{\@tempc}}}

\bibitem[\protect\citeauthoryear{{Abbott} et~al.,}{{Abbott}
  et~al.}{2016a}]{2016PhRvX...6d1015A}
{Abbott} B.~P.,  et~al., 2016a, \mn@doi [Physical Review X]
  {10.1103/PhysRevX.6.041015}, \href
  {https://ui.adsabs.harvard.edu/abs/2016PhRvX...6d1015A} {6, 041015}

\bibitem[\protect\citeauthoryear{{Abbott} et~al.,}{{Abbott}
  et~al.}{2016b}]{2016PhRvL.116f1102A}
{Abbott} B.~P.,  et~al., 2016b, \mn@doi [Physical Review Letters]
  {10.1103/PhysRevLett.116.061102}, \href
  {http://adsabs.harvard.edu/abs/2016PhRvL.116f1102A} {116, 061102}

\bibitem[\protect\citeauthoryear{{Bachetti} et~al.,}{{Bachetti}
  et~al.}{2014}]{2014Natur.514..202B}
{Bachetti} M.,  et~al., 2014, \mn@doi [\nat] {10.1038/nature13791}, \href
  {http://adsabs.harvard.edu/abs/2014Natur.514..202B} {514, 202}

\bibitem[\protect\citeauthoryear{{Begelman}}{{Begelman}}{2002}]{2002ApJ...568L..97B}
{Begelman} M.~C.,  2002, \mn@doi [\apjl] {10.1086/340457}, \href
  {http://adsabs.harvard.edu/abs/2002ApJ...568L..97B} {568, L97}

\bibitem[\protect\citeauthoryear{{Bertin}}{{Bertin}}{2006}]{2006ASPC..351..112B}
{Bertin} E.,  2006, in {Gabriel} C.,  {Arviset} C.,  {Ponz} D.,   {Enrique} S.,
   eds,  Astronomical Society of the Pacific Conference Series Vol. 351,
  Astronomical Data Analysis Software and Systems XV. p.~112

\bibitem[\protect\citeauthoryear{{Bertin} \& {Arnouts}}{{Bertin} \&
  {Arnouts}}{1996}]{1996A&AS..117..393B}
{Bertin} E.,  {Arnouts} S.,  1996, \mn@doi [\aaps] {10.1051/aas:1996164}, \href
  {http://adsabs.harvard.edu/abs/1996A%26AS..117..393B} {117, 393}

\bibitem[\protect\citeauthoryear{{Bertin}, {Mellier}, {Radovich}, {Missonnier},
  {Didelon}  \& {Morin}}{{Bertin} et~al.}{2002}]{2002ASPC..281..228B}
{Bertin} E.,  {Mellier} Y.,  {Radovich} M.,  {Missonnier} G.,  {Didelon} P.,
  {Morin} B.,  2002, in {Bohlender} D.~A.,  {Durand} D.,   {Handley} T.~H.,
  eds,  Astronomical Society of the Pacific Conference Series Vol. 281,
  Astronomical Data Analysis Software and Systems XI. p.~228

\bibitem[\protect\citeauthoryear{{Bohlin}, {Savage}  \& {Drake}}{{Bohlin}
  et~al.}{1978}]{1978ApJ...224..132B}
{Bohlin} R.~C.,  {Savage} B.~D.,   {Drake} J.~F.,  1978, \mn@doi [\apj]
  {10.1086/156357}, \href
  {https://ui.adsabs.harvard.edu/abs/1978ApJ...224..132B} {224, 132}

\bibitem[\protect\citeauthoryear{{Carpano}, {Haberl}, {Maitra}  \&
  {Vasilopoulos}}{{Carpano} et~al.}{2018}]{2018MNRAS.476L..45C}
{Carpano} S.,  {Haberl} F.,  {Maitra} C.,   {Vasilopoulos} G.,  2018, \mn@doi
  [Monthly Notices of the Royal Astronomical Society] {10.1093/mnrasl/sly030},
  \href {https://ui.adsabs.harvard.edu/abs/2018MNRAS.476L..45C} {476, L45}

\bibitem[\protect\citeauthoryear{{Casares} \& {Jonker}}{{Casares} \&
  {Jonker}}{2014}]{2014SSRv..183..223C}
{Casares} J.,  {Jonker} P.~G.,  2014, \mn@doi [\ssr]
  {10.1007/s11214-013-0030-6}, \href
  {http://adsabs.harvard.edu/abs/2014SSRv..183..223C} {183, 223}

\bibitem[\protect\citeauthoryear{{Coatman}, {Hewett}, {Banerji}  \&
  {Richards}}{{Coatman} et~al.}{2016}]{2016MNRAS.461..647C}
{Coatman} L.,  {Hewett} P.~C.,  {Banerji} M.,   {Richards} G.~T.,  2016,
  \mn@doi [\mnras] {10.1093/mnras/stw1360}, \href
  {https://ui.adsabs.harvard.edu/abs/2016MNRAS.461..647C} {461, 647}

\bibitem[\protect\citeauthoryear{{Colbert}, {Heckman}, {Ptak}, {Strickland}  \&
  {Weaver}}{{Colbert} et~al.}{2004}]{2004ApJ...602..231C}
{Colbert} E.~J.~M.,  {Heckman} T.~M.,  {Ptak} A.~F.,  {Strickland} D.~K.,
  {Weaver} K.~A.,  2004, \mn@doi [\apj] {10.1086/380899}, \href
  {https://ui.adsabs.harvard.edu/abs/2004ApJ...602..231C} {602, 231}

\bibitem[\protect\citeauthoryear{{Copperwheat}, {Cropper}, {Soria}  \&
  {Wu}}{{Copperwheat} et~al.}{2005}]{2005MNRAS.362...79C}
{Copperwheat} C.,  {Cropper} M.,  {Soria} R.,   {Wu} K.,  2005, \mn@doi
  [\mnras] {10.1111/j.1365-2966.2005.09223.x}, \href
  {http://adsabs.harvard.edu/abs/2005MNRAS.362...79C} {362, 79}

\bibitem[\protect\citeauthoryear{{Copperwheat}, {Cropper}, {Soria}  \&
  {Wu}}{{Copperwheat} et~al.}{2007}]{2007MNRAS.376.1407C}
{Copperwheat} C.,  {Cropper} M.,  {Soria} R.,   {Wu} K.,  2007, \mn@doi
  [\mnras] {10.1111/j.1365-2966.2007.11551.x}, \href
  {http://adsabs.harvard.edu/abs/2007MNRAS.376.1407C} {376, 1407}

\bibitem[\protect\citeauthoryear{{Dobrzycki}, {Ebeling}, {Glotfelty},
  {Freeman}, {Damiani}, {Elvis}  \& {Calderwood}}{{Dobrzycki}
  et~al.}{1999}]{ciao}
{Dobrzycki} A.,  {Ebeling} H.,  {Glotfelty} K.,  {Freeman} P.,  {Damiani} F.,
  {Elvis} M.,   {Calderwood} T.,  1999. p.~142

\bibitem[\protect\citeauthoryear{{Drilling} \& {Landolt}}{{Drilling} \&
  {Landolt}}{2000}]{2000asqu.book..381D}
{Drilling} J.~S.,  {Landolt} A.~U.,  2000, {Normal Stars}.
p.~381

\bibitem[\protect\citeauthoryear{{Dubus} \& {Rutledge}}{{Dubus} \&
  {Rutledge}}{2002}]{2002MNRAS.336..901D}
{Dubus} G.,  {Rutledge} R.~E.,  2002, \mn@doi [\mnras]
  {10.1046/j.1365-8711.2002.05827.x}, \href
  {https://ui.adsabs.harvard.edu/abs/2002MNRAS.336..901D} {336, 901}

\bibitem[\protect\citeauthoryear{{Dubus}, {Charles}  \& {Long}}{{Dubus}
  et~al.}{2004}]{2004A&A...425...95D}
{Dubus} G.,  {Charles} P.~A.,   {Long} K.~S.,  2004, \mn@doi [\aap]
  {10.1051/0004-6361:20041253}, \href
  {https://ui.adsabs.harvard.edu/abs/2004A%26A...425...95D} {425, 95}

\bibitem[\protect\citeauthoryear{{Earnshaw}}{{Earnshaw}}{2016}]{2016AN....337..448E}
{Earnshaw} H.~M.,  2016, \mn@doi [Astronomische Nachrichten]
  {10.1002/asna.201612328}, \href
  {https://ui.adsabs.harvard.edu/abs/2016AN....337..448E} {337, 448}

\bibitem[\protect\citeauthoryear{{Elias}, {Frogel}  \& {Humphreys}}{{Elias}
  et~al.}{1985}]{1985ApJS...57...91E}
{Elias} J.~H.,  {Frogel} J.~A.,   {Humphreys} R.~M.,  1985, \mn@doi [\apjs]
  {10.1086/190997}, \href {http://adsabs.harvard.edu/abs/1985ApJS...57...91E}
  {57, 91}

\bibitem[\protect\citeauthoryear{{Fabbiano}, {Zezas}  \& {Murray}}{{Fabbiano}
  et~al.}{2001}]{2001ApJ...554.1035F}
{Fabbiano} G.,  {Zezas} A.,   {Murray} S.~S.,  2001, \mn@doi [\apj]
  {10.1086/321397}, \href {http://adsabs.harvard.edu/abs/2001ApJ...554.1035F}
  {554, 1035}

\bibitem[\protect\citeauthoryear{{Fabrika}, {Ueda}, {Vinokurov}, {Sholukhova}
  \& {Shidatsu}}{{Fabrika} et~al.}{2015}]{2015NatPh..11..551F}
{Fabrika} S.,  {Ueda} Y.,  {Vinokurov} A.,  {Sholukhova} O.,   {Shidatsu} M.,
  2015, \mn@doi [Nature Physics] {10.1038/nphys3348}, \href
  {http://cdsads.u-strasbg.fr/abs/2015NatPh..11..551F} {11, 551}

\bibitem[\protect\citeauthoryear{{Farrell} et~al.,}{{Farrell}
  et~al.}{2011}]{2011AN....332..392F}
{Farrell} S.~A.,  et~al., 2011, \mn@doi [Astronomische Nachrichten]
  {10.1002/asna.201011507}, \href
  {https://ui.adsabs.harvard.edu/abs/2011AN....332..392F} {332, 392}

\bibitem[\protect\citeauthoryear{Fitzpatrick}{Fitzpatrick}{1999}]{Fitzpatrick_1999}
Fitzpatrick E.~L.,  1999, \mn@doi [Publications of the Astronomical Society of
  the Pacific] {10.1086/316293}, 111, 63

\bibitem[\protect\citeauthoryear{{Fraternali}, {van Moorsel}, {Sancisi}  \&
  {Oosterloo}}{{Fraternali} et~al.}{2002}]{2002AJ....123.3124F}
{Fraternali} F.,  {van Moorsel} G.,  {Sancisi} R.,   {Oosterloo} T.,  2002,
  \mn@doi [\aj] {10.1086/340358}, \href
  {https://ui.adsabs.harvard.edu/abs/2002AJ....123.3124F} {123, 3124}

\bibitem[\protect\citeauthoryear{{F{\"u}rst} et~al.,}{{F{\"u}rst}
  et~al.}{2016}]{2016ApJ...831L..14F}
{F{\"u}rst} F.,  et~al., 2016, \mn@doi [\apjl] {10.3847/2041-8205/831/2/L14},
  \href {http://adsabs.harvard.edu/abs/2016ApJ...831L..14F} {831, L14}

\bibitem[\protect\citeauthoryear{{Gaia Collaboration} et~al.,}{{Gaia
  Collaboration} et~al.}{2018}]{2018A&A...616A...1G}
{Gaia Collaboration} et~al., 2018, \mn@doi [\aap]
  {10.1051/0004-6361/201833051}, \href
  {https://ui.adsabs.harvard.edu/abs/2018A&A...616A...1G} {616, A1}

\bibitem[\protect\citeauthoryear{{Gladstone}, {Roberts}  \& {Done}}{{Gladstone}
  et~al.}{2009}]{2009MNRAS.397.1836G}
{Gladstone} J.~C.,  {Roberts} T.~P.,   {Done} C.,  2009, \mn@doi [\mnras]
  {10.1111/j.1365-2966.2009.15123.x}, \href
  {http://adsabs.harvard.edu/abs/2009MNRAS.397.1836G} {397, 1836}

\bibitem[\protect\citeauthoryear{{Gladstone}, {Copperwheat}, {Heinke},
  {Roberts}, {Cartwright}, {Levan}  \& {Goad}}{{Gladstone}
  et~al.}{2013}]{2013ApJS..206...14G}
{Gladstone} J.~C.,  {Copperwheat} C.,  {Heinke} C.~O.,  {Roberts} T.~P.,
  {Cartwright} T.~F.,  {Levan} A.~J.,   {Goad} M.~R.,  2013, \mn@doi [\apjs]
  {10.1088/0067-0049/206/2/14}, \href
  {http://adsabs.harvard.edu/abs/2013ApJS..206...14G} {206, 14}

\bibitem[\protect\citeauthoryear{{Gris{\'e}}, {Pakull}, {Soria}  \&
  {Motch}}{{Gris{\'e}} et~al.}{2009}]{2009AIPC.1126..201G}
{Gris{\'e}} F.,  {Pakull} M.~W.,  {Soria} R.,   {Motch} C.,  2009, in
  {Rodriguez} J.,  {Ferrando} P.,  eds,  American Institute of Physics
  Conference Series Vol. 1126, American Institute of Physics Conference Series.
  pp 201--203 (\mn@eprint {arXiv} {0902.4431}), \mn@doi{10.1063/1.3149412}

\bibitem[\protect\citeauthoryear{{Guti{\'e}rrez} \&
  {L{\'o}pez-Corredoira}}{{Guti{\'e}rrez} \&
  {L{\'o}pez-Corredoira}}{2006}]{2006IAUS..230..310G}
{Guti{\'e}rrez} C.~M.,  {L{\'o}pez-Corredoira} M.,  2006, in {Meurs} E.~J.~A.,
  {Fabbiano} G.,  eds,  IAU Symposium Vol. 230, Populations of High Energy
  Sources in Galaxies. pp 310--311, \mn@doi{10.1017/S1743921306008556}

\bibitem[\protect\citeauthoryear{{HI4PI Collaboration} et~al.,}{{HI4PI
  Collaboration} et~al.}{2016}]{2016AA...594A.116H}
{HI4PI Collaboration} et~al., 2016, \mn@doi [\aap]
  {10.1051/0004-6361/201629178}, \href
  {https://ui.adsabs.harvard.edu/abs/2016A&A...594A.116H} {594, A116}

\bibitem[\protect\citeauthoryear{{Heida} et~al.,}{{Heida}
  et~al.}{2014}]{2014MNRAS.442.1054H}
{Heida} M.,  et~al., 2014, \mn@doi [\mnras] {10.1093/mnras/stu928}, \href
  {http://adsabs.harvard.edu/abs/2014MNRAS.442.1054H} {442, 1054}

\bibitem[\protect\citeauthoryear{{Heida} et~al.,}{{Heida}
  et~al.}{2015a}]{2015MNRAS.453.3510H}
{Heida} M.,  et~al., 2015a, \mn@doi [\mnras] {10.1093/mnras/stv1853}, \href
  {http://adsabs.harvard.edu/abs/2015MNRAS.453.3510H} {453, 3510}

\bibitem[\protect\citeauthoryear{{Heida}, {Jonker}  \& {Torres}}{{Heida}
  et~al.}{2015b}]{2015MNRAS.454L..26H}
{Heida} M.,  {Jonker} P.~G.,   {Torres} M.~A.~P.,  2015b, \mn@doi [\mnras]
  {10.1093/mnrasl/slv121}, \href
  {https://ui.adsabs.harvard.edu/abs/2015MNRAS.454L..26H} {454, L26}

\bibitem[\protect\citeauthoryear{{Heida}, {Jonker}, {Torres}, {Roberts},
  {Walton}, {Moon}, {Stern}  \& {Harrison}}{{Heida}
  et~al.}{2016}]{2016MNRAS.459..771H}
{Heida} M.,  {Jonker} P.~G.,  {Torres} M.~A.~P.,  {Roberts} T.~P.,  {Walton}
  D.~J.,  {Moon} D.~S.,  {Stern} D.,   {Harrison} F.~A.,  2016, \mn@doi
  [\mnras] {10.1093/mnras/stw695}, \href
  {https://ui.adsabs.harvard.edu/abs/2016MNRAS.459..771H} {459, 771}

\bibitem[\protect\citeauthoryear{{Heida} et~al.,}{{Heida}
  et~al.}{2019}]{2019ApJ...883L..34H}
{Heida} M.,  et~al., 2019, \mn@doi [\apjl] {10.3847/2041-8213/ab4139}, \href
  {https://ui.adsabs.harvard.edu/abs/2019ApJ...883L..34H} {883, L34}

\bibitem[\protect\citeauthoryear{{Hu} et~al.,}{{Hu}
  et~al.}{2018}]{2018ApJ...854...68H}
{Hu} N.,  et~al., 2018, \mn@doi [\apj] {10.3847/1538-4357/aaa6ca}, \href
  {https://ui.adsabs.harvard.edu/abs/2018ApJ...854...68H} {854, 68}

\bibitem[\protect\citeauthoryear{{Israel} et~al.,}{{Israel}
  et~al.}{2017a}]{2017Sci...355..817I}
{Israel} G.~L.,  et~al., 2017a, \mn@doi [Science] {10.1126/science.aai8635},
  \href {https://ui.adsabs.harvard.edu/abs/2017Sci...355..817I} {355, 817}

\bibitem[\protect\citeauthoryear{{Israel} et~al.,}{{Israel}
  et~al.}{2017b}]{2017MNRAS.466L..48I}
{Israel} G.~L.,  et~al., 2017b, \mn@doi [\mnras] {10.1093/mnrasl/slw218}, \href
  {http://adsabs.harvard.edu/abs/2017MNRAS.466L..48I} {466, L48}

\bibitem[\protect\citeauthoryear{{Jonker}, {Torres}, {Fabian}, {Heida},
  {Miniutti}  \& {Pooley}}{{Jonker} et~al.}{2010}]{2010MNRAS.407..645J}
{Jonker} P.~G.,  {Torres} M.~A.~P.,  {Fabian} A.~C.,  {Heida} M.,  {Miniutti}
  G.,   {Pooley} D.,  2010, \mn@doi [\mnras]
  {10.1111/j.1365-2966.2010.16943.x}, \href
  {http://adsabs.harvard.edu/abs/2010MNRAS.407..645J} {407, 645}

\bibitem[\protect\citeauthoryear{{Kaaret}, {Feng}  \& {Roberts}}{{Kaaret}
  et~al.}{2017}]{2017ARA&A..55..303K}
{Kaaret} P.,  {Feng} H.,   {Roberts} T.~P.,  2017, \mn@doi [Annual Review of
  Astronomy and Astrophysics] {10.1146/annurev-astro-091916-055259}, \href
  {https://ui.adsabs.harvard.edu/abs/2017ARA&A..55..303K} {55, 303}

\bibitem[\protect\citeauthoryear{{Kalirai} et~al.,}{{Kalirai}
  et~al.}{2009}]{2009wfc..rept...21K}
{Kalirai} J.~S.,  et~al., 2009, Technical report, {WFC3 SMOV Proposal 11450:
  The Photometric Performance and Calibration of WFC3/UVIS}

\bibitem[\protect\citeauthoryear{{Kalogera} \& {Baym}}{{Kalogera} \&
  {Baym}}{1996}]{1996ApJ...470L..61K}
{Kalogera} V.,  {Baym} G.,  1996, \mn@doi [\apjl] {10.1086/310296}, \href
  {http://adsabs.harvard.edu/abs/1996ApJ...470L..61K} {470, L61}

\bibitem[\protect\citeauthoryear{{Kausch} et~al.,}{{Kausch}
  et~al.}{2015}]{2015A&A...576A..78K}
{Kausch} W.,  et~al., 2015, \mn@doi [\aap] {10.1051/0004-6361/201423909}, \href
  {https://ui.adsabs.harvard.edu/abs/2015A&A...576A..78K} {576, A78}

\bibitem[\protect\citeauthoryear{{King}, {Davies}, {Ward}, {Fabbiano}  \&
  {Elvis}}{{King} et~al.}{2001}]{2001ApJ...552L.109K}
{King} A.~R.,  {Davies} M.~B.,  {Ward} M.~J.,  {Fabbiano} G.,   {Elvis} M.,
  2001, \mn@doi [\apjl] {10.1086/320343}, \href
  {http://adsabs.harvard.edu/abs/2001ApJ...552L.109K} {552, L109}

\bibitem[\protect\citeauthoryear{{Kiss}, {Szab{\'o}}  \& {Bedding}}{{Kiss}
  et~al.}{2006}]{2006MNRAS.372.1721K}
{Kiss} L.~L.,  {Szab{\'o}} G.~M.,   {Bedding} T.~R.,  2006, \mn@doi [\mnras]
  {10.1111/j.1365-2966.2006.10973.x}, \href
  {https://ui.adsabs.harvard.edu/abs/2006MNRAS.372.1721K} {372, 1721}

\bibitem[\protect\citeauthoryear{{Kormendy} \& {Ho}}{{Kormendy} \&
  {Ho}}{2013}]{2013ARA&A..51..511K}
{Kormendy} J.,  {Ho} L.~C.,  2013, \mn@doi [\araa]
  {10.1146/annurev-astro-082708-101811}, \href
  {https://ui.adsabs.harvard.edu/abs/2013ARA&A..51..511K} {51, 511}

\bibitem[\protect\citeauthoryear{{Lan{\c{c}}on}, {Hauschildt}, {Ladjal}  \&
  {Mouhcine}}{{Lan{\c{c}}on} et~al.}{2007}]{2007A&A...468..205L}
{Lan{\c{c}}on} A.,  {Hauschildt} P.~H.,  {Ladjal} D.,   {Mouhcine} M.,  2007,
  \mn@doi [\aap] {10.1051/0004-6361:20065824}, \href
  {https://ui.adsabs.harvard.edu/abs/2007A&A...468..205L} {468, 205}

\bibitem[\protect\citeauthoryear{{Lau} et~al.,}{{Lau}
  et~al.}{2019}]{2019ApJ...878...71L}
{Lau} R.~M.,  et~al., 2019, \mn@doi [\apj] {10.3847/1538-4357/ab1b1c}, \href
  {https://ui.adsabs.harvard.edu/abs/2019ApJ...878...71L} {878, 71}

\bibitem[\protect\citeauthoryear{{Levesque}, {Massey}, {Olsen}, {Plez},
  {Josselin}, {Maeder}  \& {Meynet}}{{Levesque}
  et~al.}{2005}]{2005ApJ...628..973L}
{Levesque} E.~M.,  {Massey} P.,  {Olsen} K.~A.~G.,  {Plez} B.,  {Josselin} E.,
  {Maeder} A.,   {Meynet} G.,  2005, \mn@doi [\apj] {10.1086/430901}, \href
  {https://ui.adsabs.harvard.edu/abs/2005ApJ...628..973L} {628, 973}

\bibitem[\protect\citeauthoryear{{Li} et~al.,}{{Li}
  et~al.}{2010}]{2010ApJ...721.1368L}
{Li} Z.,  et~al., 2010, \mn@doi [\apj] {10.1088/0004-637X/721/2/1368}, \href
  {https://ui.adsabs.harvard.edu/abs/2010ApJ...721.1368L} {721, 1368}

\bibitem[\protect\citeauthoryear{{Lin}, {Webb}  \& {Barret}}{{Lin}
  et~al.}{2012}]{2012ApJ...756...27L}
{Lin} D.,  {Webb} N.~A.,   {Barret} D.,  2012, \mn@doi [\apj]
  {10.1088/0004-637X/756/1/27}, \href
  {https://ui.adsabs.harvard.edu/abs/2012ApJ...756...27L} {756, 27}

\bibitem[\protect\citeauthoryear{{Liu}}{{Liu}}{2011}]{2011ApJS..192...10L}
{Liu} J.,  2011, \mn@doi [\apjs] {10.1088/0067-0049/192/1/10}, \href
  {http://adsabs.harvard.edu/abs/2011ApJS..192...10L} {192, 10}

\bibitem[\protect\citeauthoryear{{Liu}, {Bregman}, {Miller}  \& {Kaaret}}{{Liu}
  et~al.}{2007}]{2007ApJ...661..165L}
{Liu} J.-F.,  {Bregman} J.,  {Miller} J.,   {Kaaret} P.,  2007, \mn@doi [\apj]
  {10.1086/516624}, \href
  {https://ui.adsabs.harvard.edu/abs/2007ApJ...661..165L} {661, 165}

\bibitem[\protect\citeauthoryear{{Long}, {Dodorico}, {Charles}  \&
  {Dopita}}{{Long} et~al.}{1981}]{1981ApJ...246L..61L}
{Long} K.~S.,  {Dodorico} S.,  {Charles} P.~A.,   {Dopita} M.~A.,  1981,
  \mn@doi [\apjl] {10.1086/183553}, \href
  {https://ui.adsabs.harvard.edu/abs/1981ApJ...246L..61L} {246, L61}

\bibitem[\protect\citeauthoryear{{L{\'o}pez}, {Heida}, {Jonker}, {Torres},
  {Roberts}, {Walton}, {Moon}  \& {Harrison}}{{L{\'o}pez}
  et~al.}{2017}]{2017MNRAS.469..671L}
{L{\'o}pez} K.~M.,  {Heida} M.,  {Jonker} P.~G.,  {Torres} M.~A.~P.,  {Roberts}
  T.~P.,  {Walton} D.~J.,  {Moon} D.-S.,   {Harrison} F.~A.,  2017, \mn@doi
  [\mnras] {10.1093/mnras/stx857}, \href
  {http://cdsads.u-strasbg.fr/abs/2017MNRAS.469..671L} {469, 671}

\bibitem[\protect\citeauthoryear{{L{\'o}pez}, {Jonker}, {Heida}, {Torres},
  {Roberts}, {Walton}, {Moon}  \& {Harrison}}{{L{\'o}pez}
  et~al.}{2019}]{2019MNRAS.489.1249L}
{L{\'o}pez} K.~M.,  {Jonker} P.~G.,  {Heida} M.,  {Torres} M.~A.~P.,  {Roberts}
  T.~P.,  {Walton} D.~J.,  {Moon} D.~S.,   {Harrison} F.~A.,  2019, \mn@doi
  [\mnras] {10.1093/mnras/stz2127}, \href
  {https://ui.adsabs.harvard.edu/abs/2019MNRAS.489.1249L} {489, 1249}

\bibitem[\protect\citeauthoryear{{Manchado} et~al.,}{{Manchado}
  et~al.}{1998}]{1998SPIE.3354..448M}
{Manchado} A.,  et~al., 1998, in {Fowler} A.~M.,  ed.,  Society of
  Photo-Optical Instrumentation Engineers (SPIE) Conference Series Vol. 3354,
  \procspie. pp 448--455, \mn@doi{10.1117/12.317270}

\bibitem[\protect\citeauthoryear{{Markert} \& {Rallis}}{{Markert} \&
  {Rallis}}{1983}]{1983ApJ...275..571M}
{Markert} T.~H.,  {Rallis} A.~D.,  1983, \mn@doi [\apj] {10.1086/161555}, \href
  {https://ui.adsabs.harvard.edu/abs/1983ApJ...275..571M} {275, 571}

\bibitem[\protect\citeauthoryear{{Massey} \& {Evans}}{{Massey} \&
  {Evans}}{2016}]{2016ApJ...826..224M}
{Massey} P.,  {Evans} K.~A.,  2016, \mn@doi [\apj]
  {10.3847/0004-637X/826/2/224}, \href
  {https://ui.adsabs.harvard.edu/abs/2016ApJ...826..224M} {826, 224}

\bibitem[\protect\citeauthoryear{{McLean} et~al.,}{{McLean}
  et~al.}{2010}]{2010SPIE.7735E..1EM}
{McLean} I.~S.,  et~al., 2010, in \procspie. p. 77351E,
  \mn@doi{10.1117/12.856715}

\bibitem[\protect\citeauthoryear{{McLean} et~al.,}{{McLean}
  et~al.}{2012}]{2012SPIE.8446E..0JM}
{McLean} I.~S.,  et~al., 2012, in \procspie. p. 84460J,
  \mn@doi{10.1117/12.924794}

\bibitem[\protect\citeauthoryear{{McQuinn}, {Skillman}, {Dolphin}, {Berg}  \&
  {Kennicutt}}{{McQuinn} et~al.}{2017}]{2017AJ....154...51M}
{McQuinn} K. B.~W.,  {Skillman} E.~D.,  {Dolphin} A.~E.,  {Berg} D.,
  {Kennicutt} R.,  2017, \mn@doi [\aj] {10.3847/1538-3881/aa7aad}, \href
  {https://ui.adsabs.harvard.edu/abs/2017AJ....154...51M} {154, 51}

\bibitem[\protect\citeauthoryear{{Merritt}, {Schnittman}  \&
  {Komossa}}{{Merritt} et~al.}{2009}]{2009ApJ...699.1690M}
{Merritt} D.,  {Schnittman} J.~D.,   {Komossa} S.,  2009, \mn@doi [\apj]
  {10.1088/0004-637X/699/2/1690}, \href
  {http://cdsads.u-strasbg.fr/abs/2009ApJ...699.1690M} {699, 1690}

\bibitem[\protect\citeauthoryear{{Mezcua}, {Roberts}, {Sutton}  \&
  {Lobanov}}{{Mezcua} et~al.}{2013}]{2013MNRAS.436.3128M}
{Mezcua} M.,  {Roberts} T.~P.,  {Sutton} A.~D.,   {Lobanov} A.~P.,  2013,
  \mn@doi [Monthly Notices of the Royal Astronomical Society]
  {10.1093/mnras/stt1794}, \href
  {https://ui.adsabs.harvard.edu/abs/2013MNRAS.436.3128M} {436, 3128}

\bibitem[\protect\citeauthoryear{{Mineo}, {Gilfanov}  \& {Sunyaev}}{{Mineo}
  et~al.}{2012}]{2012MNRAS.419.2095M}
{Mineo} S.,  {Gilfanov} M.,   {Sunyaev} R.,  2012, \mn@doi [\mnras]
  {10.1111/j.1365-2966.2011.19862.x}, \href
  {https://ui.adsabs.harvard.edu/abs/2012MNRAS.419.2095M} {419, 2095}

\bibitem[\protect\citeauthoryear{{Moon}, {Eikenberry}  \& {Wasserman}}{{Moon}
  et~al.}{2003}]{2003ApJ...586.1280M}
{Moon} D.-S.,  {Eikenberry} S.~S.,   {Wasserman} I.~M.,  2003, \mn@doi [\apj]
  {10.1086/367826}, \href {http://adsabs.harvard.edu/abs/2003ApJ...586.1280M}
  {586, 1280}

\bibitem[\protect\citeauthoryear{{Moorwood}, {Cuby}  \& {Lidman}}{{Moorwood}
  et~al.}{1998}]{1998Msngr..91....9M}
{Moorwood} A.,  {Cuby} J.-G.,   {Lidman} C.,  1998, The Messenger, \href
  {https://ui.adsabs.harvard.edu/abs/1998Msngr..91....9M} {91, 9}

\bibitem[\protect\citeauthoryear{{Motch}, {Pakull}, {Gris{\'e}}  \&
  {Soria}}{{Motch} et~al.}{2011}]{2011AN....332..367M}
{Motch} C.,  {Pakull} M.~W.,  {Gris{\'e}} F.,   {Soria} R.,  2011, \mn@doi
  [Astronomische Nachrichten] {10.1002/asna.201011501}, \href
  {https://ui.adsabs.harvard.edu/abs/2011AN....332..367M} {332, 367}

\bibitem[\protect\citeauthoryear{{Motch}, {Pakull}, {Soria}, {Gris{\'e}}  \&
  {Pietrzy{\'n}ski}}{{Motch} et~al.}{2014}]{2014Natur.514..198M}
{Motch} C.,  {Pakull} M.~W.,  {Soria} R.,  {Gris{\'e}} F.,   {Pietrzy{\'n}ski}
  G.,  2014, \mn@doi [\nat] {10.1038/nature13730}, \href
  {http://adsabs.harvard.edu/abs/2014Natur.514..198M} {514, 198}

\bibitem[\protect\citeauthoryear{{Mucciarelli}, {Zampieri}, {Falomo}, {Turolla}
   \& {Treves}}{{Mucciarelli} et~al.}{2005}]{2005ApJ...633L.101M}
{Mucciarelli} P.,  {Zampieri} L.,  {Falomo} R.,  {Turolla} R.,   {Treves} A.,
  2005, \mn@doi [\apjl] {10.1086/498448}, \href
  {https://ui.adsabs.harvard.edu/abs/2005ApJ...633L.101M} {633, L101}

\bibitem[\protect\citeauthoryear{{Osterbrock}}{{Osterbrock}}{1989}]{1989agna.book.....O}
{Osterbrock} D.~E.,  1989, {Astrophysics of gaseous nebulae and active galactic
  nuclei}

\bibitem[\protect\citeauthoryear{{Pakull} \& {Gris{\'e}}}{{Pakull} \&
  {Gris{\'e}}}{2008}]{2008AIPC.1010..303P}
{Pakull} M.~W.,  {Gris{\'e}} F.,  2008, in {Bandyopadhyay} R.~M.,  {Wachter}
  S.,  {Gelino} D.,   {Gelino} C.~R.,  eds,  American Institute of Physics
  Conference Series Vol. 1010, A Population Explosion: The Nature \& Evolution
  of X-ray Binaries in Diverse Environments. pp 303--307 (\mn@eprint {arXiv}
  {0803.4345}), \mn@doi{10.1063/1.2945062}

\bibitem[\protect\citeauthoryear{{Pakull}, {Gris{\'e}}  \& {Motch}}{{Pakull}
  et~al.}{2006}]{2006IAUS..230..293P}
{Pakull} M.~W.,  {Gris{\'e}} F.,   {Motch} C.,  2006, in {Meurs} E.~J.~A.,
  {Fabbiano} G.,  eds,  IAU Symposium Vol. 230, Populations of High Energy
  Sources in Galaxies. pp 293--297 (\mn@eprint {arXiv} {astro-ph/0603771}),
  \mn@doi{10.1017/S1743921306008489}

\bibitem[\protect\citeauthoryear{{Patruno} \& {Zampieri}}{{Patruno} \&
  {Zampieri}}{2008}]{2008MNRAS.386..543P}
{Patruno} A.,  {Zampieri} L.,  2008, \mn@doi [\mnras]
  {10.1111/j.1365-2966.2008.13063.x}, \href
  {http://adsabs.harvard.edu/abs/2008MNRAS.386..543P} {386, 543}

\bibitem[\protect\citeauthoryear{{Poutanen}, {Fabrika}, {Valeev}, {Sholukhova}
  \& {Greiner}}{{Poutanen} et~al.}{2013}]{2013MNRAS.432..506P}
{Poutanen} J.,  {Fabrika} S.,  {Valeev} A.~F.,  {Sholukhova} O.,   {Greiner}
  J.,  2013, \mn@doi [\mnras] {10.1093/mnras/stt487}, \href
  {http://adsabs.harvard.edu/abs/2013MNRAS.432..506P} {432, 506}

\bibitem[\protect\citeauthoryear{{Ptak}, {Colbert}, {van der Marel}, {Roye},
  {Heckman}  \& {Towne}}{{Ptak} et~al.}{2006}]{2006ApJS..166..154P}
{Ptak} A.,  {Colbert} E.,  {van der Marel} R.~P.,  {Roye} E.,  {Heckman} T.,
  {Towne} B.,  2006, \mn@doi [\apjs] {10.1086/505218}, \href
  {http://adsabs.harvard.edu/abs/2006ApJS..166..154P} {166, 154}

\bibitem[\protect\citeauthoryear{{Qiu} et~al.,}{{Qiu}
  et~al.}{2019}]{2019ApJ...877...57Q}
{Qiu} Y.,  et~al., 2019, \mn@doi [The Astrophysical Journal]
  {10.3847/1538-4357/ab16e7}, \href
  {https://ui.adsabs.harvard.edu/abs/2019ApJ...877...57Q} {877, 57}

\bibitem[\protect\citeauthoryear{{Ramsey}, {Williams}, {Gruendl}, {Chen}, {Chu}
   \& {Wang}}{{Ramsey} et~al.}{2006}]{2006ApJ...641..241R}
{Ramsey} C.~J.,  {Williams} R.~M.,  {Gruendl} R.~A.,  {Chen} C.-H.~R.,  {Chu}
  Y.-H.,   {Wang} Q.~D.,  2006, \mn@doi [\apj] {10.1086/499070}, \href
  {https://ui.adsabs.harvard.edu/abs/2006ApJ...641..241R} {641, 241}

\bibitem[\protect\citeauthoryear{{Roberts}, {Warwick}, {Ward}  \&
  {Murray}}{{Roberts} et~al.}{2002}]{2002MNRAS.337..677R}
{Roberts} T.~P.,  {Warwick} R.~S.,  {Ward} M.~J.,   {Murray} S.~S.,  2002,
  \mn@doi [\mnras] {10.1046/j.1365-8711.2002.05950.x}, \href
  {http://cdsads.u-strasbg.fr/abs/2002MNRAS.337..677R} {337, 677}

\bibitem[\protect\citeauthoryear{{Roberts}, {Gladstone}, {Goulding},
  {Swinbank}, {Ward}, {Goad}  \& {Levan}}{{Roberts}
  et~al.}{2011}]{2011AN....332..398R}
{Roberts} T.~P.,  {Gladstone} J.~C.,  {Goulding} A.~D.,  {Swinbank} A.~M.,
  {Ward} M.~J.,  {Goad} M.~R.,   {Levan} A.~J.,  2011, \mn@doi [Astronomische
  Nachrichten] {10.1002/asna.201011508}, \href
  {https://ui.adsabs.harvard.edu/abs/2011AN....332..398R} {332, 398}

\bibitem[\protect\citeauthoryear{{Rodr{\'\i}guez Castillo}
  et~al.,}{{Rodr{\'\i}guez Castillo} et~al.}{2019}]{2019arXiv190604791R}
{Rodr{\'\i}guez Castillo} G.~A.,  et~al., 2019, arXiv e-prints, \href
  {https://ui.adsabs.harvard.edu/abs/2019arXiv190604791R} {p. arXiv:1906.04791}

\bibitem[\protect\citeauthoryear{{Sathyaprakash} et~al.,}{{Sathyaprakash}
  et~al.}{2019}]{2019MNRAS.488L..35S}
{Sathyaprakash} R.,  et~al., 2019, \mn@doi [\mnras] {10.1093/mnrasl/slz086},
  \href {https://ui.adsabs.harvard.edu/abs/2019MNRAS.488L..35S} {488, L35}

\bibitem[\protect\citeauthoryear{{Schirmer}}{{Schirmer}}{2013}]{2013ApJS..209...21S}
{Schirmer} M.,  2013, \mn@doi [\apjs] {10.1088/0067-0049/209/2/21}, \href
  {http://adsabs.harvard.edu/abs/2013ApJS..209...21S} {209, 21}

\bibitem[\protect\citeauthoryear{{Shortridge}}{{Shortridge}}{1993}]{1993ASPC...52..219S}
{Shortridge} K.,  1993, in {Hanisch} R.~J.,  {Brissenden} R.~J.~V.,   {Barnes}
  J.,  eds,  Astronomical Society of the Pacific Conference Series Vol. 52,
  Astronomical Data Analysis Software and Systems II. p.~219

\bibitem[\protect\citeauthoryear{{Sirianni} et~al.,}{{Sirianni}
  et~al.}{2005}]{2005PASP..117.1049S}
{Sirianni} M.,  et~al., 2005, \mn@doi [\pasp] {10.1086/444553}, \href
  {https://ui.adsabs.harvard.edu/abs/2005PASP..117.1049S} {117, 1049}

\bibitem[\protect\citeauthoryear{{Skrutskie} et~al.,}{{Skrutskie}
  et~al.}{2006}]{2006AJ....131.1163S}
{Skrutskie} M.~F.,  et~al., 2006, \mn@doi [\aj] {10.1086/498708}, \href
  {http://adsabs.harvard.edu/abs/2006AJ....131.1163S} {131, 1163}

\bibitem[\protect\citeauthoryear{{Smette} et~al.,}{{Smette}
  et~al.}{2015}]{2015A&A...576A..77S}
{Smette} A.,  et~al., 2015, \mn@doi [\aap] {10.1051/0004-6361/201423932}, \href
  {https://ui.adsabs.harvard.edu/abs/2015A&A...576A..77S} {576, A77}

\bibitem[\protect\citeauthoryear{{Stocke}, {Wang}, {Perlman}, {Donahue}  \&
  {Schachter}}{{Stocke} et~al.}{1995}]{1995AJ....109.1199S}
{Stocke} J.~T.,  {Wang} Q.~D.,  {Perlman} E.~S.,  {Donahue} M.~E.,
  {Schachter} J.~F.,  1995, \mn@doi [\aj] {10.1086/117352}, \href
  {https://ui.adsabs.harvard.edu/abs/1995AJ....109.1199S} {109, 1199}

\bibitem[\protect\citeauthoryear{{Swartz}, {Ghosh}, {Tennant}  \&
  {Wu}}{{Swartz} et~al.}{2004}]{2004ApJS..154..519S}
{Swartz} D.~A.,  {Ghosh} K.~K.,  {Tennant} A.~F.,   {Wu} K.,  2004, \mn@doi
  [\apjs] {10.1086/422842}, \href
  {https://ui.adsabs.harvard.edu/abs/2004ApJS..154..519S} {154, 519}

\bibitem[\protect\citeauthoryear{{The LIGO Scientific Collaboration}
  et~al.,}{{The LIGO Scientific Collaboration}
  et~al.}{2018}]{2018arXiv181112907T}
{The LIGO Scientific Collaboration} et~al., 2018, arXiv e-prints, \href
  {https://ui.adsabs.harvard.edu/abs/2018arXiv181112907T} {p. arXiv:1811.12907}

\bibitem[\protect\citeauthoryear{{Tikhonov}, {Lebedev}  \&
  {Galazutdinova}}{{Tikhonov} et~al.}{2015}]{2015AstL...41..239T}
{Tikhonov} N.~A.,  {Lebedev} V.~S.,   {Galazutdinova} O.~A.,  2015, \mn@doi
  [Astronomy Letters] {10.1134/S1063773715060080}, \href
  {http://adsabs.harvard.edu/abs/2015AstL...41..239T} {41, 239}

\bibitem[\protect\citeauthoryear{{Tully} et~al.,}{{Tully}
  et~al.}{2013}]{2013AJ....146...86T}
{Tully} R.~B.,  et~al., 2013, \mn@doi [\aj] {10.1088/0004-6256/146/4/86}, \href
  {http://cdsads.u-strasbg.fr/abs/2013AJ....146...86T} {146, 86}

\bibitem[\protect\citeauthoryear{{Vinokurov}, {Fabrika}  \&
  {Atapin}}{{Vinokurov} et~al.}{2018}]{2018ApJ...854..176V}
{Vinokurov} A.,  {Fabrika} S.,   {Atapin} K.,  2018, \mn@doi [The Astrophysical
  Journal] {10.3847/1538-4357/aaaa6c}, \href
  {https://ui.adsabs.harvard.edu/abs/2018ApJ...854..176V} {854, 176}

\bibitem[\protect\citeauthoryear{Wang, Liu, Qiu, Bai, Yang, Guo  \& Zhang}{Wang
  et~al.}{2016}]{Wang_2016}
Wang S.,  Liu J.,  Qiu Y.,  Bai Y.,  Yang H.,  Guo J.,   Zhang P.,  2016,
  \mn@doi [The Astrophysical Journal Supplement Series]
  {10.3847/0067-0049/224/2/40}, 224, 40

\bibitem[\protect\citeauthoryear{{Wenger} et~al.,}{{Wenger}
  et~al.}{2000}]{2000A&AS..143....9W}
{Wenger} M.,  et~al., 2000, \mn@doi [\aaps] {10.1051/aas:2000332}, \href
  {http://adsabs.harvard.edu/abs/2000A%26AS..143....9W} {143, 9}

\bibitem[\protect\citeauthoryear{{Wiktorowicz}, {Sobolewska}, {Lasota}  \&
  {Belczynski}}{{Wiktorowicz} et~al.}{2017}]{2017ApJ...846...17W}
{Wiktorowicz} G.,  {Sobolewska} M.,  {Lasota} J.-P.,   {Belczynski} K.,  2017,
  \mn@doi [\apj] {10.3847/1538-4357/aa821d}, \href
  {https://ui.adsabs.harvard.edu/abs/2017ApJ...846...17W} {846, 17}

\bibitem[\protect\citeauthoryear{Yuan, Liu  \& Xiang}{Yuan
  et~al.}{2013}]{10.1093/mnras/stt039}
Yuan H.~B.,  Liu X.~W.,   Xiang M.~S.,  2013, \mn@doi [Monthly Notices of the
  Royal Astronomical Society] {10.1093/mnras/stt039}, 430, 2188

\bibitem[\protect\citeauthoryear{{Zampieri} \& {Roberts}}{{Zampieri} \&
  {Roberts}}{2009}]{2009MNRAS.400..677Z}
{Zampieri} L.,  {Roberts} T.~P.,  2009, \mn@doi [Monthly Notices of the Royal
  Astronomical Society] {10.1111/j.1365-2966.2009.15509.x}, \href
  {https://ui.adsabs.harvard.edu/abs/2009MNRAS.400..677Z} {400, 677}

\bibitem[\protect\citeauthoryear{{Zampieri}, {Mucciarelli}, {Falomo}, {Kaaret},
  {Di Stefano}, {Turolla}, {Chieregato}  \& {Treves}}{{Zampieri}
  et~al.}{2004}]{2004ApJ...603..523Z}
{Zampieri} L.,  {Mucciarelli} P.,  {Falomo} R.,  {Kaaret} P.,  {Di Stefano} R.,
   {Turolla} R.,  {Chieregato} M.,   {Treves} A.,  2004, \mn@doi [\apj]
  {10.1086/381541}, \href
  {https://ui.adsabs.harvard.edu/abs/2004ApJ...603..523Z} {603, 523}

\makeatother
\end{thebibliography}

%%%%%%%%%%%%%%%%% APPENDICES %%%%%%%%%%%%%%%%%%%%%

\appendix

\section{Images}

\begin{figure*}
  \begin{minipage}{0.5\textwidth}
  \includegraphics[width=\textwidth]{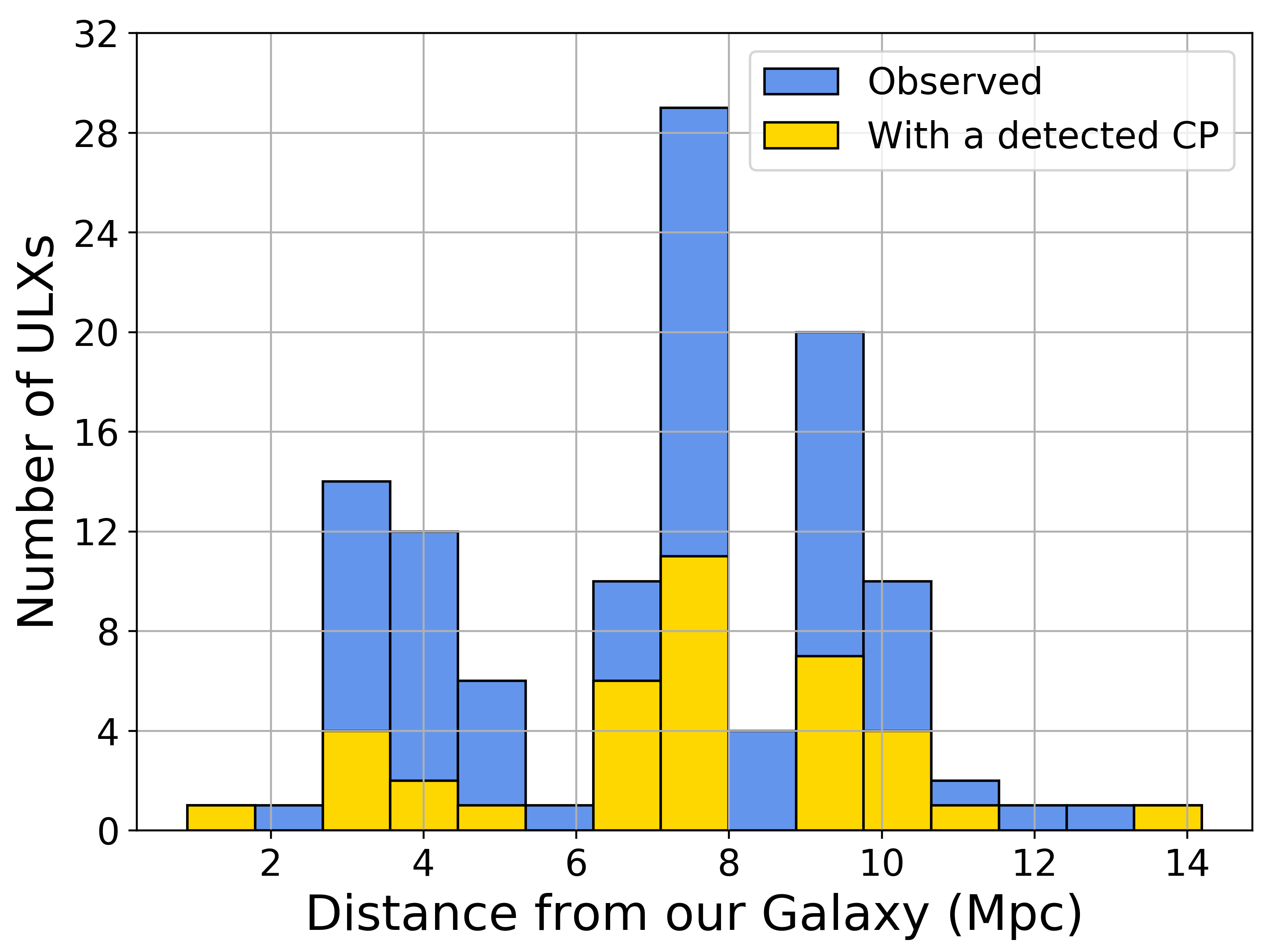}%
  \subcaption{} 
  \label{fig:dist}
  \vspace{0.5cm}   
 \end{minipage}%
 \begin{minipage}{0.5\textwidth}
  \includegraphics[width=\textwidth]{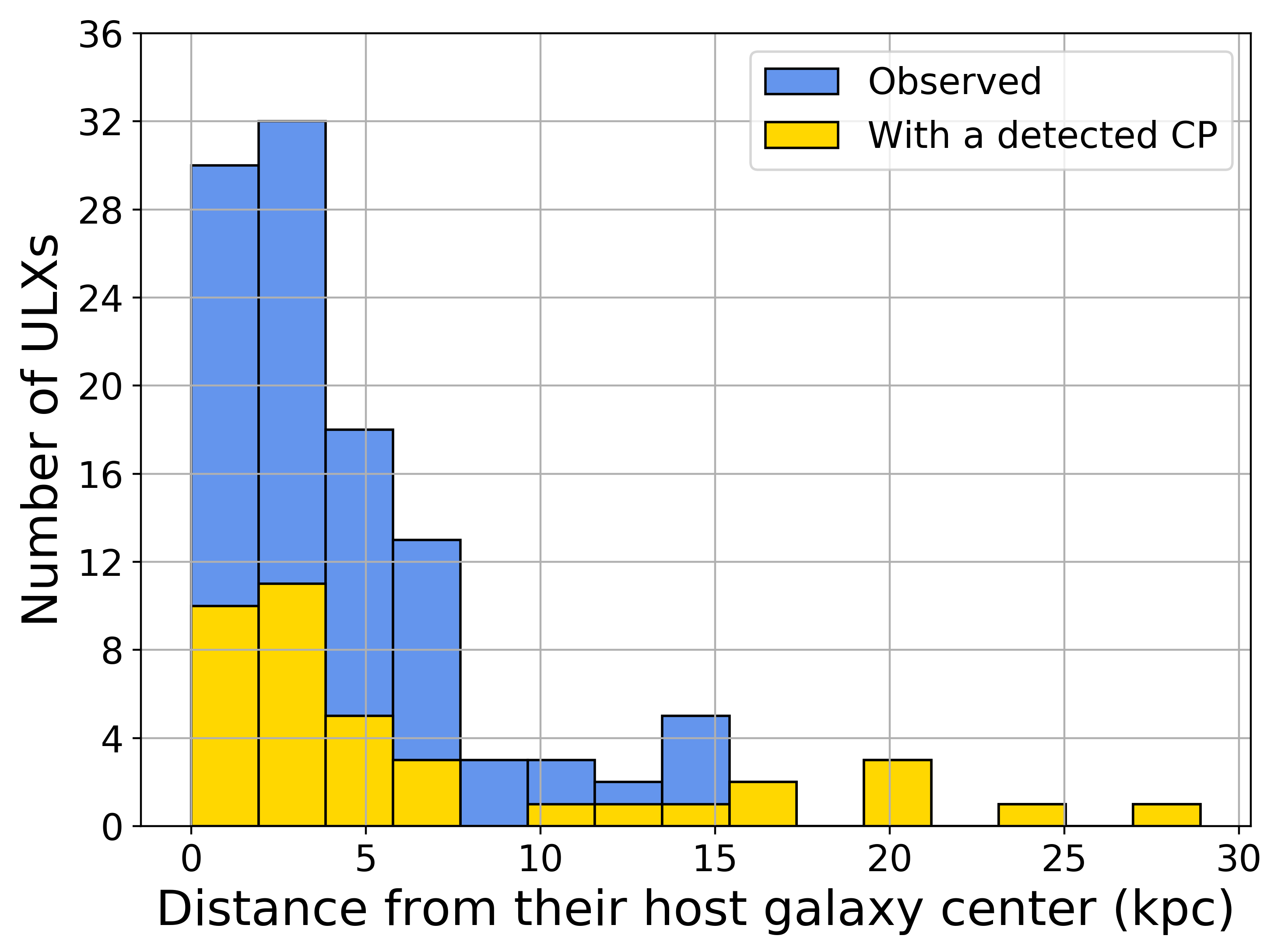}%
  \subcaption{}
  \label{fig:center} 
  \vspace{0.5cm}   
 \end{minipage}
\caption{(a) Histogram of the distribution of distances from our Galaxy to the 113 observed ULXs (in blue) and to the 38 ULXs for which we detected a NIR candidate counterpart (in yellow). A K--S test between both distributions reveals that they are not different. (b) Histogram of the distribution of distances to the 113 observed ULXs (in blue) and to the 38 ULXs for which we detected a NIR candidate counterpart (in yellow) from the centre of their respective host galaxies. A K--S test between both distributions reveals that they are not different.}
\label{fig:distances}
\end{figure*}

\begin{figure*}
  \begin{minipage}{0.5\textwidth}
  \includegraphics[width=\textwidth]{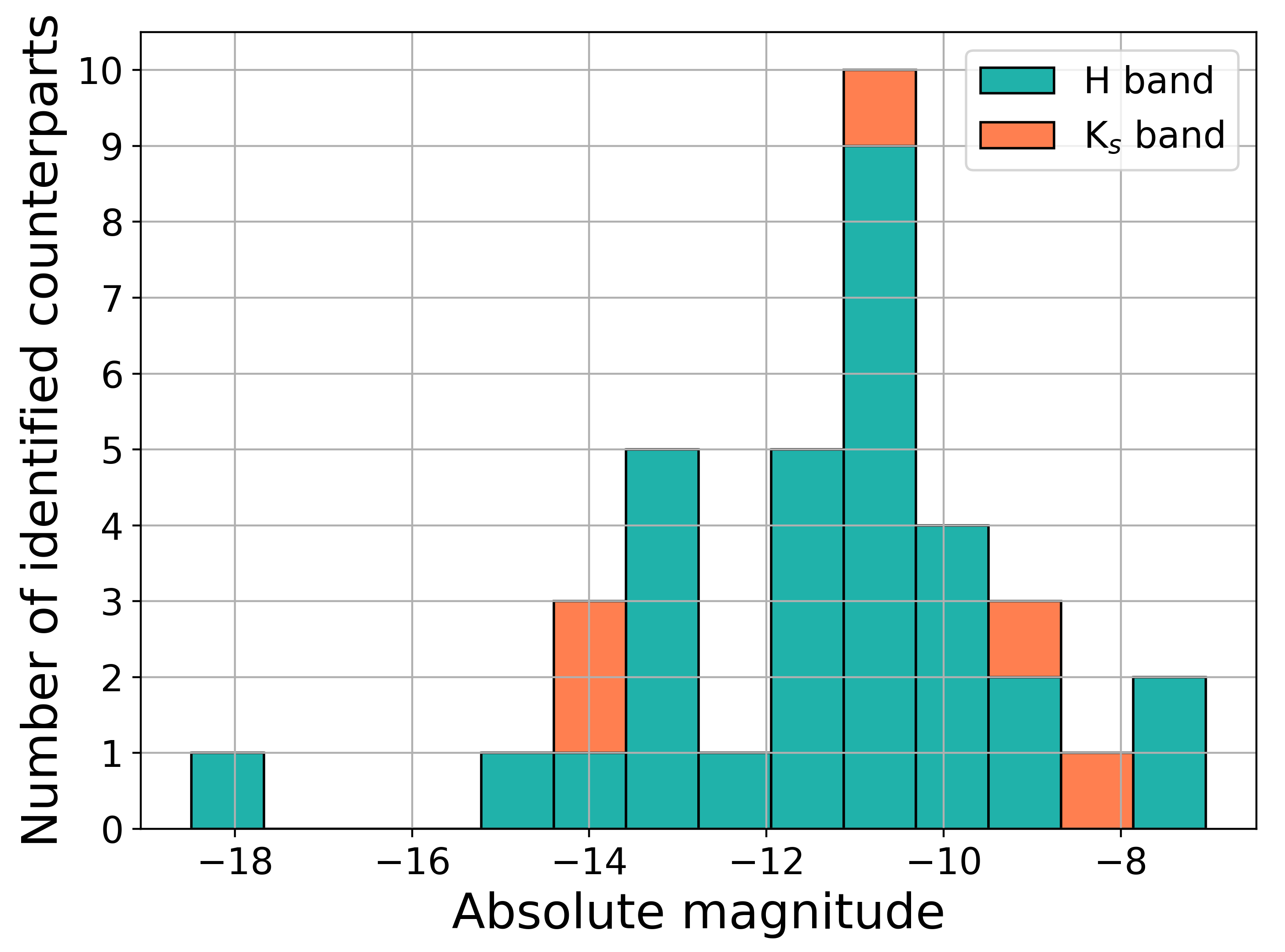}%
  \subcaption{} 
  \label{fig:histomags1}
  \vspace{0.5cm}   
 \end{minipage}%
  \begin{minipage}{0.5\textwidth}
  \includegraphics[width=\textwidth]{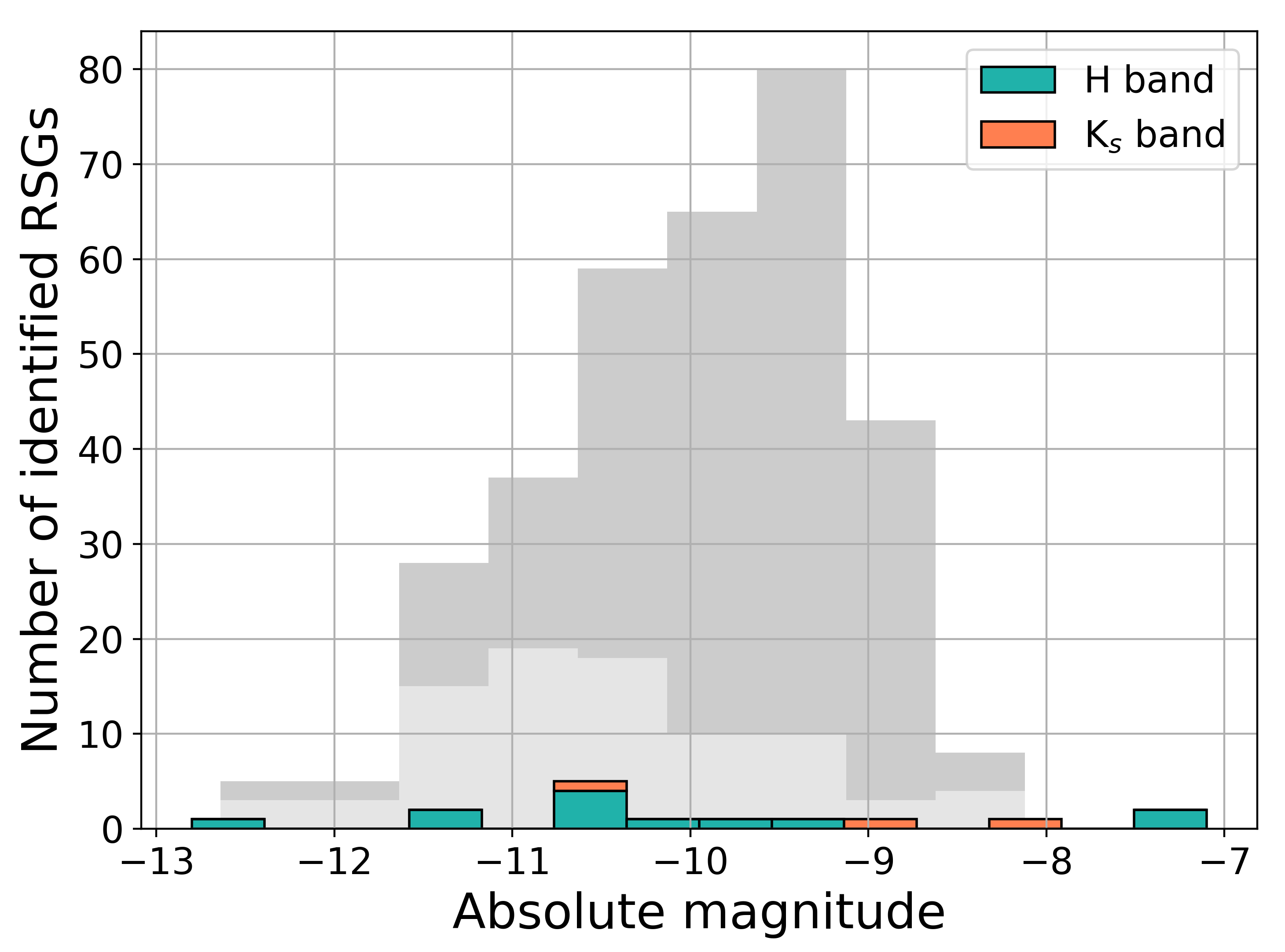}%
  \subcaption{}
  \label{fig:both} 
  \vspace{0.5cm}   
 \end{minipage}
\caption{Histogram of the distribution of absolute magnitudes of (a) the detected NIR candidate counterparts, indicating in which band they were observed and detected; and (b) the detected (candidates or confirmed) RSGs counterparts compared with the RSGs present in the MW and LMC (light grey), and in M31 (dark grey).}
  \label{fig:histomags}
\end{figure*}

\begin{figure*}
  \begin{minipage}{0.5\textwidth}
  \includegraphics[width=\textwidth]{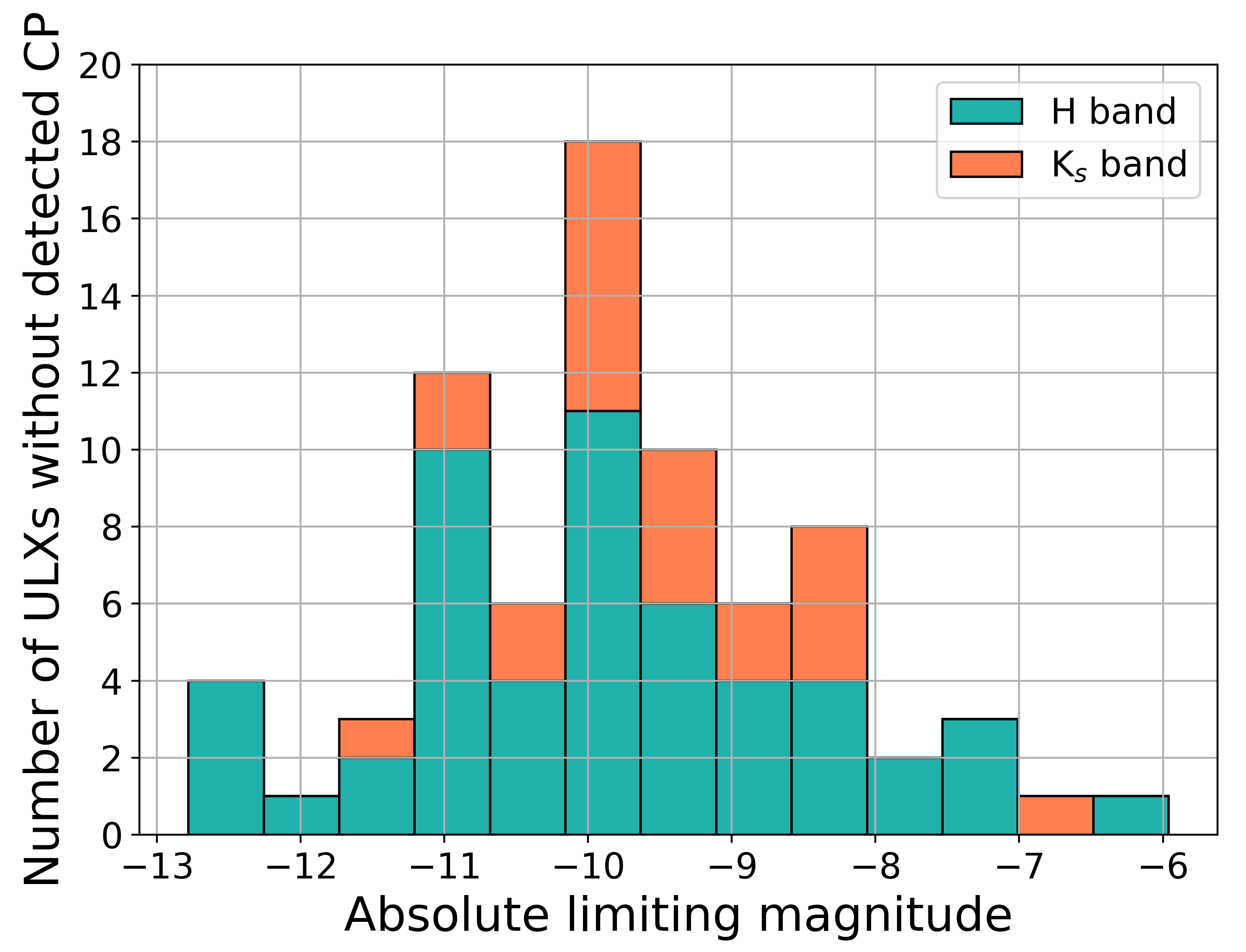}%
  \subcaption{} 
  \label{fig:limhistomags1}
  \vspace{0.5cm}   
 \end{minipage}%
 \begin{minipage}{0.5\textwidth}
  \includegraphics[width=\textwidth]{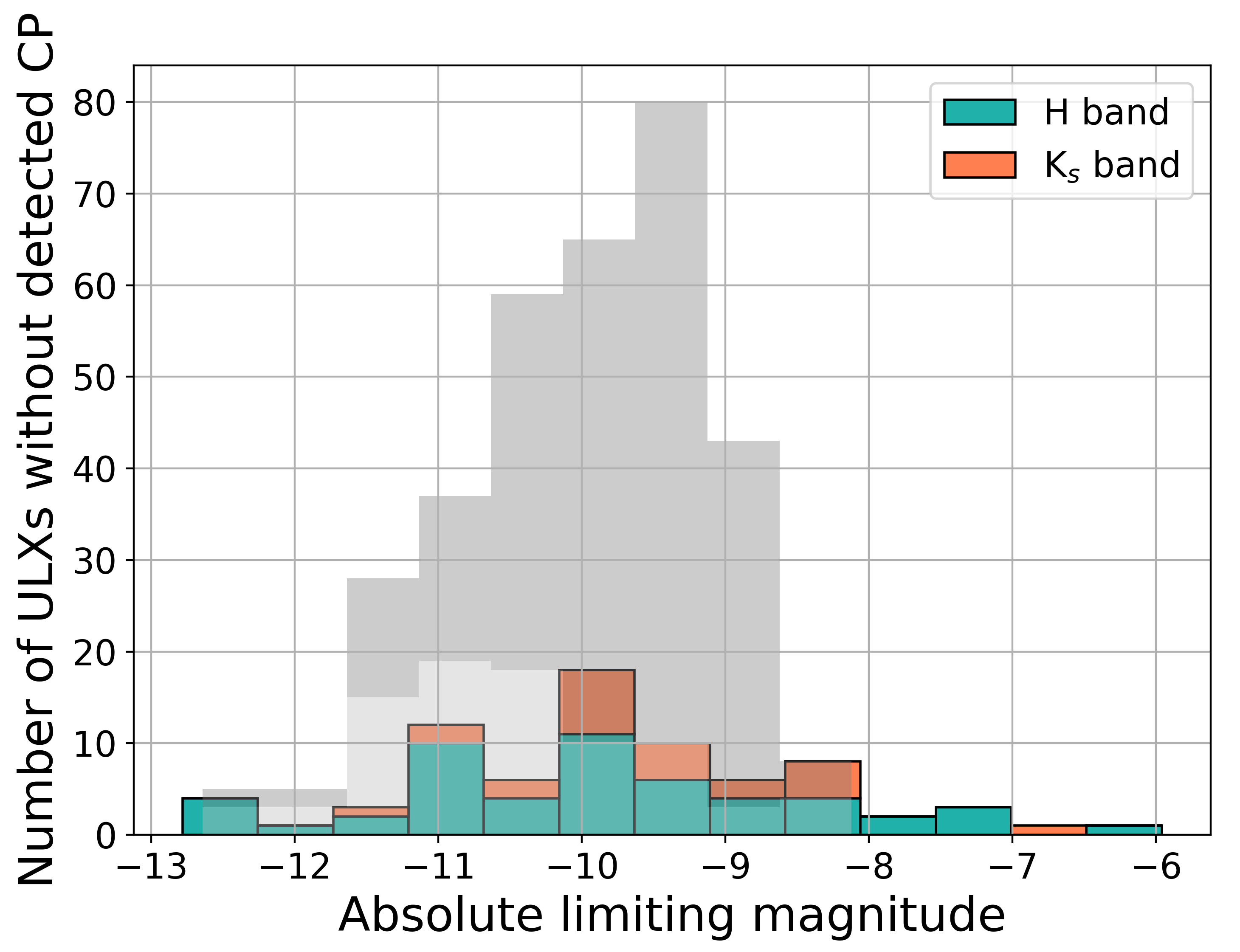}%
  \subcaption{}
  \label{fig:limhistomags2} 
  \vspace{0.5cm}   
 \end{minipage}
\caption{(a) Histogram of the distribution of limiting absolute magnitudes at the position of the ULXs for which we did not detect any counterpart, indicating in which band they were observed. (b) Same as (a) but compared with a histogram of the distribution of absolute magnitudes of the RSGs in the MW and LMC (light grey), and in M31 (dark grey).}
  \label{fig:limhistomags}
\end{figure*}

\section{Tables}

\begin{table*}
\vspace{5mm}
\begin{center}
\caption{Updated coordinates for the NIR candidate counterparts detected by \citet{2014MNRAS.442.1054H} and \citet{2017MNRAS.469..671L}, as they used 2MASS sources for the astrometric correction of their NIR images.. We improved the astrometry on those NIR images by finding a local solution around the position of the ULXs using isolated Gaia DR2 sources. We used point sources with a proper motion pm $< 5$ mas yr$^{-1}$, and, if necessary (i.e. not sufficient sources satisfying that constraint), pm $< 10$ mas yr$^{-1}$.}
\label{tab:allcp}
%\resizebox{\textwidth}{!}{
\begin{tabular}{|llccc|}
\hline\hline
Galaxy & ULX name & R.A. & Dec. & 3-$\sigma$ positional \\
 & (in SIMBAD) & (hh:mm:ss) & (dd:mm:ss) & uncertainty (\arcsec) \\
\hline\hline
NGC 253 & RX J$004722.4-252051$ & 00:47:22.6 & $-$25:20:51.3 & 1.5\\
NGC 925 & $[$SST2011$]$ J$022721.52+333500.7$ & 02:27:21.5 & +33:35:00.7 & 1.7\\
NGC 925 & $[$SST2011$]$ J$022727.53+333443.0$ & 02:27:27.6 & +33:34:43.4 & 1.7\\
NGC 1058 & XMMU J$024323.5+372038$ & 02:43:23.4 & +37:20:42.3 & 1.5\\
NGC 1637 & $[$IWL2003 68$]$ & 04:41:32.9 & $-$02:51:26.3 & 2.0\\
NGC 2403 & RX J$073655.7+653542$ & 07:36:55.4 & +65:35:41.6 & 0.7\\
NGC 2500 & CXO J$080157.8+504339$ & 08:01:57.9 & +50:43:39.1 & 1.4\\
Holmberg I & $[$WMR2006$]$ Ho I XMM1 & 09:41:30.2 & +71:12:35.6 & 0.6\\
Holmberg II & Holmberg II X-1 & 08:19:29.0 & +70:42:19.1 & 1.3\\
NGC 3521 & $[$SST2011$]$ J$110545.62+000016.2$ & 11:05:45.6 & +00:00:17.5 & 0.4\\
NGC 3627 & $[$SST2011$]$ J$112018.32+125900.8$ & 11:20:18.3 & +12:59:01.2 & 1.1\\
NGC 4136 & CXOU J$120922.6+295551$ & 12:09:22.6 & +29:55:50.5 & 1.8\\
NGC 4136 & $[$SST2011$]$ J$120922.18+295559.7$ & 12:09:22.1 & +29:55:59.0 & 1.5\\
NGC 4258 & RX J$121844.0+471730$ & 12:18:43.9 & +47:17:31.9 & 1.3\\
NGC 4258 & 3XMM J$121847.6+472054$ & 12:18:47.7 & +47:20:52.8 & 3.9\\
NGC 4258 & $[$LB2005$]$ NGC 4258 X9 & 12:19:23.3 & +47:09:40.7 & 1.1\\
NGC 4485 & RX J$1230.5+4141$ & 12:30:30.4 & +41:41:42.7 & 0.6\\
NGC 4490 & $[$SST2011$]$ J$123029.55+413927.6$ & 12:30:29.6 & +41:49:27.0 & 0.6\\
NGC 4490 & CXO J$123038.4+413831$ & 12:30:38.2 & +41:38:31.5 & 0.6\\
NGC 4490 & 2XMM J$123043.1+413819$ & 12:30:43.1 & +41:38:19.2 & 0.6\\
NGC 4559 & $[$SST2011$]$ J$123557.79+275807.4$ & 12:35:57.7 & +27:58:08.0 & 1.5\\
 & & 12:35:57.8 & +27:58:07.2 & 1.5\\
NGC 4559 & RX J$123558+27577$ & 12:35:58.6 & +27:57:41.7 & 1.5\\
NGC 4594 & $[$LB2005$]$ NGC 4594 X5 & 12:40:22.7 & $-$11:39:24.8 & 0.5\\
NGC 4631 & $[$SST2011$]$ J$124211.13+323235.9$ & 12:43:11.2 & +32:32:36.7 & 1.0\\
NGC 5194 & XMMU J$132953.3+471040$ & 13:29:53:3 & +47:10:42.9 & 2.5\\
NGC 5194 & RX J$132947+47096$ & 13:29:47.5 & +47:09:40.8 & 0.8\\
NGC 5408 & NGC 5408 X-1 & 14:03:19.7 & $-$41:22:58.7 & 	2.1\\
NGC 5457 & 2E $1402.4+5440$ & 14:04:14.3 & +54:26:02:4 & 1.4\\
NGC 5457 & 2XMM J$140248.0+541350$ & 14:02:48.2 & +54:13:50.4 & 1.3\\
NGC 5457 & CXOU J$140314.3+541807$ & 14:03:14.4 & +54:18:07.2 & 1.4\\
NGC 5457 & $[$LB2005$]$ NGC 5457 X32 & 14:02:28.2 & +54:16:26.3 & 0.8\\
NGC 5457 & $[$LB2005$]$ NGC 5457 X26 & 14:04:29.2 & +54:23:53.4 & 0.7\\
\hline\hline
\end{tabular}%}
\end{center}
\end{table*}

%%%%%%%%%%%%%%%%%%%%%%%%%%%%%%%%%%%%%%%%%%%%%%%%%%

% Don't change these lines
\bsp	% typesetting comment
\label{lastpage}
\end{document}